\documentclass[singlecolumn,floatfix]{revtex4-1}
\usepackage{amsfonts}
\usepackage{amsmath}
\usepackage{amssymb}
\usepackage{amsthm}
\usepackage{array}
\usepackage{epsfig}
\usepackage{rotating,graphicx}
\usepackage{stix}
\usepackage{bm}%
\usepackage{verbatim}
\usepackage{tensor}
\usepackage{xcolor}
\usepackage{tikz}
\usetikzlibrary{arrows,shapes,positioning}
\usetikzlibrary{decorations.markings}
\usetikzlibrary{arrows.meta,decorations.pathmorphing,backgrounds,fit,petri}
\usetikzlibrary{calc}
\tikzstyle arrowstyle=[scale=1]
\tikzstyle directed=[postaction={decorate,decoration={markings,
    mark=at position .7 with {\arrow[arrowstyle]{stealth}}}}]
\tikzstyle reverse directed=[postaction={decorate,decoration={markings,
    mark=at position .65 with {\arrowreversed[arrowstyle]{stealth};}}}]
\usepackage{longtable}
\usepackage{textgreek}
\usepackage{tikz-cd}
\usepackage{multirow}
\tikzcdset{
arrow style=tikz,
diagrams={>={Straight Barb[scale=0.8]}}
}
\theoremstyle{plain}
\usepackage{amsthm}

\theoremstyle{definition}

\newtheorem{exatitle}{Example}

\newenvironment{myexample}[2]%
{\begin{exatitle} \label{#2} #1 \end{exatitle}}%
{\hfill $\Box$ \\}

\newcommand{\ket}[1]{| #1 \rangle}
\newcommand{\bra}[1]{\langle #1 |}
\newcommand{\braket}[2]{\langle #1 | #2 \rangle}
\newcommand{\pket}[1]{[ #1 ]}
\newcommand{\ev}[1]{\langle #1 \rangle}
\newcommand{\td}{\text{d}}
\newcommand{\Tr}{\text{Tr}}

\newcommand{\eg}{\hbox{\em e.g.{}}}
\newcommand{\etc}{\hbox{\em etc.{}}}
\newcommand{\ie}{\hbox{\em i.e.{}}}
\newcommand{\wrt}{\hbox{w.r.t.{}}}
\newcommand{\lhs}{\hbox{l.h.s.{}}}
\newcommand{\rhs}{\hbox{r.h.s.{}}}

\newcommand{\Hs}{\mathcal{H}}

\newcommand{\rha}[1]{\rightharpoonaccent{#1}}
\newcommand{\ip}[2]{\langle #1|#2 \rangle}

\newcommand{\ipp}[2]{\left( #1,#2 \right)}

\makeatletter
\g@addto@macro\bfseries{\boldmath}
\makeatother

\newcommand{\Rr}{\mathsf{R}}
\newcommand{\al}{\alpha}
\def\nn{\nonumber}

\def \CP {{\cal P}}
\newcommand{\co}{\mathcal{C}}

\usepackage{bm}
\newcommand{\diff}{\mathop{}\!\mathrm{d}}
\newcommand{\Fc}{\mathcal{F}}
\newcommand{\da}{\dagger}
\begin{document}

\title{Symmetric Multiqudit States: Stars, Entanglement, Rotosensors}

\author{Chryssomalis Chryssomalakos}
\email{chryss@nucleares.unam.mx}
\affiliation{Instituto de Ciencias Nucleares,
	Universidad Nacional Aut\'onoma de M\'exico\\
	PO Box 70-543, 04510, CDMX, M\'exico.}

\author{Louis Hanotel}
\email{hanotel@correo.nucleares.unam.mx}
\affiliation{Instituto de Ciencias Nucleares,
	Universidad Nacional Aut\'onoma de M\'exico\\
	PO Box 70-543, 04510, CDMX, M\'exico.}

\author{Edgar Guzm\'an-Gonz\'alez}
\email{edgar.guzman@correo.nucleares.unam.mx}
\affiliation{Instituto de Ciencias Nucleares,
	Universidad Nacional Aut\'onoma de M\'exico\\
	PO Box 70-543, 04510, CDMX, M\'exico.}

\author{Daniel Braun}
\email{daniel.braun@uni-tuebingen.de}
\affiliation{Institut f\"ur Theoretische Physik,
Universit\"at T\"ubingen \\
72076 T\"ubingen Germany}

\author{Eduardo Serrano-Ens\'astiga}
\email{edensastiga@ens.cnyn.unam.mx}
\affiliation{Centro de Nanociencias y Nanotecnolog\'{\i}a,
Universidad Nacional Aut\'onoma de M\'exico\\
PO Box 14, 22800, Ensenada, Baja California, M\'exico
}

\author{Karol \.Zyczkowski}
\email{karol.zyczkowski@uj.edu.pl}
\affiliation{Institute of Theoretical Physics,
Jagiellonian University\\
30-348 Krak\'ow, Poland and\\
Center for Theoretical Physics,
Polish Academy of Sciences\\
02-668 Warsaw, Poland
}

\begin{abstract}
	\noindent
A constellation of $N=d-1$ Majorana stars represents an arbitrary pure quantum state of dimension $d$ or a permutation-symmetric state of a system consisting of $n$ qubits. We generalize the latter construction to represent in a similar way an arbitrary symmetric pure state of $k$ subsystems with $d$ levels each. For $d\geq 3$, such states are equivalent, as far as rotations are concerned, to a collection of various spin states, with definite relative complex weights. Following Majorana's lead,  we introduce a multiconstellation, consisting of the Majorana constellations of  the above spin states, augmented by an auxiliary, ``spectator'' constellation, encoding the complex weights. Examples of stellar representations of symmetric states of four qutrits, and two spin-3/2 systems, are presented.  We revisit the Hermite and Murnaghan isomorphisms, which relate multipartite states of various spins, number of parties, and even symmetries. We show how the tools introduced can be used to analyze multipartite entanglement and to identify optimal quantum rotosensors, \ie,  pure states which are maximally sensitive to rotations around a specified axis, or averaged over all axes. 
\end{abstract}

\maketitle

\tableofcontents
\newsavebox{\Potb}
\section{Introduction}
\label{Intro}
Quantum states form basic mathematical tools of quantum mechanics, 
used to calculate the probability of a given outcome of any measurement.
A pure spin-$s$ quantum state is represented by a vector 
in the  complex $d=2s+1$-dimensional Hilbert space. 
Any non-zero state is assumed to be normalized,
$|\psi|^2=\braket{\psi}{\psi} =1$,
so the set of all pure states in the Hilbert space forms a unit sphere.
As the global phase is not measurable,
a \emph{physical} state is defined as an equivalence class
with respect to multiplication by a complex phase,
$\ket{\psi} \sim \ket{\psi'}$ if $\ket{\psi'}=e^{i \varphi} \ket{\psi}$.
With this identification, physical  pure states form a complex projective space,
 $\mathbb{C}P^{d-1} \equiv \mathbb{P}$ (we omit the superscript denoting dimension when no confusion arises), and each such state is 
 determined by a collection of $d-1$ complex numbers.
In the simplest case,  $d=2$, 
the set of all (physical) pure quantum states forms the Bloch sphere 
$S^2 \sim \mathbb{C}P^{1}$,
 so one-qubit pure states can be visualized as points on the sphere.
 
Any pure state can be represented in an orthonormal basis,
$|\psi\rangle = \sum_{i=0}^{d-1} c_i |i\rangle$.
The normalization condition implies that $\sum_{i=0}^{d-1} |c_i|^2=1$
and the first nonzero component can be taken real and positive,
due to the freedom of the global phase $\varphi$ --- this leaves $2d-2$ real coordinates to be specified. However, 
representing $|\psi\rangle$ by this set of variables 
is not always particularly illuminating. It is often more convenient to apply
 Majorana's \emph{stellar representation} \cite{Maj:32}, in which any state of size $d$
is encoded by $d-1$ indistinguishable points on the unit sphere, that rotate rigidly when the state is transformed by an $SU(2)$ group element. 
This notion was popularized by Roger Penrose \cite{Pen.Rin:90},
 who discussed it also in his books dedicated to a larger audience \cite{Pen:90,Pen:94,Pen:07}. 
 It might be interesting to know that,
independently of his  seminal papers on
general relativity, the structure of the universe,
 the formation of black holes,  and twistor theory,
the Nobel laureate occupied himself working  
on the problem of the stellar representation of two orthogonal pure states~\cite{Pen:up}.

A particular, degenerate constellation, in which all stars
coalesce in a single point on the sphere corresponds to
\emph{spin coherent} states \cite{Per:86,Rad:71,Are.Cou.Gil.Tho:72,Gil:08,Chr.Guz.Ser:18}.
These states, of a given dimension $d$, labeled by a point on the sphere,  
$|\theta,  \phi\rangle$,
can be obtained by the action of
the Wigner rotation matrix on the maximal weight (\ie, maximal $S_z$-eigenvalue) state,
so they are also called $SU(2)$-coherent states.
Projectors on spin coherent states
satisfy the following identity resolution,
$\frac{1}{4 \pi}\int_{S^2} |\theta,  \varphi\rangle \langle \theta,  \varphi |
d \Omega ={\mathbb 1}/d$,
where $d \Omega$ stands for the area form on the unit sphere.
Due to this relation
any pure state $|\psi\rangle \in {\cal H}_s \equiv \mathbb{C}^d$, $s = (d-1)/2$
(note that the subscript of $\mathcal{H}$ denotes the value $s$ of spin, not the dimension $d$),
can be represented by its overlap with coherent states,
$Q_{\psi}(\theta,\varphi)= | \langle \psi |  \theta,  \varphi \rangle |^2$,
also called its {\sl Husimi function}, furnishing what is known as the  \emph{Q--representation} of $\ket{\psi}$ (note that $Q_\psi$ only retains the information about the modulus of $\braket{\theta,\phi}{\psi}$, but this is enough to recover $\ket{\psi}$, up to an overall phase).

It is easy to show that for a generic pure state  $|\psi\rangle \in \mathcal{H}_s$,
the function   $f(z)= \braket{\psi}{\theta,\phi}$, considered as a function of the complex
variable $z=\tan (\theta/2) e^{i \phi}$,
is given, apart from a positive overall factor, by a polynomial of order $d-1$.
The $d-1$, possibly coincident, complex
zeros of this polynomial, projected on a sphere, determine the above mentioned 
stellar representation of $\ket{\psi}$ --- in fact, the zeros project to the antipodes of the stars of the Majorana constellation.
In particular, if the  state $|\psi\rangle$ is spin-coherent,
pointing in the direction $(\theta,  \phi)$,
all the (coincident) zeros project to the antipodal point $(\pi-\theta,\phi+\pi \mod 2\pi)$
 on the sphere.  Such states are characterized by the fact that they maximize the modulus of their spin expectation value. 
Naturally, one expects that a state represented by a
configuration of stars clustered close to a given point on the sphere
is close to be coherent, in an appropriate sense.

On the other hand, one may define a random pure state of size $d$ 
 by a configuration of  
$d-1$ randomly chosen  points on the sphere \cite{Leb:91,Bog.Boh.Leb:92,Han:96}.
The modulus of the spin expectation value of such states is, generally, much smaller than the maximal value $s$. In the special case where this modulus vanishes, so that the state does not ``point'' in any particular direction in space, one talks of \emph{anticoherent} states~\cite{Zim:06} --- these are often represented by particularly symmetric 
configurations of points on the sphere, and tend to lie far away from the subset of coherent states, in the standard Fubini-Study metric on $\mathbb{P}$~\cite{Bag.Dam.Gir.Mar:15,Bag.Mar:17}.
Since spin-coherent states display several classical properties \cite{Gir.Bra.Bra:08},
their anticoherent  cousins are often considered as key
examples of states with exclusively quantum properties 
\cite{Boh.Bra.Gir:16},
and appropriate refinements  of the concept have been termed  \emph{queens of quantum} and \emph{kings of quantum} 
\cite{Gir.Bra.Bra:10,Bjo.Kli.Hoz.Gra.Leu.San:15,Gol.Kli.Gra.Leu.San:20}.

Interestingly, a given constellation of $N$ stars on the sphere
can represent a pure state of a monopartite spin-$s$ system of size $d=2s+1=N+1$,
or a permutation symmetric state of a system composed of $N$ qubits 
 \cite{Mar:11,Bjo.Gra.Hoz.Leu.San:15,Mat.Kri.God.Lam.Sol.Bas:10,Ben.Zyc:17}.
The spin-coherent state in ${\cal H}_s$
 corresponds then to a separable state in ${\cal H}_{1/2}^{\otimes N}$.
At the other extreme, anticoherent states represented by stars 
forming a particular geometric design on the sphere, 
correspond to highly entangled multipartite state of $N$ qubits
\cite{Mar.Gir.Bra.Bra.Bas:10,Aul.Mar.Mur:10,Aul:12,Bag.Mar:17}.
In the simplest case of $N=2$ two stars, placed at antipodal poles 
of the sphere,
one gets  the eigenstate $|j,m\rangle=|1,0\rangle \in {\cal H}_1$
 of the spin operators $\mathbf{S}^2$, $S_z$,
 or the maximally entangled,  symmetric Bell state 
 $|\phi_+\rangle=(|01\rangle +|10\rangle)/\sqrt{2}$ 
 of a two-qubit system.

The stellar representation also allows one to define a simplified Monge distance
between two quantum states, proportional to the minimal total 
distance needed to shift the constellation of stars representing the
first state into the second one \cite{Zyc.Slo:01,Ben.Zyc:17}. 
Furthermore,
the distance of the barycenter of a constellation of  $N$
stars from the center of the sphere, 
can be used as a simple measure of coherence of a
state of a simple system of size $N+1$,
or as an inverse measure of entanglement of a 
symmetric state of $N$ qubits \cite{Gan.Kus.Zyc:12}.

The stellar representation of Majorana, originally designed for pure states 
 of a monopartite system \cite{Maj:32}, 
 can be also generalized for mixed quantum states
\cite{Ser.Bra:20}.
In the case of multipartite systems, the stellar representation
is directly applicable to symmetric states of a system composed of several qubits,
but a generalization applicable to antisymmetric states 
 has also been established \cite{Chr.Guz.Han.Ser:21}.
 
The first aim of this work is to extend the stellar representation
to the case of permutation-symmetric states of a system consisting of $k$ identical subsystems with $d$ levels each.
A symmetrized product of $2s$ qubit states transforms like an irreducible spin-$s$ state, when each factor state is transformed by the same $SU(2)$ element. In the case of symmetrized products of states with spin higher than 1/2, the situation gets more complicated, as the states available to the system separate in several irreducible components, of distinct, in general spins. We are thus led to represent such states by a multiconstellation, consisting of the Majorana constellations of the above irreducible components, augmented by a spectator constellation, which carries the information of their relative (complex) weights. The construction provides a 1-to-1 mapping between such states and multiconstellations, maintaining the desirable ``covariance'' property of the original Majorana representation, as the multiconstellation rotates rigidly on the sphere, when the individual parts are transformed in the spin-$s$ representation of $SU(2)$. 

Our second aim in this work is to revisit several aspects of permutation-symmetric spin-$s$ states, and see what new insights can be had, from our novel point of view. Thus, we explore the geometrical measure of entanglement of a class of such states, and find an intriguing correlation with the eigenvalues of the corresponding Gram matrix. We also examine various concepts of quantum rotosensors, and identify the corresponding optimal symmetric states. 

This paper is organized as follows. In Section~\ref{kpMvtPE} we analyze $k$-symmetric states of spin $s$ and introduce the notion of a multiconstellation, generalizing Majorana's construction.
Some illustrative examples of stellar representation of symmetric states
of systems of several qutrits are presented in Section~\ref{Acoe}.
An algebraic approach to the problem is presented in Section~\ref{AclatHr},
which includes discussion
of the Hermite and  the Murnaghan isomorphisms \cite{Mur:38,Mur:51}.
Possible applications of the representation proposed
 include investigations of measures of entanglement of symmetric states of 
multipartite systems \cite{Wei.Gol:03,Hub.Kle.Wei.Gon.Guh:09,Aul.Mar.Mur:10,Mar.Gir.Bra.Bra.Bas:10,Enr.Win.Zyc:16}
and designing optimal quantum rotosensors
\cite{Chr.Her:17,Mar.Wei.Gir:19} --- these are discussed in Sections~\ref{Gmoefvfs} and \ref{Sqr}, respectively. The final Section~\ref{Conc} summarizes our findings.
\vskip 0.4cm

\section{$k$-symmetric spin-$s$ states and their multiconstellations}
\label{kpMvtPE}
\subsection{The $SU(2)$ action on $\mathcal{H}^{\vee k}$}
\label{TSao}
A spin-$s$ quantum state $\ket{\psi}$ lives in the Hilbert space $\mathcal{H}_s=\mathbb{C}^{d}$, $d \equiv N+1 \equiv 2s+1$ (we drop the index $s$ when no confusion arises). Its image in the projective space $\mathbb{P} \equiv \mathbb{C}P^N$ will be denoted by
$\pket{\psi}$.
Consider  the $k$-th symmetric power of $\mathcal{H}$, $\vee^k \mathcal{H} \equiv \mathcal{H}^{\vee k}$, which, given a basis $\{e_1,\ldots,e_{\tilde{N}}\}$ of $\mathcal{H}$, inherits naturally the \emph{induced} basis $\{ e_{\rha{A}} =e_{a_1}\vee \ldots \vee e_{a_k} \}$ , $a_1 \leq a_2 \leq \ldots \leq a_k$, where the symmetrized tensor product $\vee$ is defined by
\begin{equation}
\label{symmtp}
v_1 \vee v_2 \vee \ldots \vee v_k
=
\frac{1}{k!} \sum_{\sigma \in S_k}
v_{\sigma(1)} \otimes \ldots \otimes v_{\sigma(k)}
\, ;
\end{equation}
it is easily shown that  $\text{dim}\mathcal{H}^{\vee k}=\binom{d+k-1}{k}$.

The Hilbert space inner product $\ip{\cdot}{\cdot}$ in $\mathcal{H}$ induces one in $\mathcal{H}^{\otimes k}$,
\begin{equation}
\label{ipinduced}
\ip{v_1 \otimes \ldots \otimes v_k}{w_1 \otimes \ldots \otimes w_k}=\ip{v_1}{w_1} \ldots \ip{v_k}{w_k}
\, ,
\end{equation}
which descends to $\mathcal{H}^{\vee k}$,
\begin{align}
\ip{v_1 \vee \ldots \vee v_k}{w_1 \vee \ldots \vee w_k}
&=
\frac{1}{(k!)^2}
\sum_{\sigma,\sigma' \in S_k}
\ip{
v_{\sigma(1)} \otimes \ldots \otimes v_{\sigma(k)}
}{
w_{\sigma'(1)}  \otimes \ldots \otimes w_{\sigma'(k)}}
\nonumber
\\
&=
\frac{1}{k!}
\sum_{\sigma \in S_k}
\ip{
v_1 \otimes \ldots \otimes v_k
}{
w_{\sigma(1)}  \otimes \ldots \otimes w_{\sigma(k)}}
\nonumber
\\
&=
\frac{1}{k!}
\sum_{\sigma \in S_k}
\ip{v_1}{w_{\sigma(1)}} \ldots \ip{v_k}{w_{\sigma(k)}}
\label{ipinduced2}
\\
&=
\frac{1}{k!}
\,
\text{perm}\left( (\ip{v_i}{w_j}) \right)
\nonumber
\, .
\end{align}
 If $\{e_i \}$, $i=1,\ldots,d$, is an orthonormal basis in $\mathcal{H}$, then the induced basis $\{e_{\vec{A}} \}$ mentioned above is orthogonal but not orthonormal, when $k \geq 2$, in the induced inner product of~(\ref{ipinduced2}). For a $k$-multiindex $\vec{A}=(a_1,\ldots,a_k)$, denote by $\{A_1,\ldots,A_r\}$ the associated partition of $k$, where $A_i$ is the number of times the $i$-th distinct index in $\vec{A}$ repeats itself (\eg, for $k=7$ and
 $\vec{A}=(1,1,2,2,2,2,3)$, we have $r=3$ (because there are three distinct indices in $\vec{A}$) and the corresponding partition of 7 is $\{A_1,A_2,A_3\}=\{2,4,1\}$). Then an orthonormal \emph{standard} basis is given by
 $\hat{e}_{\vec{A}}= \sqrt{\frac{k!}{A_1! \ldots A_r!}} e_{\vec{A}}$. We will have use for both bases in the examples we consider, so we adopt the following notation to distinguish the corresponding components,
 \begin{equation}
 \label{Psiexpbasis}
 \ket{\boldsymbol{\Psi}}
 =\sum_{\rha{I}}\Psi^{\rha{I}} e_{\rha{I}}
=\sum_{\rha{A}}\hat{\Psi}^{\rha{A}} \hat{e}_{\vec{A}}
 \, .
 \end{equation}
 Note that the multiindices $\vec{I}$, $\vec{A}$, range over the same values --- the two bases are only differentiated by the use of $\hat{\cdot}$ for both the components and basis vectors of the orthonormal (standard) one (for $k=1$ the two bases coincide).
 
$g \in SU(2)$ acts on $\mathcal{H}$ via
\begin{equation}
\label{gactH}
\left( g \triangleright \ket{\psi} \right)_\text{S}= D^{(s)}(g)\ket{\psi}_\text{S}
\, ,
\end{equation}
where $D^{(s)}(g)$ is its $d$-dimensional irreducible representation (irrep), with rotations around $\hat{z}$ represented by diagonal matrices, and $\ket{\psi}_\text{S}$ is the column vector of the components of $\ket{\psi}$ in the standard basis of $S^2$, $S_z$ eigenvectors. This is an appropriate point to establish our notation: we reserve the ket symbol, as in $\ket{\psi}$, for the abstract vector in Hilbert space, and add to it a subscript in capitals, as in $\ket{\psi}_\text{I}$, or $\ket{\psi}_\text{S}$,  to refer to the column vector of its components in a particular basis (in this case, the Induced $\{e_{\vec{I}}\}$, or Standard one $\{\hat{e}_{\vec{A}}\}$, respectively). We use the same subscripts, with the same meaning, in representation matrices, but omit them in some cases --- see below.

 The above action of $SU(2)$ on $\mathcal{H}$ extends naturally to
$\mathcal{H}^{\otimes k}$,
\begin{equation}
\label{reptensork}
\left( g \triangleright  \ket{\psi_1} \otimes \ldots \otimes \ket{\psi_k} \right)_\text{S} = D^{(s)}(g) \ket{\psi_1}_\text{S} \otimes \ldots \otimes D^{(s)}(g) \ket{\psi_k}_\text{S}
\, .
\end{equation}
The totally symmetric subspace $\mathcal{H}^{\vee k} \subset \mathcal{H}^{\otimes k}$ is invariant under this action, resulting in an induced representation,
\begin{equation}
\label{repwedgek}
\left( g \triangleright  \ket{\psi_1} \vee \ldots \vee \ket{\psi_k} \right)_\text{S} = D^{(s)}(g) \ket{\psi_1}_\text{S} \vee \ldots \vee D^{(s)}(g) \ket{\psi_k}_\text{S}
\equiv
\mathfrak{D}^{(s,k)}_\text{S}(g) \ket{\boldsymbol{\Psi}}_\text{S}
\, ,
\end{equation}
where  $\ket{\boldsymbol{\Psi}}_\text{S}$ on the right hand side stands for the column vector of the components of
$\ket{\psi_1}\vee \ldots \vee \ket{\psi_n}$ in the $\hat{e}_{\vec{A}}$-basis, and $\mathfrak{D}^{(s,k)}_\text{S}(g)$ is the unitary matrix representing $g$ on $\mathcal{H}^{\vee k}$, in the standard basis. This representation may be  block-diagonalized by a suitable change of basis in $\mathcal{H}^{\vee k}$. Its block-diagonal (BD) form, which we denote by $\mathfrak{D}^{(s,k)}_\text{BD}$, contains $\mu^{(s,k)}_j$ copies of $D^{(j)}$, $j=s_\text{min}, \ldots, s_{\text{max}}$,
\begin{equation}
\label{Dskmulti}
\mathfrak{D}^{(s,k)}_\text{BD}=\bigoplus_{j=s_\text{min}}^{s_{\text{max}}} \mu^{(s,k)}_j D^{(j)}
\, ,
\end{equation}
where
\begin{equation}
\label{smincs}
s_\text{min}=
\begin{cases}
0 & \text{if } k \text{ even}
\\
\frac{1}{2} & \text{if } s \text{ half integral,} \, \, k \text{ odd}
\end{cases}
\, ,
\end{equation}
with the lower bound for $j$ not necessarily saturated (see Ex.~\ref{seq1sp}).
Note that, when $k=1$, \ie, for the original Hilbert space $\mathcal{H}$, the standard basis results in $D^{(s)}_\text{S}$ being automatically in BD form (with a single irreducible block), so we omit the S (or BD) subscript in this case, as in~(\ref{gactH}). We extend these conventions to the representation matrices of the Lie algebra $\mathfrak{su}(2)$ below.
We turn now to the determination of $s_{\text{max}}$ and the multiplicities $\mu^{(s,k)}_j$.
\subsection{Irreducible components}
\label{Ic}
\subsubsection{Elementary considerations}
\label{Ecs}
Consider the abstract $SU(2)$ element $g=e^{-i t \hat{n} \cdot \mathbf{S}}$, where $S_a$, $a=1,2,3$,  (or $x$, $y$, $z$) are the
$\mathfrak{su}(2)$ Lie algebra generators, satisfying
\begin{equation}
\label{su2rels}
[S_a,S_b]=i\epsilon_{abc} S_c
\, .
\end{equation}
Corresponding to the action of $g$ on $\mathcal{H}$, $\mathcal{H}^{\otimes k}$, and $\mathcal{H}^{\vee k}$, given in~(\ref{gactH}), (\ref{reptensork}), and (\ref{repwedgek}), respectively, we have the following action of the generators,
\begin{align}
\left( S_a \triangleright \ket{\psi} \right)_\text{S}
&=
S_a^{(s)}\ket{\psi}_\text{S}
\\
\left( S_a \triangleright  \ket{\psi_1} \otimes \ldots \otimes \ket{\psi_k} \right)_\text{S}
&=
\sum_{m=1}^k
\ket{\psi_1}_\text{S} \otimes \ldots \otimes S_a^{(s)}\ket{\psi_m}_\text{S} \otimes \ldots \otimes \ket{\psi_k}_\text{S}
\\
\left( S_a \triangleright  \ket{\psi_1} \vee \ldots \vee \ket{\psi_k} \right)_\text{S}
&=
\sum_{m=1}^k
\ket{\psi_1}_\text{S} \vee \ldots \vee S_a^{(s)}\ket{\psi_m}_\text{S} \vee \ldots \vee \ket{\psi_k}_\text{S}
\label{SLeibnizvee}
\\
&\equiv
S_{a\, \text{S}}^{(s,k)} \ket{\boldsymbol{\Psi}}_\text{S}
\nonumber
\, ,
\end{align}
with $\ket{\boldsymbol{\Psi}}_\text{S}$ as in~(\ref{repwedgek}), and where $S_a^{(s)}$, $S_{a \, \text{S}}^{(s,k)}$ are hermitean matrices of dimension $\text{dim}\mathcal{H}$, $\text{dim} \mathcal{H}^{\vee k}$, respectively, satisfying~(\ref{su2rels}), and related to the corresponding representations of $g$ via
\begin{alignat}{2}
\label{Sskdef}
D^{(s)}(g)
&=
e^{-i t \hat{n} \cdot \mathbf{S}^{(s)}}
\, ,
\qquad
&
\hat{n} \cdot \mathbf{S}^{(s)}
&=
\left.
i \frac{\partial}{\partial t} D^{(s)}(g)\right|_{t=0}
\\
\mathfrak{D}^{(s,k)}_\text{S}(g)
&=
e^{-i t \hat{n} \cdot \mathbf{S}_\text{S}^{(s,k)}}
\, ,
\qquad
&
\hat{n} \cdot \mathbf{S}^{(s,k)}_\text{S}
&=
\left.
i \frac{\partial}{\partial t} \mathfrak{D}^{(s,k)}_\text{S}(g)\right|_{t=0}
\, .
\end{alignat}
As a result of~(\ref{SLeibnizvee}),
\begin{equation}
\label{Szev}
\left( S_z \triangleright
\ket{s,m_1} \vee \ldots \vee \ket{s,m_k}\right)_\text{S}
=
(m_1+\ldots+m_k) \left(
\ket{s,m_1} \vee \ldots \vee \ket{s,m_k}\right)_\text{S}
\, ,
\end{equation}
\ie, a symmetric product of  $S^{(s)}_z$ eigenvectors is an $S^{(s,k)}_z$ eigenvector, with eigenvalue equal to the sum of the eigenvalues of the factors. The maximal possible $S^{(s,k)}_z$-eigenvalue, which is also the maximal value $s_\text{max}$ of the spin $j$ in the decomposition of $\mathfrak{D}_\text{BD}^{(s,k)}$ in~(\ref{Dskmulti}), is clearly attained by the ``top'' spin-$s$ $k$-symmetric product $\ket{s,s} \vee \ldots \vee \ket{s,s}\equiv \ket{s_\text{max},s_\text{max}}$, satisfying
\begin{equation}
\label{smaxeqs}
(\mathbf{S}^{(s,k)})^2 \ket{s_\text{max},s_\text{max}}=s_\text{max}(s_\text{max}+1) \ket{s_\text{max},s_\text{max}}
\, ,
\qquad
S_z^{(s,k)} \ket{s_\text{max},s_\text{max}}=s_\text{max} \ket{s_\text{max},s_\text{max}}
\, ,
\end{equation}
with $s_\text{max}=ks$. Applying to this ket the lowering operator $S_-$ one obtains the (projectively) unique ket with eigenvalue $s_\text{max}-1$, given by the (appropriately normalized) state
\begin{equation}
\label{smaxm1}
\ket{s_\text{max},s_\text{max}-1}=\sqrt{k} \, \ket{s,s}\vee \ket{s,s}\vee \ldots \ldots \ket{s,s}\vee \ket{s,s-1}
\, .
\end{equation}
When $s \geq 1$, for the $S_z$-eigenvalue $s_\text{max}-2$ one finds a 2-dimensional subspace, spanned by
\begin{equation}
\label{smaxm2s}
\ket{s,s}\vee \ldots \vee \ket{s,s}\vee\ket{s,s-1}\vee\ket{s,s-1}
\, ,
\qquad
\ket{s,s}\vee \ldots \vee \ket{s,s}\vee\ket{s,s-2}
\, .
\end{equation}
A linear combination of these may be obtained as
$S_-\ket{s_\text{max},s_\text{max}-1}$ and, hence, belongs to the same irrep, with $j=s_{\text{max}}$, as the two kets found above, \emph{viz.} $\ket{s_{\text{max}},s_{\text{max}}}$, $\ket{s_{\text{max}},s_{\text{max}}-1}$, while the combination orthogonal to $S_-\ket{s_\text{max},s_\text{max}-1}$ serves as the highest weight vector of a new, $j=s_\text{max}-2$, irreducible multiplet. Continuing in the same way, one may construct the $\mathcal{H}^{\vee k}$-basis that block diagonalizes $\mathfrak{D}^{(s,k)}$. An immediate conclusion of the process is that the value $j=s_\text{max}-1$ does not appear in the decomposition in~(\ref{Dskmulti}), \ie, $\mu_{s_\text{max}-1}^{(s,k)}=0$, while
$\mu_{s_\text{max}}^{(s,k)}=1=\mu_{s_\text{max}-2}^{(s,k)}$.
It is also clear that integral and half-integral spins do not mix in the decomposition, as stated already in~(\ref{Dskmulti}), where the summation over $j$ is in unit steps.
\subsubsection{Characters}
\label{chars}
To compute the multiplicities $\mu^{(s,k)}_j$, the standard character machinery may be employed. From~(\ref{Dskmulti}) one concludes that
\begin{align}
\label{charsk}
\xi^{(s,k)}(\alpha)
&
\equiv
\Tr \, \mathfrak{D}_\text{BD}^{(s,k)}(R_{\hat{n},\alpha})
=
 \sum_{j=s_{\text{min}}}^{s_{\text{max}}} \mu^{(s,k)}_j \chi^{(j)}(\alpha)
 \, ,
 \end{align}
where $\hat{n}$, $\alpha$ denote the rotation axis and angle, respectively, and $\chi^{(j)}$ denotes the character of the irrep $D^{(j)}$,
\begin{equation}
\label{irredjchar}
\chi^{(j)}(\alpha) \equiv \Tr \, D^{(j)}(R_{\hat{n},\alpha}) 
= 
\frac{\sin \big( (j+\frac{1}{2})\alpha \big)}{\sin \frac{\alpha}{2}}
\, .
\end{equation}
The irrep characters are orthonormal, $\ipp{\chi^{(m)}}{\chi^{(n)}}=\delta_{mn}$, where for two class functions $f$, $h$ on $SU(2)$ (\ie, functions that satisfy $f(hgh^{-1})=f(g)$, $\forall g,h \in SU(2)$), an inner product is defined by
\begin{equation}
\label{ippdef}
\ipp{f}{h}
\equiv
\frac{1}{\pi}\int_0^{2\pi} \td \alpha \, \sin^2 \frac{\alpha}{2} f^*(\alpha) h(\alpha)
\, .
\end{equation}
Then the multiplicities may be extracted as
\begin{equation}
\label{multires}
\mu^{(s,k)}_j=\frac{1}{\pi} \int_0^{2\pi} \td \alpha \,  \sin^2 \frac{\alpha}{2} \xi^{(s,k)}(\alpha) \chi^{(j)}(\alpha)
\, ,
\end{equation}
\ie, they are are known once the characters $\xi^{(s,k)}(\alpha)$ are determined.
The latter can be shown to satisfy the recursion formula
\begin{equation}
\label{chiskrec}
\xi^{(s,k)}(\alpha)=
\frac{1}{k} \sum_{m=1}^k  \chi^{(s)}(m\alpha) \xi^{(s,k-m)}(\alpha)
\, ,
\end{equation}
with $\xi^{(s,0)}(\alpha)\equiv 1$, so that, for example,
\begin{align}
\xi^{(s,2)}(\alpha)
&=
\frac{1}{2}
\left(
\chi^{(s)}(\alpha)^2
+
\chi^{(s)}(2\alpha)
\right)
\label{chis2}
\\
\xi^{(s,3)}(\alpha)
&=
\frac{1}{6}
\left(
\chi^{(s)}(\alpha)^3
+
3\chi^{(s)}(\alpha) \chi^{(s)}(2\alpha)
+
2\chi^{(s)}(3\alpha)
\right)
\label{chis3}
\\
\xi^{(s,4)}(\alpha)
&=
\frac{1}{24}
\left(
\chi^{(s)}(\alpha)^4
+6\chi^{(s)}(\alpha)^2\chi^{(s)}(2\alpha)
+3\chi^{(s)}(2\alpha)^2
+8\chi^{(s)}(\alpha)\chi^{(s)}(3\alpha)
+6\chi^{(s)}(4\alpha)
\right)
\, .
\label{chis4}
\end{align}
A closed general expression for $\xi^{(s,k)}$ may be obtained as follows. Denote by $\lambda_m \equiv e^{i m \alpha}$ , $-s \leq m \leq s$, the eigenvalues of the matrix $D^{(s)}(R_{\hat{n},\alpha})$. Then, the eigenvalues of
$\mathfrak{D}^{(s,k)}(R_{\hat{n},\alpha})$ are given by all possible monomials $\lambda_{m_1}\ldots \lambda_{m_k}$, with $m_1 \leq \ldots \leq m_k$, so that
\begin{equation}
\label{xilambdarel}
\xi^{(s,k)}=\sum_{m_1\leq \ldots \leq m_k} \lambda_{m_1}\ldots \lambda_{m_k}
\equiv H_k(\boldsymbol{\lambda})
\end{equation}
where $H_k(\boldsymbol{\lambda})$ is known as the $k$-th complete symmetric polynomial in the $2s+1$ variables
$\boldsymbol{\lambda}=\{\lambda_m\}$, $-s \leq m \leq s$. Also relevant in our discussion are the Newton polynomials $P_r(\boldsymbol{\lambda})=\sum_{m=-s}^s \lambda_m^r=\chi^{(s)}(r\alpha)$, their relation with the $H_k$'s being given by (see, \eg, appendix A of~\cite{Ful.Har:04})
\begin{equation}
\label{PHrel}
H_k=\sum_{M}\frac{1}{z(M)}P^{(M)}
\, ,
\end{equation}
where $M=(m_1,\ldots,m_k)$ ranges over all $k$-tuples of nonnegative integers satisfying $\sum_{r=1}^k r m_r=k$, and
\begin{equation}
\label{zMPM}
z(M)\equiv m_1! \, 1^{m_1} \, m_2! \, 2^{m_2} \ldots m_k! \, k^{m_k}
\, ,
\qquad
P^{(M)}\equiv P_1^{m_1}\ldots P_k^{m_k}
\, .
\end{equation}
Putting everything together we get
\begin{equation}
\label{xiskP}
\xi^{(s,k)} (\alpha)
=
\sum_M
\frac{1}{z(M)}
\left( \chi^{(s)}(\alpha)\right)^{m_1}
\left( \chi^{(s)}(2\alpha)\right)^{m_2}
\ldots
\left( \chi^{(s)}(k\alpha)\right)^{m_k}
\, ,
\end{equation}
which, when substituted in~(\ref{multires}), together with $\chi^{(j)}$ from~(\ref{irredjchar}), yield any desired multiplicity $\mu^{(s,k)}_j$.  Only integer (half-integer) values of $j$ need be considered in~(\ref{multires})  when $s_\text{max}$ (and, hence, $s_\text{min}$) is integer (half-integer) (see~(\ref{smincs})). As an example, for $k=3$, the possible values of $M$ in~(\ref{PHrel}) are $(3,0,0)$, $(1,1,0)$, $(0,0,1)$, corresponding, in that same order, to the three terms in the \rhs{} of~(\ref{chis3}).
\subsubsection{Generating functions}
\label{Gfs}
The result in~(\ref{xiskP}) for the characters $\xi^{(s,k)}$ is interesting in its own right, but if all that is needed are the multiplicities $\mu^{(s,k)}_j$, alternative approaches, based on generating functions, are possible. By their definition in~(\ref{xilambdarel}), the complete symmetric polynomials satisfy
\begin{equation}
\label{Hkprop1}
\frac{1}{(1-z \lambda_s)(1-z \lambda_{s-1})\ldots (1-z \lambda_{-s})} = \sum_{k=0}^\infty H_k(\boldsymbol{\lambda}) z^k
\, ,
\end{equation}
since each factor in the \lhs{} is the sum of all powers of a particular $\lambda_r$,  $(1-z\lambda_r)^{-1}=1+z\lambda_r+z^2 \lambda_r^2+\ldots$, and their product clearly gives all possible monomials, the role of $z$ being to group the latter by polynomial degree. The denominator in the \lhs{} of~(\ref{Hkprop1}) is just the product of the eigenvalues of the matrix $I-zD^{(s)}(g)$, \ie, its determinant, so that a generating function for the multiplicities $\mu^{(s,k)}_j$, for fixed $s$, $j$, and all values of $k$, may be obtained in the form of an integral formula,
\begin{equation}
\label{HMf}
Z^{(s)}_j \equiv
\sum_{k=0}^\infty \mu^{(s,k)}_j z^k =\frac{1}{\pi} \int_0^{2\pi} \frac{\chi^{(j)}(\alpha) \, \sin^2\frac{\alpha}{2} \, \td \alpha}{\det \! \left(
I-zD^{(s)}(g)
\right)}
\, .
\end{equation}
The particular case of $j=0$ (known as the \emph{Molien-Weyl formula}~\cite{Mol:97,Wey:68}) deserves special attention, as the multiplicity $\mu_0^{(s,k)}$ gives the number of linearly independent spin-$s$ invariants of polynomial order $k$. $Z^{(s)}_0$ is then the \emph{Poincar\'e series} of the algebra $A_s$ of spin-$s$ invariants, and is known to converge in the disc $|z|<1$, representing a rational function of the form
\begin{equation}
\label{Zform}
Z^{(s)}_0=\frac{h(z)}{(1-z^{m_1}) \ldots (1-z^{m_r})}
\, ,
\end{equation}
where $m_1, \ldots , m_r$ are the degrees of homogeneous generators of $A_s$ and $h(z)$ is a polynomial with integral coefficients (see \S 3.10 of~\cite{Pop.Vin:94}). An explicit (but complicated) formula for $Z_0^{(s)}$ appears in~\cite{Spr:80}, and the results up to spin 8 are generated by computer in~\cite{Bro.Coh:79} (Sylvester had computed some of these functions by hand~\cite{Syl:73}) --- we quote the first few cases,
\begin{equation}
\label{Pss52}
Z^{(\frac{1}{2})}_0=1
\, ,
\quad
Z^{(1)}_0=\frac{1}{1-z^2}
\, ,
\quad
Z^{(\frac{3}{2})}_0=\frac{1}{1-z^4}
\, ,
\quad
Z^{(2)}_0=\frac{1}{(1-z^2)(1-z^3)}
\, ,
\quad
Z^{(\frac{5}{2})}_0=\frac{1+z^{18}}{(1-z^4)(1-z^8)(1-z^{12})}
\, .
\end{equation}

Another generating function, for fixed $s$, $k$, appears in~\cite{Pol.Sfe:16}, bypassing the costly integrations in~(\ref{multires}), (\ref{HMf}): $\mu^{(s,k)}_j$ is given by the coefficient of $x^j$, $0 \leq j \leq s_\text{max}$, in the Laurent expansion, around $x=0$, of the function
\begin{equation}
\label{genfmu}
\tilde{\zeta}_{s,k}(x)=(1-x^{-1})\prod_{r=1}^{2s} \frac{x^{\frac{k}{2}+r}-x^{-\frac{k}{2}}}{x^r-1}
\, .
\end{equation}
Note that the Laurent expansion mentioned involves also powers of $x$ higher than $s_\text{max}$, as well as negative ones, which are to be ignored.
\begin{myexample}{Irreducible components of spin-1, spin-3/2,  and spin-2 symmetric states}{seq1sp}
For spin-1 symmetrized states we find (spin-$s$ irreducible components are denoted by their dimension $2s+1$ in bold)
\begin{align}
\label{s1ss}
\mathbf{3}^{\vee 2}
&=
\mathbf{5} \oplus \mathbf{1}
\\
\label{s1ss2}
\mathbf{3}^{\vee 3}
&=
\mathbf{7} \oplus \mathbf{3}
\\
\label{s1ss3}
\mathbf{3}^{\vee 4}
&=
\mathbf{9} \oplus \mathbf{5} \oplus \mathbf{1}
\\
\label{s1ss4}
\mathbf{3}^{\vee 5}
&=
\mathbf{11} \oplus \mathbf{7} \oplus \mathbf{3}
\, ,
\end{align}
suggesting an easy pattern: the decomposition starts at $j=s_\text{max}$ and descends in steps of two, all nonzero multiplicities being equal to 1 --- this can be shown to be true in general (see below).  For spin-3/2, the corresponding results are
\begin{align}
\label{s32ss}
\mathbf{4}^{\vee 2}
&=
\mathbf{7} \oplus \mathbf{3}
\\
\label{s32ss2}
\mathbf{4}^{\vee 3}
&=
\mathbf{10} \oplus \mathbf{6} \oplus \mathbf{4}
\\
\mathbf{4}^{\vee 4}
&=
\mathbf{13} \oplus \mathbf{9} \oplus \mathbf{7} \oplus \mathbf{5} \oplus \mathbf{1}
\\
\mathbf{4}^{\vee 5}
&=
\mathbf{16} \oplus \mathbf{12} \oplus \mathbf{10}\oplus \mathbf{8} \oplus \mathbf{6} \oplus \mathbf{4}
\\
\mathbf{4}^{\vee 6}
&=
\mathbf{19} \oplus \mathbf{15} \oplus \mathbf{13}\oplus \mathbf{11} \oplus \mathbf{9} \oplus 2 \times \mathbf{7}
\oplus \mathbf{3}
\, ,
\end{align}
where we included the $k=6$ case, as it is the first one where a nonbinary multiplicity shows up (the irrep $\mathbf{7}$ appears twice).
Finally, for spin-2 symmetrized states we find
\begin{align}
\label{s2ss}
\mathbf{5}^{\vee 2}
&=
\mathbf{9} \oplus \mathbf{5} \oplus \mathbf{1}
\\
\mathbf{5}^{\vee 3}
&=
\mathbf{13} \oplus \mathbf{9} \oplus \mathbf{7} \oplus \mathbf{5} \oplus \mathbf{1}
\\
\mathbf{5}^{\vee 4}
&=
\mathbf{17} \oplus \mathbf{13} \oplus \mathbf{11} \oplus 2 \times \mathbf{9} \oplus 2 \times \mathbf{5} \oplus \mathbf{1}
\\
\mathbf{5}^{\vee 5}
&=
\mathbf{21} \oplus \mathbf{17} \oplus \mathbf{15} \oplus 2 \times \mathbf{13} \oplus \mathbf{11} \oplus
2 \times \mathbf{9} \oplus \mathbf{7} \oplus 2 \times \mathbf{5} \oplus \mathbf{1}
\, ,
\end{align}
which is already a bit more involved.
\end{myexample}

An interesting observation made in~\cite{Pol.Sfe:16} is the ``duality relation''
\begin{equation}
\label{zetaduality}
\tilde{\zeta}_{s,k}(x)=\tilde{\zeta}_{\frac{k}{2},2s}(x)
\, .
\end{equation}
This implies, for example, that $\tilde{\zeta}_{s,2}(x)=\tilde{\zeta}_{1,2s}(x)$, which explains why
 the \rhs{} of ~(\ref{s32ss}), (\ref{s2ss}) coincides with that of~(\ref{s1ss2}), (\ref{s1ss3}), respectively.
This last relation also yields a simple proof of the pattern mentioned above for $s=1$. Indeed, instead of considering a symmetrized state of $k$ qutrits, due to the above duality, we may consider one of two spin-$k/2$ particles. We know, from standard spin addition, that the tensor product of two such states decomposes into a direct sum of spins, from a maximum value $k$ down to zero. The tensor product of two factors though decomposes into a direct sum of symmetric and antisymmetric subspaces, and it can be shown, with the help of the Pieri formula, that the symmetric subspace contains  spins $k$, $k-2$, $k-4$, $\ldots$ (which are the values that appear in the \rhs{} of~(\ref{s1ss})--(\ref{s1ss4})), while the antisymmetric subspace contains the rest.

 Although~(\ref{zetaduality}) can be checked directly given the explicit form of $\tilde{\zeta}_{s,k}$ in~(\ref{genfmu}), it is conceptually advantageous to realize that it is a consequence of ``Hermite reciprocity'', which states that for a 2-dimensional representation space $V$, $\vee^m(\vee^n V) \sim \vee^n(\vee^m V)$, with ``$\sim$'' denoting a vector space  isomorphism which commutes with the action of $SU(2)$~\cite{Her:54,Ful.Har:04}. In our case, a spin-$s$ state is considered as a vector in the $2s$-fold symmetric product of the spin-1/2 Hilbert space, and Hermite reciprocity says that $\vee^k(\vee^{2s} V) \sim \vee^{2s}(\vee^k V)$, which implies~(\ref{zetaduality}).

A second duality relation is also mentioned in~\cite{Pol.Sfe:16}, namely
$\zeta_{s,k}(x)=\zeta_{s,2s+1-k}(x)$, where $\zeta_{s,k}(x)$ is the generating function for the multiplicities in the fully antisymmetric case. This latter relation is made obvious if one realizes that totally antisymmetric $k$-partite $\wedge$-factorizable spin-$s$ states (\ie, $k$-fold wedge products of spin-$s$ states) are in 1-to-1 correspondence with $k$-planes in the spin-$s$ Hilbert space, and the fact that a $k$-plane and its orthogonal complement, a $(2s+1-k)$-plane, carry the same geometrical information (for more details see, \eg,~\cite{Chr.Guz.Han.Ser:21}). Since a general antisymmetric state is a linear combination of $\wedge$-factorizable ones, the above isomorphism, which we denote by $\varphi$,  extends to the entire $\mathcal{H}_s^{\wedge k}$,
\begin{equation}
\label{dualkplanes}
\varphi \colon \mathcal{H}_s^{\wedge k} \rightarrow \mathcal{H}_s^{\wedge \,  2s+1-k}
\, .
\end{equation}
 It would be interesting to find an analogous, geometrical, interpretation of~(\ref{zetaduality}) (in this respect, see section~\ref{Hics12p}). Meanwhile, it is instructive to pursue the Hermite isomorphism, its generalizations, and their physical consequences --- we do that in section~\ref{AclatHr} below.

The problem of the decomposition of the $k$-fold tensor product of the spin-$s$ irrep of $SU(2)$ has been considered before, both by physicists and mathematicians. Related material, from a physical perspective, can be found in~\cite{Zac:92,Cur.Kor.Zac:90}, the analogous problem for fully antisymmetric states is examined in~\cite{Chr.Guz.Han.Ser:21}, while a more mathematical approach is undertaken in~\cite{Gya.Bar:18}. Both the fully symmetric and antisymmetric cases considered here and in~\cite{Chr.Guz.Han.Ser:21} are particular cases of the general \emph{plethysm} problem, which is still open (see, \eg,~\cite{Lit:50,Wyb:70}).
\subsection{The multiconstellation of a $k$-symmetric spin-$s$ state}
\label{Tmoaksss}
Following the procedure described in section~\ref{Ic}, one may determine the unitary matrix $U$ that transforms spinors from the standard normalized induced basis $\{ \hat{e}_{\vec{A}} \}$ to the BD one,
$\ket{\boldsymbol{\Psi}}_\text{BD}=U \ket{\boldsymbol{\Psi}}_\text{S}$, where $\ket{\boldsymbol{\Psi}}_\text{BD}$ is the direct sum of spin-$j$ spinors, $0 \leq j \leq s_\text{max}$,
\begin{equation}
\label{PsiDdef}
\ket{\boldsymbol{\Psi}}_\text{BD}^T
=\left(
\ket{\psi^{(s_\text{max})}}^T, \,
\ket{\psi^{(s_\text{max}-2)}}^T, \,
\ldots
\ket{\psi^{(j)}_\alpha}^T,
\ldots
\right)
\, ,
\end{equation}
with spin $j$ appearing $\mu^{(s,k)}_j$ times, labeled by the index $\alpha$ above --- the various irreducible multiplets are ordered in decreasing spin value. Note that for the spinors in the \rhs{} of~(\ref{PsiDdef}) we could use a subscript S or BD, since in this case the standard basis coincides with the BD one --- accordingly, we just omit the subscript. Note also that $\braket{\boldsymbol{\Psi}}{\boldsymbol{\Psi}}=\sum_{j\alpha} \braket{\psi^{(j)}_\alpha}{\psi^{(j)}_\alpha}$.  Under the above change of basis, representation matrices transform by conjugation,
\begin{equation}
\label{DSxform}
\mathfrak{D}^{(s,k)}_\text{BD}=U \mathfrak{D}^{(s,k)}_\text{S} U^\dagger
\, ,
\qquad
S^{(s,k)}_\text{BD}=U S^{(s,k)}_\text{S} U^\dagger
\, ,
\end{equation}
both group and Lie algebra BD matrices being block-diagonal, with structure corresponding to that of
$\ket{\boldsymbol{\Psi}}_\text{BD}$ in~(\ref{PsiDdef}).
\subsubsection{The canonical section}
\label{Tcs}
We saw in the previous section that a multipartite state can be considered as a collection of various spin-$j$ states, with definite relative (complex) weights. Each of these states, following Majorana's 1932 construction~\cite{Maj:32}, can be represented by an unordered, possibly coincident, set of $2j$ points on the sphere. The gist of the method is to use the components of the state in the standard $S_z$ basis to form a polynomial of degree $2j$ in an auxiliary complex variable $\zeta$. The $2j$ roots of this polynomial are then mapped stereographically onto the unit sphere, giving rise to the \emph{Majorana  stars} of the state, collectively referred to as its \emph{Majorana constellation}. The essential property of this construction is that $SU(2)$ transformations of the state result in (corresponding) rigid rotations of the constellation --- more details can be consulted 
in~\cite{Ben.Zyc:17,Chr.Her:17,Chr.Guz.Ser:18,Chr.Guz.Han.Ser:21}. 

We denote by $C_{j\alpha}$ the constellation of the ket
$\ket{\psi^{(j)}_\alpha}$ and by $\mathcal{C}$ the collection of all these constellations, $\mathcal{C}=\{C_{j\alpha}\}$. Each constellation $C_{j\alpha}$ in $\mathcal{C}$ misses the information of the normalization and phase of $\ket{\psi^{(j)}_\alpha}$, so that knowledge of $\mathcal{C}$ is not enough to recover $\ket{\boldsymbol{\Psi}}_\text{BD}$. Suppose now that we devise a way to assign to each possible constellation $C_{j\alpha}$ a \emph{reference} normalized ket
$\ket{\psi_{C_{j\alpha}}}$, which has $C_{j\alpha}$ as its Majorana constellation, essentially lifting (locally) the projective space $\mathbb{P}$ into the Hilbert space $\mathcal{H}$ (\ie, in fibre bundle parlance, choosing a local section in the $U(1)$ bundle over $\mathbb{P}$, which embeds in $\mathcal{H}$ as the unit sphere). Then $\ket{\psi^{(j)}_\alpha}$ (in the \rhs{} of~(\ref{PsiDdef})) and $\ket{\psi_{C_{j\alpha}}}$ can only differ by an overall complex constant (since they have the same constellation),
\begin{equation}
\label{psijzj}
\ket{\psi^{(j)}_\alpha}=z_{j\alpha} \ket{\psi_{C_{j\alpha}}}
\, .
\end{equation}
The comparison between all $\ket{\psi^{(j)}_\alpha}$ and the corresponding (arbitrarily chosen) reference $\ket{\psi_{C_{j\alpha}}}$ produces a collection $Z$ of complex numbers, $Z=\{z_{j\alpha}\}$, which, together with $\mathcal{C}$, now completely determine $\pket{\boldsymbol{\Psi}}_\text{BD}$ (we write $\pket{\boldsymbol{\Psi}}_\text{BD}$, rather than $\ket{\boldsymbol{\Psi}}_\text{BD}$ because the overall phase information, as well as the normalization of $\ket{\boldsymbol{\Psi}}_\text{BD}$ is still missing). Just so that our description of $\pket{\boldsymbol{\Psi}}_\text{BD}$ becomes entirely visual, we may pretend $Z$ to be a \emph{spectator} ``spinor'', and assign to it a (spectator) constellation $C_Z$ \emph{\`a la} Majorana, so that
$\pket{\boldsymbol{\Psi}}_\text{BD}$ is fully coded in the \emph{multiconstellation} $\{ C_Z,\mathcal{C} \}$. It should be emphasized though that $Z$ is not a spinor, in the sense that its transformation properties under rotations are not those of a spinor, and the assignment to it of a constellation relies on the completely arbitrary ordering of the ``components'' $z_{j\alpha}$, so there is no underlying geometrical content in $C_Z$, in contrast to the $C_{j\alpha}$. For an alternative approach to the problem of encoding the complex weights $z_{j\alpha}$ see~\cite{Ser.Bra:20}.

By an argument identical to the one given  in~\cite{Chr.Guz.Han.Ser:21} for the case of permutation antisymmetric states, it can be shown that a particular choice of the reference kets $\ket{\psi_{C_{j\alpha}}}$, which we will call the \emph{canonical section}, results in $C_Z$ being invariant under rotations (except for a subset of states in $\mathcal{H}^{\vee k}$ of measure zero), while $C_{j\alpha}$ rotate as Majorana constellations do. Since we will follow that recipe in the examples that follow, we summarize the procedure here, referring the
reader to~\cite{Chr.Guz.Han.Ser:21} for further details and proofs (keep in mind though that the notation and the exposition here is slightly different, aiming at improved clarity).

We simplify, for the moment,  the presentation assuming that
$\ket{\boldsymbol{\Psi}}_{\text{BD}}$ is such that none of the corresponding $C_{j\alpha}$ have any rotational symmetries. For every constellation $C_{j\alpha}$, define a \emph{reference constellation} $\tilde{C}_{j\alpha}$, such that the former is obtained from the latter by a rotation in $SO(3)$, which, by our simplifying assumption above, is unique,
\begin{equation}
\label{CCCtrot}
C_{j\alpha}=R_{C_{j\alpha}}(\tilde{C}_{j\alpha})
\, .
\end{equation}
Note that the choice of  $\tilde{C}_{j\alpha}$ is arbitrary, but it should be the same for all constellations related by a rotation. Another way to state this is to define an equivalence relation $\sim$  among constellations, with $C\sim C'$ iff there is $R \in SO(3)$ such that $C'=R(C)$. Then the  reference constellation $\tilde{C}_{j\alpha}$ is a particular representative of the equivalence class of $C_{j\alpha}$. Given such a choice of representative,  define for the corresponding reference kets
\begin{equation}
\label{psijadef}
\ket{\psi_{C_{j\alpha}}} \equiv D^{(j)}(e^{-i\eta \hat{n}\cdot \mathbf{S}})\ket{\psi_{\tilde{C}_{j\alpha}}}
\, ,
\end{equation}
where the axis $\hat{n}$ and the angle $\eta$ are those of $R_{C_{j\alpha}}$. Thus defined, the kets $\ket{\psi_{C_{j\alpha}}}$ give rise, via~(\ref{psijzj}), to constants $z_{j\alpha}$ that are invariant under rotations --- see~\cite{Chr.Guz.Han.Ser:21} for the proof.

The last remaining ingredient, in order to be able to apply the above procedure to concrete examples, is the specification of the reference constellation, and its reference ket. Following~\cite{Chr.Guz.Han.Ser:21}, for the latter, we fix the phase of the normalized $\ket{\psi_{\tilde{C}_{j\alpha}}}$ by requiring that its first nonzero component be real and positive. Then the phase of any other reference ket $\ket{\psi_{C_{j\alpha}}}$ is fixed by the rotation in~(\ref{psijadef}). For the former, we perform the following steps:
\begin{enumerate}
\item
Given a constellation $C_{j\alpha}$, compute the corresponding density matrix $\rho_{j\alpha}$ (note that there is a 1-to-1 correspondence between constellations and density matrices, both characterizing uniquely points in $\mathbb{P}$).
\item
Compute the spin expectation value $\vec{S}=\Tr(\rho_{j\alpha}\mathbf{S})$. Assuming $\vec{S}$ to be nonzero, rotate $C_{j\alpha}$ to $C'_{j\alpha}=R_1(C_{j\alpha})$, so that $R_1(\vec{S})$ points along $\hat{z}$.
\item
Compute the rotated density matrix $\rho_1=D_1 \rho_{j\alpha} D_1^{-1}$, where
$D_1 \equiv D^{(j)}(e^{-i \gamma \hat{m}\cdot \mathbf{S}})$, and $\hat{m}$, $\gamma$, parametrize $R_1$.
\item
Use the polarization tensor expansion~\cite{Var.Mos.Khe:88},
\begin{equation}
\label{rho1exp}
\rho_1=\sum_{\ell=0}^{2j} \sum_{m=-\ell}^\ell c_{\ell m} T_{\ell m}
\, ,
\end{equation}
to assign coordinates $\{c_{\ell m}\}=\{ (c_{0,0}), (c_{1,1},c_{1,0},c_{1,-1}), \ldots, (c_{2j,2j},\ldots,c_{2j,-2j}) \}$ to
$\rho_1$ (we use extra parentheses to visually group together same-$\ell$ multiplets).
\item
Identify the first nonzero coordinate $c_{\tilde{\ell} \tilde{m}}$, with $\tilde{m} \neq 0$, and rotate $C_{j\alpha}'$ around $\hat{z}$ by an angle $\delta/\tilde{m}$, where $c_{\tilde{\ell}\tilde{m}}=|c_{\tilde{\ell}\tilde{m}}|e^{i\delta}$ --- the resulting constellation is the required
$\tilde{C}_{j\alpha}=R_2(C_{j\alpha}')=(R_2 \circ R_1)(C_{j\alpha})$, with $R_2$ parametrized by $\hat{z}$, $\delta/\tilde{m}$.
\end{enumerate}
Note that the  last step above guarantees that the first nonzero coordinate of $\tilde{\rho}$ (the density matrix corresponding to $\tilde{C}_{j\alpha}$), is real and positive.
\section{A collection of examples}
\label{Acoe}
\begin{myexample}{Irreducible basis for 2-symmetric qutrit states}{mc2sq}
An orthonormal basis in the qutrit Hilbert space $\mathcal{H}_1$ is given by the eigenvectors of $S_z$, $\{e_1,e_2,e_3\}=\{ \ket{1,1},\ket{1,0},\ket{1,-1}\}$, in the $\ket{s,m}$ notation. The associated orthonormal basis in $\mathcal{H}^{\vee 2}_1$ is
\begin{equation}
\label{ehedef}
\{
\hat{e}_{11},
\hat{e}_{12},
\hat{e}_{13},
\hat{e}_{22},
\hat{e}_{23},
\hat{e}_{33}
\}
=
\{
e_{11},
\sqrt{2} e_{12},
\sqrt{2} e_{13},
e_{22},
\sqrt{2} e_{23},
e_{33}
\}
\, ,
\end{equation}
where $e_{ij}\equiv e_i \vee e_j$. The highest $S_z$-eigenvalue eigenvector is $\hat{e}_{11}$, with eigenvalue 2. Applying $S_-$ four times, one generates the entire spin-2 multiplet,
\begin{equation}
\label{s2multi}
\{
\hat{e}_{(2,2)}
, \,
\hat{e}_{(2,1)}
, \,
\hat{e}_{(2,0)}
, \,
\hat{e}_{(2,-1)}
, \,
\hat{e}_{(2,-2)}
\}
=
\{
\hat{e}_{11}
, \,
\hat{e}_{12}
, \,
\frac{1}{\sqrt{3}} (\hat{e}_{13}+\sqrt{2} \, \hat{e}_{22})
, \,
\hat{e}_{23}
, \,
\hat{e}_{33}
\}
\, ,
\end{equation}
where $\hat{e}_{(s,m)}$ denotes a spin-$s$, $z$-projection $m$ normalized vector.
The $S_z$-eigenvalue 0 is doubly degenerate, the state orthogonal to $\hat{e}_{(2,0)}$ is the spin-0 state $\hat{e}_{(0,0)}=(-\sqrt{2} \, \hat{e}_{13}+\hat{e}_{22})/\sqrt{3}$. The matrix $U_{(s=1,k=2)}$ effecting the change between the two bases, $\ket{\boldsymbol{\Psi}}_\text{BD}=U_{(1,2)} \ket{\boldsymbol{\Psi}}_\text{S}$, is
\begin{equation}
\label{Us1k2m}
U_{(1,2)}=
\left(
\begin{array}{cccccc}
1 & 0 & 0 & 0 & 0 & 0
\\
0 & 1 & 0 & 0 & 0 & 0
\\
0 & 0 & \frac{1}{\sqrt{3}} & \sqrt{\frac{2}{3}} & 0 & 0
\\
0 & 0 & 0 & 0 & 1 & 0
\\
0 & 0 & 0 & 0 & 0 & 1
\\
0 & 0 & -\sqrt{\frac{2}{3}} & \frac{1}{\sqrt{3}} & 0 & 0
\end{array}
\right)
\, .
\end{equation}
\end{myexample}
\begin{myexample}{Tetrahedral 2-qutrit state}{T2qs}
A symmetric 2-qutrit state decomposes into one spin-2 and one spin-0 component --- see Eq.~(\ref{s1ss}). We fix the spin-2 component to be the tetrahedral state, and also allow an arbitrary spin-0 component, so that in the BD basis we have (ignoring normalization)
\begin{equation}
\label{psitetraBD}
\ket{\boldsymbol{\Psi}^{\text{tetra}}}_\text{BD}
=
\left(
\big(1,0,0,\sqrt{2},0\big),\big(a\big)
\right)^T
\, ,
\end{equation}
with $a$ an arbitrary complex number.
We switch to the standard,  $\{ \hat{e}_{ij} \}$ basis by left-multiplying with the inverse of the matrix $U_{(1,2)}$ in~(\ref{Us1k2m}), to get
\begin{align*}
\ket{\boldsymbol{\Psi}^{\text{tetra}}}_\text{S}
&=
U^{-1}_{(1,2)} \ket{\boldsymbol{\Psi}^{\text{tetra}}}_\text{BD}
\\
&=
\left(
1, 0,-\sqrt{\frac{2}{3}} a, \frac{a}{\sqrt{3}}, \sqrt{2},0
\right)^T
\\
&=
\hat{e}_{11} + \sqrt{2} \hat{e}_{23} + a
\left(
-\sqrt{\frac{2}{3}} \hat{e}_{13}
+
\frac{1}{\sqrt{3}} \hat{e}_{22}
\right)
\\
&=
e_{11} + 2 e_{23}
+
a \left( -\frac{2}{\sqrt{3}} e_{13} + \frac{1}{\sqrt{3}} e_{22} \right)
\, ,
\end{align*}
where, in the last line, we switched to the non-normalized $\{e_{ij} \}$ basis, according to~(\ref{ehedef}). Expanding the $e_{ij}=e_i \vee e_j$ according to~(\ref{symmtp}), we finally get
\begin{equation}
\label{Psitetraf}
\ket{\boldsymbol{\Psi}^{\text{tetra}}}
=
\ket{1,1}+\ket{0,-1}+\ket{-\! 1,0}
+\frac{a}{\sqrt{3}}
\left(
\ket{0,0}-\ket{1,-1}-\ket{- \! 1,1}
\right)
\, ,
\end{equation}
using the $\ket{m_1,m_2}$ notation.
We kept the spin-0 component above to check if, for some value of $a$, the state is $\vee$-factorizable --- it is easily seen, by direct calculation,
that this is not the case. The calculation alluded to involves mapping symmetric states to polynomials in commuting variables, equating the  expression thus obtained to the most general factorizable form, and looking for solutions to the resulting linear system. Note that for $\vee$-factorizable states  another stellar representation is possible, consisting of the Majorana constellations of the $v$'s, each, say, with a different color. The general problem of whether a particular element of $\vee^k \mathbb{C}^n$ is $\vee$-factorizable reduces to the question of whether its components satisfy \emph{Brill's equations}~\cite{Gel.Kap.Zel:94} --- the latter are rather complicated to write out explicitly.

Putting $a=0$ in~\eqref{Psitetraf} we get the pure tetrahedral state
\begin{equation}
\label{puretetra}
\ket{\boldsymbol{\Psi}^{\text{tetra}}_0}
=
\ket{1,1}+\ket{0,-1}+\ket{-\! 1,0}
\, .
\end{equation}
That the above state is invariant under rotations by $2\pi/3$ around the $z$-axis is readily checked. That the same is true for the other axes of symmetry of the tetrahedron is not so obvious, but can also be checked to be true.
\end{myexample}
\begin{myexample}{Bell-like 2-qutrit state}{Bl2qs}
Consider the 2-qutrit state (in $\ket{m_1,m_2}$ notation)
\begin{align}
\ket{\boldsymbol{\Psi}^\text{B}}
&=
\ket{1,1}+\ket{0,0}+\ket{{-1},{-1}}
\nonumber
\\
&=
e_{11}+e_{22}+e_{33}
\nonumber
\\
\Rightarrow \ket{\boldsymbol{\Psi}^\text{B}}_\text{S}
&=
(1,0,0,1,0,1)
\label{psiB}
\, .
\end{align}
In the BD basis this becomes
\begin{align}
\ket{\boldsymbol{\Psi}^\text{B}}_\text{BD}
&=
U_{(1,2)}\ket{\boldsymbol{\Psi}^\text{B}}_\text{S}
\\
&=
\left( \big( 1,0,\sqrt{2/3},0,1 \big), \, \big(1/\sqrt{3} \big) \right)
\label{psiBBD}
\, ,
\end{align}
where the internal parentheses group the different spin components. The spin-2 component has constellation consisting of two stars at $y=1$ and two more antipodal, at $y=-1$. Accordingly (since the spin-0 component is isotropic), the state is invariant under rotations around the $z$-axis by $\pi$ (obvious), around the $x$-axis by $\pi$ (less obvious), and also around the $y$-axis by arbitrary angles. Note that in the presence of a nonzero spin-0 component, the rotational symmetries of the spin-2 constellation are not necessarily symmetries of the 2-qutrit state, because a symmetry rotation of the spin-2 constellation might still impart a phase to the spin-2 state, while the spin-0 amplitude is invariant, resulting in a change in the full state. In our case, the spin-2 state is clearly an $S_y=0$ eigenstate, so rotations around the $y$-axis leave the state invariant, without imparting any phase, so the full 2-qutrit state is invariant under such rotations. Direct computation shows that this state is non-$\vee$-factorizable.

 Note that the state
$\ket{\boldsymbol{\Psi}^{\text{B}'}} =\ket{1,1}-\ket{0,0}+\ket{{-1},{-1}}$ is similar to $\ket{\boldsymbol{\Psi}^\text{B}}$, but with its stars along $x$, while
$\ket{\boldsymbol{\Psi}^{\bar{\text{B}}}}=\ket{1,1}+\ket{0,0}-\ket{{-1},{-1}}$ gives $\ket{\boldsymbol{\Psi}^{\bar{\text{B}}}}_\text{BD}=\big( (1,0,\sqrt{2/3},0,{-1}), \, (1/\sqrt{3})\big)$, the spin-2 component of which has a tetrahedral (albeit non-regular) constellation,
\begin{align}
n_1
&=
\left(
-\sqrt{2(\sqrt{2}-1)},0,\sqrt{2}-1
\right)
\\
n_2
&=
\left(
\sqrt{2(\sqrt{2}-1)},0,\sqrt{2}-1
\right)
\\
n_3
&=
\left(
0, -\sqrt{2(\sqrt{2}-1)}, -\frac{1}{\sqrt{3+2\sqrt{2}}}
\right)
\\
n_4
&=
\left(
0, \sqrt{2(\sqrt{2}-1)}, -\frac{1}{\sqrt{3+2\sqrt{2}}}
\right)
\, .
\end{align}
 All of the above three states may be considered to be in the Schmidt form (by suitable phase redefinition of the basis states), so they are all maximally entangled, while their spin-2 constellations are all 1-anticoherent (but no higher).
 \end{myexample}

Another physical quantity of interest might be the $t$-anticoherence measure defined in Eq.{} (24) of~\cite{Bag.Mar:17}. For a $k$-qubit symmetric state $\ket{\psi}$, viewed as a vector in the $k$-fold tensor product of the single qubit Hilbert space, the authors of~\cite{Bag.Mar:17} define a $t$-anticoherence measure $\mathcal{A}_t$ as follows
\begin{equation}
\label{Atdef}
\mathcal{A}_t(\ket{\psi})=\frac{t+1}{t} \left(1-\text{Tr}(\rho_t^2) \right)
\, ,
\end{equation}
where $\rho_t$ is the reduced density matrix of $\ket{\psi}$, traced over $k-t$ bits (\ie, of dimension $2^t$). Adopting this same definition for $k$-symmetric qudit states (we will have to make sure this is ok), we compute,
\begin{align}
\label{Atqudit2}
\mathcal{A}_1(\ket{\psi_1}\vee \ket{\psi_2})
&=
2 \left( \frac{1-x}{1+x} \right)^2
\\
\mathcal{A}_2(\ket{\psi_1}\vee \ket{\psi_2}\vee \ket{\psi_3})
&=
\frac{3}{2}
\left(
1-
\frac{3+2(x^2+y^2+z^2)+4(xy+yz+zx)+12(x+y+z+w)}{9(1+x+y+z+w)^2}
\right)
\, ,
\label{Atqudit3}
\end{align}
where
\begin{equation}
\label{xyzw}
x=\sigma_{12}\sigma_{21}
\, ,
\quad
y=\sigma_{13}\sigma_{31}
\, ,
\quad
z=\sigma_{23}\sigma_{32}
\, ,
\quad
w=2 \Re (\sigma_{13}\sigma_{32} \sigma_{21} )
\, ,
\end{equation}
and $\sigma_{ij}=\braket{\psi_i}{\psi_j}$ (note that the symmetrized states $\ket{\psi_1}\vee \ket{\psi_2}$ \etc{} above are  normalized to unity). Eq.{}~(\ref{Atqudit2}) is not applicable to the Bell state considered in Example~\ref{Bl2qs}, because the latter is not $\vee$-factorizable, but we quote it here, along with~(\ref{Atqudit3}), for future reference.
\begin{myexample}{Bell-like 2-qudit states}{Bl2qds}
We generalize the Bell-like 2-qutrit state of Ex.~\ref{Bl2qs} by defining  spin-$s$ Bell states along $\hat{\mathbf{y}}$ via
\begin{equation}
\label{PsiB}
\ket{\boldsymbol{\Psi}^\text{B}_{\hat{\mathbf{y}}}}
=
\frac{1}{\sqrt{2s+1}}
\left(
\ket{s,s}+\ket{s-1,s-1}+\ldots +\ket{-s,-s}
\right)
\end{equation}
(it will soon become apparent why we denote this state by an index $\hat{\mathbf{y}}$).
Using $S_y \sim S_+ - S_-$ and the fact that $S_+ \ket{k,k}=S_-\ket{k+1,k+1}$, where $\ket{k,k}$ is a bipartite spin-$s$ state with each subsystem in a $S_z=k$ eigenstate, one concludes easily that
\begin{equation}
\label{SyPsiB}
S_y \ket{\boldsymbol{\Psi}^\text{B}_{\hat{\mathbf{y}}}} \sim \sum_{m=-s}^s
S_y \ket{s,m} \otimes \ket{s,m}+\ket{s,m}\otimes S_y \ket{s,m}=0
\, ,
\end{equation}
where we abuse slightly notation by denoting by $S_y$ both the one- and two-qudit operator. Thus, when brought in the BD form, every component $\ket{\psi^{(j)}}$ of $\ket{\boldsymbol{\Psi}^\text{B}_{\hat{\mathbf{y}}}}$ must be an $S_y=0$ eigenstate, with constellation consisting of $j$ stars at $y=1$ and another $j$ stars at $y=-1$. As a corollary, all such states are invariant under arbitrary rotations around the $y$-axis. An alternative way to show the above invariance is by writing for the rotated state (omitting the normalization factor)
\begin{align*}
\ket{\boldsymbol{\Psi}^{\text{B}'}_{\hat{\mathbf{y}}}}
& \sim
\sum_{k,m,m'=-s}^s
D^{(s)}(R)_{k,m} \ket{s,m} \otimes D^{(s)}(R)_{k,m'} \ket{s,m'}
\\
&=
\sum_{k,m,m'=-s}^s
D^{(s)}(R)^\dagger_{m,k}D^{(s)}(R)_{k,m'} \ket{s,m} \otimes  \ket{s,m'}
\\
&=
\sum_{m,m'=-s}^s
\delta_{mm'}  \ket{s,m} \otimes  \ket{s,m'}
\\
& \sim
\ket{\boldsymbol{\Psi}^{\text{B}}_{\hat{\mathbf{y}}}}
\, ,
\end{align*}
where we used the fact that, for rotations around the $y$-axis, the (unitary) rotation matrices $D^{(s)}$ are real, and, hence, orthogonal.

We may easily rotate $\ket{\boldsymbol{\Psi}^\text{B}_{\hat{\mathbf{y}}}}$ around the $z$-axis by $-\pi/2$ to obtain
$\ket{\boldsymbol{\Psi}^\text{B}_{\hat{\mathbf{x}}}}$, which has its symmetry axis along $x$,
\begin{align}
\ket{\boldsymbol{\Psi}^\text{B}_{\hat{\mathbf{x}}}}
&\sim
e^{i \frac{\pi}{2} S_z} \sum_{m=-s}^s \ket{s,m} \otimes \ket{s,m}
\nonumber
\\
&=
\sum_{m=-s}^s
e^{i \frac{\pi}{2} S_z}  \ket{s,m} \otimes e^{i \frac{\pi}{2} S_z}  \ket{s,m}
\nonumber
\\
&=
\sum_{m=-s}^s
(-1)^m  \ket{s,m} \otimes \ket{s,m}
\nonumber
\\
&=
(-1)^s \left(
\ket{s,s}-\ket{s,s-1}+\ldots +(-1)^{2s}\ket{s,-s}
\right)
\label{PsiBx}
\, .
\end{align}
Rotating $\ket{\boldsymbol{\Psi}^\text{B}_{\hat{\mathbf{y}}}}$ around the $x$-axis by $\pi/2$, to get $\ket{\boldsymbol{\Psi}^\text{B}_{\hat{\mathbf{z}}}}$, with symmetry axis along $z$, is also easy.
The above invariance of $\ket{\boldsymbol{\Psi}^\text{B}_{\hat{\mathbf{y}}}}$ may be stated as $(S_y \otimes I)\ket{\boldsymbol{\Psi}^\text{B}_{\hat{\mathbf{y}}}}
=-(I \otimes S_y)\ket{\boldsymbol{\Psi}^\text{B}_{\hat{\mathbf{y}}}}$. A calculation similar to the one outlined above shows that one also has
\begin{equation}
\label{SxPsiB}
 (S_x \otimes I)\ket{\boldsymbol{\Psi}^\text{B}_{\hat{\mathbf{y}}}}
 =
 (I \otimes S_x)\ket{\boldsymbol{\Psi}^\text{B}_{\hat{\mathbf{y}}}}
 \, .
 \end{equation}
As a consequence, we may calculate easily
\begin{align}
\ket{\boldsymbol{\Psi}^\text{B}_{\hat{\mathbf{z}}}}
&\sim
e^{-i \frac{\pi}{2} S_x} \sum_{m=-s}^s \ket{s,m} \otimes \ket{s,m}
\nonumber
\\
&=
\sum_{m=-s}^s
e^{-i \frac{\pi}{2} S_x}  \ket{s,m} \otimes e^{-i \frac{\pi}{2} S_x}  \ket{s,m}
\nonumber
\\
&=
\sum_{m=-s}^s
\ket{s,m} \otimes e^{-i \pi S_x}  \ket{s,m}
\nonumber
\\
&=
(-i)^{2s}
\sum_{m=-s}^s
\ket{s,m} \otimes   \ket{s,-m}
\label{PsiBz}
\, .
\end{align}
A further interesting consequence of~(\ref{SxPsiB}) is the following: for two unit vectors $\hat{\mathbf{n}}_1$, $\hat{\mathbf{n}}_2$ in $\mathbb{R}^3$, define the state
\begin{equation}
\label{PsiBn1n2}
\ket{\boldsymbol{\Psi}^\text{B}_{\hat{\mathbf{n}}_1\hat{\mathbf{n}}_2}}
\sim \sum_{m=-s}^s
\ket{\hat{\mathbf{n}}_1 m} \otimes \ket{\hat{\mathbf{n}}_2 m}
\, ,
\end{equation}
where $(\hat{\mathbf{n}}_i \cdot \mathbf{S})\ket{\hat{\mathbf{n}}_i m}=m \ket{\hat{\mathbf{n}}_i m}$ --- we define the phase of $\ket{\hat{\mathbf{n}}_i m}$ below.
One might think that in order to get a symmetric state (under exchange of the two tensor factors) a second term should be added, with the indices 1, 2 exchanged. In fact, the above state is already symmetric. To see this, assume for the moment that
$\hat{\mathbf{z}}$ bisects the angle between $\hat{\mathbf{n}}_1$, $\hat{\mathbf{n}}_2$, making an angle $\alpha$ with either, then
\begin{align}
\ket{\boldsymbol{\Psi}^\text{B}_{\hat{\mathbf{y}}}}
&=
\sum_{m=-s}^s
e^{-i \alpha S_x} \ket{\hat{\mathbf{z}} m} \otimes e^{i \alpha S_x} \ket{\hat{\mathbf{z}} m}
\nonumber
\\
&=
\sum_{m=-s}^s
\ket{\hat{\mathbf{n}}_1 m} \otimes  \ket{\hat{\mathbf{n}}_2 m}
\nonumber
\\
& \sim
\ket{\boldsymbol{\Psi}^\text{B}_{\hat{\mathbf{n}}_1\hat{\mathbf{n}}_2}}
\, ,
\end{align}
where the first equation is a direct consequence of~(\ref{SxPsiB}), and the second one defines the phase of $\ket{\hat{\mathbf{n}}_1 m} \equiv e^{-i \alpha S_x} \ket{\hat{\mathbf{z}} m}$, and similarly for $\ket{\hat{\mathbf{n}}_2 m}$. Thus, $\ket{\boldsymbol{\Psi}^\text{B}_{\hat{\mathbf{n}}_1\hat{\mathbf{n}}_2}}$ is indeed symmetric, because $\ket{\boldsymbol{\Psi}^\text{B}_{\hat{\mathbf{y}}}}$ is, and the general case is handled by rotating the one above.
\end{myexample}
\begin{myexample}{Multiconstellations for 3-symmetric qutrit states}{mc3sq}
The orthonormal basis in $\mathcal{H}_1^{\vee 3}$ is
\begin{equation}
\label{H1k3basis}
\begin{split}
\{
\hat{e}_{111},
\hat{e}_{112},
\hat{e}_{113},
\hat{e}_{122},
\hat{e}_{123},
\hat{e}_{133},&
\hat{e}_{222},
\hat{e}_{223},
\hat{e}_{233},
\hat{e}_{333}
\}
\\
&=
\{
e_{111},
\sqrt{3} e_{112},
\sqrt{3} e_{113},
\sqrt{3} e_{122},
\sqrt{6} e_{123},
\sqrt{3} e_{133},
e_{222},
\sqrt{3} e_{223},
\sqrt{3} e_{233},
e_{333}
\}
\end{split}
\, \, .
\end{equation}
From~(\ref{s1ss2}) we know that $\mathcal{H}_1^{\vee 3}$ decomposes into one spin-3 and one spin-1 subspace.
The highest $S_z$-eigenvalue eigenvector is $\hat{e}_{111}$, with eigenvalue 3. Applying $S_-$ six times, one generates the entire spin-3 multiplet. The $S_z$-eigenvalue 1 is doubly degenerate, the state orthogonal to $\hat{e}_{(3,1)}$ is $\hat{e}_{(1,1)}$ --- applying to it $S_-$ twice one obtains the entire spin-1 triplet. The matrix $U_{(1,3)}$ effecting the above change of basis is
\begin{equation}
\label{Us1k3m}
U_{(1,3)}=
\left(
\begin{array}{cccccccccc}
 1 & 0 & 0 & 0 & 0 & 0 & 0 & 0 & 0 & 0 \\
 0 & 1 & 0 & 0 & 0 & 0 & 0 & 0 & 0 & 0 \\
 0 & 0 & \frac{1}{\sqrt{5}} & \frac{2}{\sqrt{5}} & 0 & 0 & 0
   & 0 & 0 & 0 \\
 0 & 0 & 0 & 0 & \sqrt{\frac{3}{5}} & 0 & \sqrt{\frac{2}{5}}
   & 0 & 0 & 0 \\
 0 & 0 & 0 & 0 & 0 & \frac{1}{\sqrt{5}} & 0 &
   \frac{2}{\sqrt{5}} & 0 & 0 \\
 0 & 0 & 0 & 0 & 0 & 0 & 0 & 0 & 1 & 0 \\
 0 & 0 & 0 & 0 & 0 & 0 & 0 & 0 & 0 & 1 \\
 0 & 0 & \frac{2}{\sqrt{5}} & -\frac{1}{\sqrt{5}} & 0 & 0 & 0
   & 0 & 0 & 0 \\
 0 & 0 & 0 & 0 & \sqrt{\frac{2}{5}} & 0 & -\sqrt{\frac{3}{5}}
   & 0 & 0 & 0 \\
 0 & 0 & 0 & 0 & 0 & \frac{2}{\sqrt{5}} & 0 &
   -\frac{1}{\sqrt{5}} & 0 & 0 \\
\end{array}
\right)
\, .
\end{equation}
Consider, as an example,  the state
\begin{equation}
\label{Psis1k3ex}
\ket{\boldsymbol{\Psi}}=(\hat{e}_{111}+i \hat{e}_{113}+\hat{e}_{223})/\sqrt{3}
\, .
\end{equation}
Writing it as a column vector and left multiplying by $U_{(1,3)}$, we express it in the BD-basis,
\begin{equation}
\label{PsiBDs1k3}
\ket{\boldsymbol{\Psi}}_\text{BD}=
\left(
\ket{\psi^{(3)}}^T, \ket{\psi^{(1)}}^T
\right)^T
=
\left(
\left(
\frac{1}{\sqrt{3}},
0,
\frac{i}{\sqrt{15}},
0,
\frac{2}{\sqrt{15}},
0,
0
\right),
\left(
\frac{2i}{\sqrt{15}},
0,
-\frac{1}{\sqrt{15}}
\right)
\right)^T
\, ,
\end{equation}
where we used extra parentheses to visually define the spin-3 and spin-1 multiplets. Note that since there is no degeneracy (\ie, the multiplicities are binary-valued), we do not need the index $\alpha$ that labels the various appearances of one particular spin, and, hence, we omit it.

Each of the states $\ket{\psi^{(3)}}$, $\ket{\psi^{(1)}}$ has its own Majorana constellation. But the two states are not normalized to unity, and their constellations also miss the information about their phase. Both pieces of information are captured in the spectator spin-1/2 state $Z=(z_3,z_1)$, which we now determine.

The state $\ket{\psi^{(3)}}$ has constellation $C_3$ and modulus $|z_3|=\sqrt{\frac{2}{3}}$. Its spin expectation value is $\vec{S}=(0,0,\frac{4}{5})$ --- as it is already pointing along the positive $z$-axis, the rotation $R_1$ in the general algorithm outlined above is the identity, so, $C_3'=C_3$, and $\rho_3'=\rho_3$. The multipolar expansion of  $\rho_3'$ gives
\begin{equation}
\label{rho3multi}
\rho_3' \rightarrow
\left(
\left\{
\frac{2}{3\sqrt{7}}\right\},\left\{0,\frac{2}{5
   \sqrt{7}},0\right\},\left\{-\frac{i \left(5
   \sqrt{3}-12\right)}{45 \sqrt{14}},0,\frac{1}{3
   \sqrt{21}},0,\frac{i \left(5 \sqrt{3}-12\right)}{45
   \sqrt{14}}\right\}
   \, \ldots
\right)
\, .
\end{equation}
Note that the spin-1 component is of the form $\{ 0,\lambda,0\}$, with $\lambda>0$, as a result of the spin expectation value of $\rho_3'$ being along $z$.
The first nonzero component with $m\neq 0$ is the $m=2$ component of the spin-2 quintet, and it is complex, with phase $\pi/2$. Since this is an $m=2$ component, it will become real and positive if the state is rotated around $z$ by $\pi/4$, so that $D^{(3)}(R_2)=e^{-i \pi S^{(3)}_z/4}$. Indeed, $\tilde{\rho}_3=D^{(3)}(R_2) \rho_3' D^{(3)}(R_2)^{-1}$ expands to
\begin{equation}
\label{rho32multi}
\tilde{\rho}_3 \rightarrow
\left(
\left\{
\frac{2}{3\sqrt{7}}
\right\},
\left\{
0,
\frac{2}{5\sqrt{7}},
0
\right\},
\left\{
\frac{12-5 \sqrt{3}}{45\sqrt{14}},
0,
\frac{1}{3 \sqrt{21}},
0,
\frac{12-5\sqrt{3}}{45 \sqrt{14}}
\right\},
\ldots
\right)
\, ,
\end{equation}
and $\tilde{C}_3 \equiv R_2(C_3')=(R_2 \circ R_1)(C_3)$ has the reference orientation. We get a corresponding ket by rotating
$\ket{\psi^{(3)}}$,
\begin{equation}
\label{nrpsi3}
D^{(3)}(R_2 \circ R_1) \ket{\psi^{(3)}}=
\left(
-\frac{1+i}{\sqrt{6}},
0,
\frac{1+i}{\sqrt{30}},
0,
(1+i) \sqrt{\frac{2}{15}},
0,
0,
\right)^T
\, .
\end{equation}
We now fix the phase of this ket by demanding that the first nonzero component be real and positive, this is achieved by multiplying it by $e^{i 3\pi/4}$ --- the resulting ket is (after normalizing)
\begin{equation}
\label{psi3reforient}
\ket{\psi_{\tilde{C}_3}}
=
e^{i 3\pi/4} D^{(3)}(R_2 \circ R_1) \ket{\psi^{(3)}}/\sqrt{\frac{2}{3}}
=
\left(
\frac{1}{\sqrt{2}},
0,
-\frac{1}{\sqrt{10}},
0,
-\sqrt{\frac{2}{5}},
0,
0
\right)^T
\, .
\end{equation}
We finally obtain the canonical ket $\ket{\psi_{C_3}}$ by rotating $\ket{\psi_{\tilde{C}_3}}$,
\begin{equation}
\label{psi3C3}
\ket{\psi_{C_3}}=
D^{(3)}(R_2 \circ R_1)^{-1} \ket{\psi_{\tilde{C}_3}}
=
\left(
-\frac{1}{2}+\frac{i}{2},
0,
-\frac{\frac{1}{2}+\frac{i}{2}}{\sqrt{5}},
0,
-\frac{1-i}{\sqrt{5}},
0,
0
\right)^T
\, .
\end{equation}
Comparing with $\ket{\psi^{(3)}}$ we find $z^{(3)}=\sqrt{\frac{2}{3}}e^{-i 3\pi/4}$.

Working along the same lines we also find that $z_1=\frac{1}{\sqrt{3}}e^{i 3\pi/4}$, so that $z_1/z_3=e^{i 3\pi/2}/\sqrt{2}$ --- this is the stereographic image of the single-star spectator constellation. A plot of the constellations of $\ket{\psi^{(3)}}$, $\ket{\psi^{(1)}}$ appears in Fig.~\ref{fig:psi31StarsPlot} (left and middle), while all constellations of $\ket{\boldsymbol{\Psi}}$, including the spectator one, are shown in that same figure on the right.
\begin{figure}[h]
\includegraphics[width=.32\linewidth]{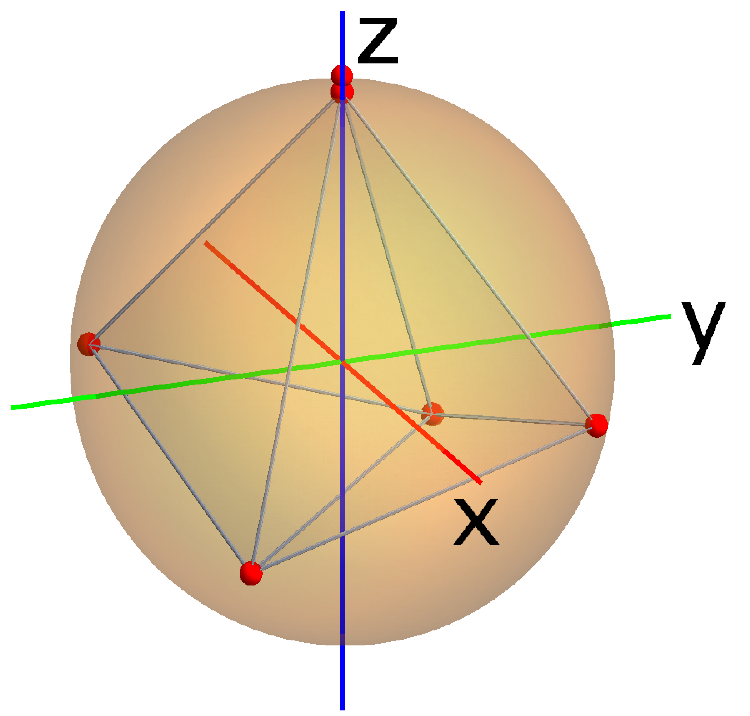}
\includegraphics[width=.32\linewidth]{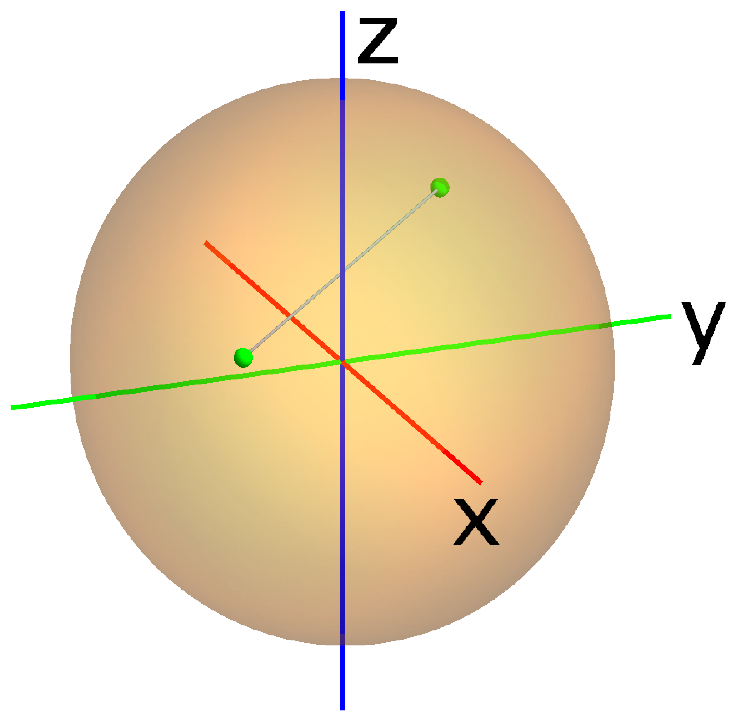}
\includegraphics[width=.32\linewidth]{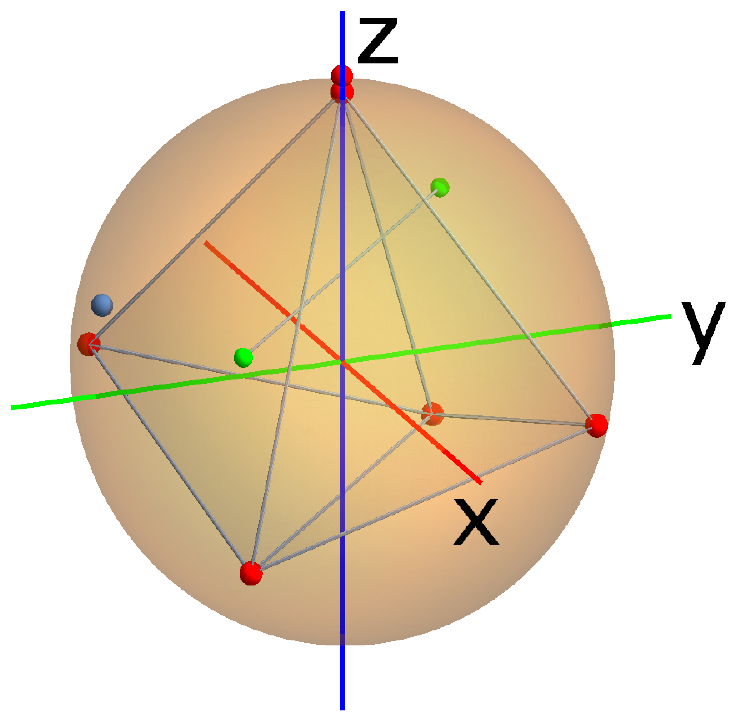}
\caption{%
Multiconstellation of the 3-qutrit symmetric state $\ket{\boldsymbol{\Psi}}$ given in~(\ref{Psis1k3ex}).
Constellations of the spin-3 component $\ket{\psi^{(3)}}$ (left) --- note the double star at the north pole, the spin-1 component $\ket{\psi^{(1)}}$ (middle)
(see Eq.~(\ref{PsiBDs1k3})), and both superimposed, together with the spectator constellation (single blue dot) (right).
}
\label{fig:psi31StarsPlot}
\end{figure}
\end{myexample}
\begin{myexample}{Octahedral 3-qutrit state}{T3qs}
A symmetric 3-qutrit state decomposes into one spin-3 and one spin-1 component (see~(\ref{s2ss})). We fix the spin-3 component to be the octahedral state, and set the spin-1 component to zero, so that in the BD basis we have (ignoring normalization)
\begin{equation}
\label{psioctaBD}
\ket{\boldsymbol{\Psi}^{\text{octa}}}_\text{BD}
=
\left(
\big(
0, -\frac{1}{\sqrt{6}}, 0, 0, 0, \frac{1}{\sqrt{6}}, 0
\big),
\big(
0, 0, 0
\big)
\right)^T
\, .
\end{equation}
We switch to the standard basis by left-multiplying with the inverse of the matrix $U_{(1,3)}$ in~(\ref{Us1k3m}), to get
\begin{align}
\label{psioctaeh}
\ket{\boldsymbol{\Psi}^{\text{octa}}}_\text{S}
&=
U^{-1}_{(1,3)} \ket{\boldsymbol{\Psi}^{\text{octa}}}_\text{BD}
\nonumber
\\
&=
\left(
0, -\frac{1}{\sqrt{6}}, 0, 0, 0, 0, 0, 0, \frac{1}{\sqrt{6}}, 0
\right)^T
\nonumber
\\
& \sim
-\hat{e}_{112}+\hat{e}_{233}
\nonumber
\\
&\sim
-e_{112} +  e_{233}
\nonumber
\\
&\sim
-\ket{1,1,0}-\ket{1,0,1}-\ket{0,1,1}+\ket{0,-1,-1}+\ket{- \! 1,0,-1}+\ket{- \! 1,-1,0}
\, .
\end{align}
Direct computation shows that this state is $\vee$-factorizable,
\begin{equation}
\label{octafact}
\ket{\boldsymbol{\Psi}^{\text{octa}}}=(1,0,-1)\vee (1,0,1) \vee (0,-1,0)
\, ,
\end{equation}
so that it could also be described by the constellations of the individual factors in~(\ref{octafact}). Interestingly, these are pairs of stars at $(\pm 1,0,0)$, $(0,\pm 1,0)$, $(0,0,\pm 1)$, respectively, so, taken together, they define the same  octahedron as the one above.
\end{myexample}
\begin{myexample}{Cubic 4-qutrit state}{C4qs}
A symmetric 4-qutrit state decomposes in spin-4, spin-2 and spin-0 components. We fix the spin-4 component to be the cubic one, and set the others to zero, so that, in the BD basis we have
\begin{equation}
\label{pscubicBD}
\ket{\boldsymbol{\Psi}^{\text{cube}}}_\text{BD}
=
\left(
\big(
1,0,0,0, \sqrt{\frac{14}{5}}, 0, 0, 0, 1
\big),
\big(
0, 0, 0, 0, 0
\big),
\big(
0
\big)
\right)^T
\, ,
\end{equation}
giving, in the induced basis,
\begin{equation}
\label{pscubic}
\ket{\boldsymbol{\Psi}^{\text{cube}}}_\text{I}
=
\left(
1,0,0,0, 0, \frac{1}{5}, 0, \frac{2}{5}, 0, 0, \frac{4}{5}, 0, 0, 0, 1
\right)^T
\, ,
\end{equation}
\ie,
\begin{equation}
\label{pscubicie}
\ket{\boldsymbol{\Psi}^{\text{cube}}}
=
e_{1111}+\frac{1}{5}e_{1133}+\frac{2}{5} e_{1223}+\frac{4}{5} e_{2222}+e_{3333}
\, ,
\end{equation}
which can be shown to be non $\vee$-factorizable.
\end{myexample}
\begin{myexample}{2-symmetric spin-3/2 states}{tssths}
The standard basis in $\mathcal{H}^{\vee 2}$ is
\begin{equation}
\label{}
\{
\hat{e}_{11},
\hat{e}_{12},
\hat{e}_{13},
\hat{e}_{14},
\hat{e}_{22},
\hat{e}_{23},
\hat{e}_{24},
\hat{e}_{33},
\hat{e}_{34},
\hat{e}_{44}
\}
=
\{
e_{11},
\sqrt{2} e_{12},
\sqrt{2} e_{13},
\sqrt{2} e_{14},
e_{22},
\sqrt{2} e_{23},
\sqrt{2} e_{24},
e_{33},
\sqrt{2} e_{34},
e_{44}
\}
\, .
\end{equation}
The highest $S_z$-eigenvalue eigenvector is $\hat{e}_{11}$, with eigenvalue 3. Applying $S_-$ six times, one generates the entire spin-3 multiplet. The $S_z$-eigenvalue 1 is doubly degenerate, the state orthogonal to $\hat{e}_{(3,1)}$ is $\hat{e}_{(1,1)}$ --- applying to it $S_-$ twice one obtains the entire spin-1 triplet. The matrix $U_{(\frac{3}{2},2)}$ effecting the above change of basis is
\begin{equation}
\label{Us32k2m}
U_{(\frac{3}{2}, 2)}=
\left(
\begin{array}{cccccccccc}
 1 & 0 & 0 & 0 & 0 & 0 & 0 & 0 & 0 &
   0 \\
 0 & 1 & 0 & 0 & 0 & 0 & 0 & 0 & 0 &
   0 \\
 0 & 0 & \sqrt{\frac{2}{5}} & 0 &
   \sqrt{\frac{3}{5}} & 0 & 0 & 0 & 0
   & 0 \\
 0 & 0 & 0 & \frac{1}{\sqrt{10}} & 0
   & \frac{3}{\sqrt{10}} & 0 & 0 & 0
   & 0 \\
 0 & 0 & 0 & 0 & 0 & 0 &
   \sqrt{\frac{2}{5}} &
   \sqrt{\frac{3}{5}} & 0 & 0 \\
 0 & 0 & 0 & 0 & 0 & 0 & 0 & 0 & 1 &
   0 \\
 0 & 0 & 0 & 0 & 0 & 0 & 0 & 0 & 0 &
   1 \\
 0 & 0 & \sqrt{\frac{3}{5}} & 0 &
   -\sqrt{\frac{2}{5}} & 0 & 0 & 0 &
   0 & 0 \\
 0 & 0 & 0 & \frac{3}{\sqrt{10}} & 0
   & -\frac{1}{\sqrt{10}} & 0 & 0 & 0
   & 0 \\
 0 & 0 & 0 & 0 & 0 & 0 &
   \sqrt{\frac{3}{5}} &
   -\sqrt{\frac{2}{5}} & 0 & 0 \\
\end{array}
\right)
\, .
\end{equation}
\end{myexample}
\section{A closer look at Hermite reciprocity and its generalizations}
\label{AclatHr}
\subsection{The Hermite and Murnaghan isomorphisms}
\label{Hics12p}
The original appearance of Hermite reciprocity was in~\cite{Her:54}, which dealt with covariants of homogeneous forms in two variables, and built upon earlier work by Cayley and Sylvester --- in modern parlance, and adapted to our context, it states that
\begin{equation}
\label{Hr}
\vee^k \left( \vee^{2s} \mathcal{H}_{\frac{1}{2}} \right)
\sim
\vee^{2s} \left( \vee^k \mathcal{H}_{\frac{1}{2}} \right)
\, ,
\end{equation}
where $\mathcal{H}_{\frac{1}{2}}$ is the spin-1/2 Hilbert space (see, \eg,~\cite{Wyb:69,Wyb:70} for concrete applications). Taking into account that the symmetrized $2s$-fold tensor product of $\mathcal{H}_{\frac{1}{2}}$ is isomorphic to the spin-$s$ Hilbert space,
\begin{equation}
\label{HsymmHs}
\mathcal{H}_{s} \sim \vee^{2s} \mathcal{H}_{\frac{1}{2}}
\, ,
\end{equation}
 (\ref{Hr}) implies that there exists an isomorphism $h$,
 \begin{equation}
 \label{Hrbis}
h \colon \vee^k \mathcal{H}_{s} \rightarrow \vee^{2s} \mathcal{H}_{\frac{k}{2}}
\, .
\end{equation}
What this means is that $h$ connects states that transform identically under rotations, \ie, it intertwines the action of $SU(2)$,
\begin{equation}
\label{interh}
h\left( \mathfrak{D}^{(s,k)}(g) \ket{\boldsymbol{\Psi}} \right)
=
\mathfrak{D}^{(\frac{k}{2},2s)}(g) h(\ket{\boldsymbol{\Psi}})
\, ,
\end{equation}
for all $g$ in $SU(2)$ and all $\ket{\boldsymbol{\Psi}}$ in $\mathcal{H}_s^{\vee k}$.
Thus, for every $k$-symmetric spin-$s$ state, there exists a $2s$-symmetric spin-$k/2$ state from which it is indistinguishable, as far as rotations are concerned --- this statement generalizes~(\ref{HsymmHs}), to which it reduces for $k=1$. The \emph{raison d'\^etre} (or, depending on the viewpoint, a consequence) of this isomorphism is that the two sides of (\ref{Hrbis}) have identical irreducible components, so that a vector in $\vee^k \mathcal{H}_{s}$, written in the BD basis, may automatically be considered an element of $\vee^{2s} \mathcal{H}_{\frac{k}{2}}$, also in the BD basis, with the exact same components. In other words, when the BD basis is used in both vector spaces involved, the isomorphism is implemented by the unit matrix,
\begin{equation}
\label{hBD}
h(\ket{\boldsymbol{\Psi}})^{(\frac{k}{2},2s)}_\text{BD}=\ket{\boldsymbol{\Psi}}^{(s,k)}_\text{BD}
\, .
\end{equation}
We also have
\begin{equation}
\label{hS}
\ket{\boldsymbol{\Psi}}^{(s,k)}_\text{BD}=U_{(s,k)} \ket{\boldsymbol{\Psi}}^{(s,k)}_\text{S}
\, ,
\qquad
h(\ket{\boldsymbol{\Psi}})^{(\frac{k}{2},2s)}_\text{S}= U_{(\frac{k}{2},2s)}^\dagger h(\ket{\boldsymbol{\Psi}})^{(\frac{k}{2},2s)}_\text{BD}
\, ,
\end{equation}
which, together with~(\ref{hBD}), gives for the matrix $H^{(s,k)}_{\text{S}}$ implementing $h$, when the standard basis is used in both vector spaces,
\begin{equation}
\label{HM}
h(\ket{\boldsymbol{\Psi}})^{(\frac{k}{2},2s)}_\text{S}=H^{(s,k)}_{\text{S}} \ket{\boldsymbol{\Psi}}^{(s,k)}_\text{S}
\, ,
\qquad
H^{(s,k)}_{\text{S}} \equiv U_{(\frac{k}{2},2s)}^\dagger U_{(s,k)}
\, .
\end{equation}
\begin{myexample}{Hermite isomorphism $\mathbf{3}^{\vee 3} \sim \mathbf{4}^{\vee 2}$}{Hiso}
We put together the results of examples~\ref{mc3sq}, \ref{tssths}, to find the unitary matrix implementing the isomorphism involved in the Hermite reciprocity $\mathcal{H}_1^{\vee 3} \sim \mathcal{H}_{\frac{3}{2}}^{\vee 2}$. Given a vector in $\mathcal{H}_1^{\vee 3}$, in the standard  basis, multiplication by $U_{(1,3)}$ of~(\ref{Us1k3m}) transforms it in the BD basis. Regarding this transformed spinor as an element of $\mathcal{H}_{\frac{3}{2}}^{\vee 2}$, in the BD basis, multiplication by $U_{(\frac{3}{2}, 2)}^\dagger$ (see~(\ref{Us32k2m})) transforms it in the standard basis of $\mathcal{H}_{\frac{3}{2}}^{\vee 2}$. Thus, the matrix
\begin{align}
H^{(1,3)}_{\text{S}}
&=
U_{(\frac{3}{2},2)}^\dagger U_{(1,3)}
\nonumber
\\
&=
\left(
\begin{array}{cccccccccc}
 1 & 0 & 0 & 0 & 0 & 0 & 0 & 0 & 0 & 0
   \\
 0 & 1 & 0 & 0 & 0 & 0 & 0 & 0 & 0 & 0
   \\
 0 & 0 & \frac{1}{5} \left(\sqrt{2}+2
   \sqrt{3}\right) & \frac{1}{5}
   \left(2 \sqrt{2}-\sqrt{3}\right) &
   0 & 0 & 0 & 0 & 0 & 0 \\
 0 & 0 & 0 & 0 & \frac{1}{10}
   \left(6+\sqrt{6}\right) & 0 &
   \frac{1}{10} \left(2-3
   \sqrt{6}\right) & 0 & 0 & 0 \\
 0 & 0 & \frac{1}{5} \left(-2
   \sqrt{2}+\sqrt{3}\right) &
   \frac{1}{5} \left(\sqrt{2}+2
   \sqrt{3}\right) & 0 & 0 & 0 & 0 & 0
   & 0 \\
 0 & 0 & 0 & 0 & \frac{1}{10}
   \left(-2+3 \sqrt{6}\right) & 0 &
   \frac{1}{10}
   \left(6+\sqrt{6}\right) & 0 & 0 & 0
   \\
 0 & 0 & 0 & 0 & 0 & \frac{1}{5}
   \left(\sqrt{2}+2 \sqrt{3}\right) &
   0 & \frac{1}{5} \left(2
   \sqrt{2}-\sqrt{3}\right) & 0 & 0 \\
 0 & 0 & 0 & 0 & 0 & \frac{1}{5}
   \left(-2 \sqrt{2}+\sqrt{3}\right) &
   0 & \frac{1}{5} \left(\sqrt{2}+2
   \sqrt{3}\right) & 0 & 0 \\
 0 & 0 & 0 & 0 & 0 & 0 & 0 & 0 & 1 & 0
   \\
 0 & 0 & 0 & 0 & 0 & 0 & 0 & 0 & 0 & 1
   \\
\end{array}
\right)
\label{HrM}
\, ,
\end{align}
implements the Hermite isomorphism $h \colon \mathcal{H}_{1}^{\vee 3} \rightarrow \mathcal{H}_{\frac{3}{2}}^{\vee 2}$  in the standard normalized bases. For example, looking at the first and third column of the above matrix we deduce, respectively,  that
\begin{equation}
\label{Hisoex}
h(\hat{e}_{111})=\hat{e}_{11}
\, ,
\qquad
h(\hat{e}_{113})=
\frac{1}{5} \left(\sqrt{2}+2\sqrt{3}\right)
\hat{e}_{13}
+
\frac{1}{5} \left(-2 \sqrt{2}+\sqrt{3}\right)
\hat{e}_{22}
\, .
\end{equation}
For example, for the octahedral 3-qutrit state considered in Example \ref{T3qs}, we have
\begin{align}
h\left( \ket{\boldsymbol{\Psi}^{\text{octa}}} \right) & \sim  h(\left( -\hat{e}_{112}+\hat{e}_{233} \right) 
\nonumber
\\
& \sim 
-\hat{e}_{12}+\hat{e}_{34}
\nonumber
\\
\label{h.octa}
& \sim
-\ket{3/2 , 1/2}- \ket{1/2 , 3/2}+\ket{ \! - \! 1/2 , -3/2}+ \ket{\! - \! 3/2 , -1/2} \, ,
\end{align}
the last line above using $\ket{m_1,m_2}$-notation.
Although $h\left( \ket{\boldsymbol{\Psi}^{\text{octa}}} \right)$ has the same multiconstellation as 
$\ket{\boldsymbol{\Psi}^{\text{octa}}}$, it is not $\vee-$factorizable.  In fact, $h\left( \ket{\boldsymbol{\Psi}^{\text{octa}}} \right)$ is maximally entangled, with~\eqref{h.octa} giving explicitly its Schmidt decomposition. Note that, in general, the Hermite  isomorphism does not respect $\vee$-factorizability. 
\end{myexample}

A  Hermite-like isomorphism $m$ has been proven by Murnaghan in~\cite{Mur:51,Mur:62} --- what sets it apart is that it relates $k$-fold spin-$s$ symmetric products with $k$-fold \emph{antisymmetric} ones of a higher spin,
\begin{equation}
\label{Murnaghan1}
m \colon \mathcal{H}^{\vee k}_s
\rightarrow
\mathcal{H}^{\wedge k}_{s+\frac{k-1}{2}}
\, ,
\end{equation}
where the $\wedge$-product stands for an antisymmetrized tensor product, \eg, for $v$, $w \in \mathcal{H}$, $v \wedge w=v \otimes w-w \otimes v \in \mathcal{H}^{\wedge 2}$.
An interesting consequence is that the \rhs{} of the above relation may be further mapped isomorphically to the complementary orthogonal subspace, via $\varphi$ of~(\ref{dualkplanes}), and then, via the inverse Murnaghan isomorphism, back to a $\vee$-product, thus giving a sequence of isomorphisms (dotted path in the diagram below),
\begin{equation}
\label{seqiso}
\begin{tikzcd}[row sep=large, column sep=large]
\mathcal{H}^{\vee k}_s
\arrow[d, dotted, "m"']
\arrow[r,"h"]
&
\mathcal{H}^{\vee 2s}_{\frac{k}{2}}
\\
\mathcal{H}^{\wedge k}_{s+\frac{k-1}{2}}
\arrow[r, dotted, "\varphi"']
&
\mathcal{H}^{\wedge 2s}_{s+\frac{k-1}{2}}
\arrow[u, dotted, "m^{-1}"']
\end{tikzcd}
\, ,
\end{equation}
which is equivalent to the Hermite isomorphism (horizontal solid path in the diagram), $h=m^{-1} \circ \varphi \circ m$.
Thus, the question of an intuitive interpretation of $h$, geometric or not, may be rephrased in terms of that of $m$.

\begin{myexample}{Murnaghan isomorphism $\mathbf{3}^{\vee 2} \sim \mathbf{4}^{\wedge 2}$}{Miso}
It has been shown in~\cite{Chr.Guz.Han.Ser:21} that the unitary matrix $\tilde{U}_{(\frac{3}{2},2)}$ that block-diagonalizes the standard basis in $\mathcal{H}^{\wedge 2}_{\frac{3}{2}}$ is given by
\begin{equation}
\label{Ut32w2}
\tilde{U}_{(\frac{3}{2},2)}
=
\left(
\begin{array}{cccccc}
 1 & 0 & 0 & 0 & 0 & 0 \\
 0 & 1 & 0 & 0 & 0 & 0 \\
 0 & 0 & \frac{1}{\sqrt{2}} &
   \frac{1}{\sqrt{2}} & 0 & 0 \\
 0 & 0 & 0 & 0 & 1 & 0 \\
 0 & 0 & 0 & 0 & 0 & 1 \\
 0 & 0 & \frac{1}{\sqrt{2}} &
   -\frac{1}{\sqrt{2}} & 0 & 0 \\
\end{array}
\right)
\, .
\end{equation}
The corresponding matrix for $\mathcal{H}^{\vee 2}_1$, $U_{(1,2)}$, was found in Ex.~\ref{mc2sq} (see Eq.~\ref{Us1k2m}).
Thus, an argument analogous to that used in Ex.~\ref{Hiso} shows that the Murnaghan isomorphism $m\colon \mathcal{H}_1^{\vee 2} \rightarrow \mathcal{H}_{\frac{3}{2}}^{\wedge 2}$, when the standard bases are used in both spaces, is implemented by the matrix
\begin{equation}
\label{Ms32s1}
M_\text{S}^{(1,2)}
=
\tilde{U}_{(\frac{3}{2},2)}^{-1}U_{(1,2)}
=
\left(
\begin{array}{cccccc}
 1 & 0 & 0 & 0 & 0 & 0 \\
 0 & 1 & 0 & 0 & 0 & 0 \\
 0 & 0 &
   -\frac{1}{\sqrt{3}}+\frac{1}{\sqrt
   {6}} &
   \frac{1}{\sqrt{3}}+\frac{1}{\sqrt{
   6}} & 0 & 0 \\
 0 & 0 &
   \frac{1}{\sqrt{3}}+\frac{1}{\sqrt{
   6}} &
   \frac{1}{\sqrt{3}}-\frac{1}{\sqrt{
   6}} & 0 & 0 \\
 0 & 0 & 0 & 0 & 1 & 0 \\
 0 & 0 & 0 & 0 & 0 & 1 \\
\end{array}
\right)
\, .
\end{equation}
Looking, for example, at the first column of this matrix, we infer that $m(\ket{1,1} \vee \ket{1,1})=\frac{1}{\sqrt{2}}\ket{\frac{3}{2},\frac{3}{2}} \wedge \ket{\frac{3}{2},\frac{1}{2}}$. It is instructive to verify the isomorphism considering all spin  states as made up of spin-1/2 ones, we get then the embeddings
\begin{align}
\label{embedM}
\ket{1,1} \vee \ket{1,1}
&=
\ket{++} \otimes \ket{++}
\nonumber
\\
&\equiv
\ket{++++}
\nonumber
\\
&
\equiv \ket{2}_{1234}
\, ,
\\
\frac{1}{\sqrt{2}}\ket{\frac{3}{2},\frac{3}{2}} \wedge \ket{\frac{3}{2},\frac{1}{2}}
&=
\frac{1}{\sqrt{2}}
\left(
\ket{\frac{3}{2},\frac{3}{2}} \otimes \ket{\frac{3}{2},\frac{1}{2}}
-
\ket{\frac{3}{2},\frac{1}{2}} \otimes \ket{\frac{3}{2},\frac{3}{2}}
\right)
\nonumber
\\
&=
\frac{1}{\sqrt{6}} \left(
\ket{+++} \otimes (\ket{++-}+\ket{+-+}+\ket{-++}) \right.
\nonumber
\\
& \qquad \qquad \left. -(\ket{++-}+\ket{+-+}+\ket{-++})\otimes\ket{+++}
\right)
\nonumber
\\
 &\equiv
 \frac{1}{\sqrt{6}}
 \left(
 \ket{+++++-}+\ket{++++-+}+\ket{+++-++}
 \right.
 \nonumber
 \\
 &
 \qquad \qquad
 \left.
 -\ket{++-+++}-\ket{+-++++}-\ket{-+++++}
 \right)
 \nonumber
 \\
 &=
 \frac{1}{\sqrt{3}}
 \left(
 \ket{0}_{16}\ket{2}_{2345}+\ket{0}_{25} \ket{2}_{1346}+\ket{0}_{34}\ket{2}_{1256}
 \right)
 \label{embedM2}
 \, ,
 \end{align}
 where we use the abbreviation $\ket{0}_{ij}$ to denote the spin-0 state $(\ket{\! +-}-\ket{-+})/\sqrt{2}$ embedded in tensor factors $i$, $j$, and, similarly, $\ket{2}_{ijkl}$ denotes the spin-2 state $\ket{++++}$, embedded in tensor factors $i$, $j$, $k$, $l$. It is clear that the two states in the \rhs{} of (\ref{embedM}), (\ref{embedM2}), respectively, transform identically under the action of $SU(2)$, since they only differ by spin-0 factors, and inconsequential permutations of the tensor factors. Similar comments hold true for states related by the Hermite isomorphism. A notable difference between the two cases though is that both sides of~(\ref{Hrbis}) get embedded in $\mathcal{H}_{\frac{1}{2}}^{\otimes 2ks}$, while, in~(\ref{Murnaghan1}), the \lhs{} embeds in $\mathcal{H}_{\frac{1}{2}}^{\otimes 2ks}$, but the \rhs{} embeds in $\mathcal{H}_{\frac{1}{2}}^{\otimes k(2s+k-1)}$.
\end{myexample}
\subsection{The constituent spin-1/2 picture}
\label{Tcs12p}
Example \ref{Miso} suggests that considering the constituent spin-1/2 picture, in which a spin-$s$ state is represented as a symmetrized tensor product of $2s$ spin-1/2 states, might shed some light on the origin of the Hermite and Murnaghan isomorphisms. We may depict the spin-1/2 states as vertices in a graph, their tensor product by simple juxtaposition of the vertices, the symmetrization of any number of them by enclosing the vertices in a circular blob, and the antisymmetrization of any pair of them by a directed segment connecting  them,
\begin{equation}
\label{psi12vw}
\ket{\psi_1} \mapsto
\text{\raisebox{-.2ex}{
\begin{tikzpicture}
\filldraw[black] (0,0) circle (1pt) node[anchor=south] {\small{1}};
\end{tikzpicture}
}},
\qquad
\ket{\psi_1} \otimes \ket{\psi_2}\mapsto
\text{
\raisebox{-.2ex}{
\begin{tikzpicture}
\filldraw[black] (0,0) circle (1pt)
node[anchor=south] {\small{1}};
\filldraw[black] (.3,0) circle (1pt)
node[anchor=south] {\small{2}};
\end{tikzpicture}
}},
\qquad
\ket{\psi_1} \vee \ket{\psi_2}  \mapsto
\text{
\raisebox{-1.4ex}{
\begin{tikzpicture}
\coordinate (x) at (0,.03);
\coordinate (y) at (1,.03);
\draw[black] (x) circle (9pt);
\filldraw[black] (-.1,-.1) circle (1pt) node[anchor=south]{\small{1}};
\filldraw[black] (.1,-.1) circle (1pt) node[anchor=south]{\small{2}};
\end{tikzpicture}
}},
\qquad
\ket{\psi_1} \wedge \ket{\psi_2}  \mapsto
\text{
\raisebox{-.2ex}{
\begin{tikzpicture}
\draw[black, thick, directed] (0,0) -- (.4,0);
\filldraw[black] (0,0) circle (1pt)
node[anchor=south] {\small{1}};
\filldraw[black] (.4,0) circle (1pt)
node[anchor=south] {\small{2}};
\end{tikzpicture}
}},
\end{equation}
where overall normalization factors are unimportant (this allows omission of arrows, when flipping them  corresponds to an overall sign change). Note that a blob containing $2s$ vertices represents a spin-$s$ state, and \emph{vice-versa}. We denote (anti)symmetrization of blobs explicitly by $\vee (\wedge)$, so that, \eg, for two qutrit states
\begin{equation}
\label{tqs}
\ket{\psi_a} \sim \ket{\hat{\mathbf{n}}_1} \vee \ket{\hat{\mathbf{n}}_2}
\mapsto
\text{
\raisebox{-1.4ex}{
\begin{tikzpicture}
\coordinate (x) at (0,.03);
\coordinate (y) at (1,.03);
\draw[black] (x) circle (9pt);
\filldraw[black] (-.1,-.1) circle (1pt) node[anchor=south]{\small{1}};
\filldraw[black] (.1,-.1) circle (1pt) node[anchor=south]{\small{2}};
\end{tikzpicture}
}},
\qquad
\ket{\psi_b} \sim \ket{\hat{\mathbf{n}}_3} \vee \ket{\hat{\mathbf{n}}_4}
\mapsto
\text{
\raisebox{-1.4ex}{
\begin{tikzpicture}
\coordinate (x) at (0,.03);
\coordinate (y) at (1,.03);
\draw[black] (y) circle (9pt);
\filldraw[black] (.9,-.1) circle (1pt) node[anchor=south]{\small{3}};
\filldraw[black] (1.1,-.1) circle (1pt) node[anchor=south]{\small{4}};
\end{tikzpicture}
}},
\end{equation}
(with $\ket{\hat{\mathbf{n}}_i}$ denoting spin-1/2 states) the symmetrized 2-qutrit state $\ket{\Psi} \sim \ket{\psi_a} \vee \ket{\psi_b}$ is denoted by
\begin{equation}
\label{Psiex}
\ket{\Psi} \mapsto
\text{
\raisebox{-1.4ex}{
\begin{tikzpicture}
\coordinate (x) at (0,.03);
\coordinate (y) at (1,.03);
\draw[black] (x) circle (9pt);
\filldraw[black] (-.1,-.1) circle (1pt) node[anchor=south]{\small{1}};
\filldraw[black] (.1,-.1) circle (1pt) node[anchor=south]{\small{2}};
\end{tikzpicture}
}}
\vee
\text{
\raisebox{-1.4ex}{
\begin{tikzpicture}
\coordinate (x) at (0,.03);
\coordinate (y) at (1,.03);
\draw[black] (y) circle (9pt);
\filldraw[black] (.9,-.1) circle (1pt) node[anchor=south]{\small{3}};
\filldraw[black] (1.1,-.1) circle (1pt) node[anchor=south]{\small{4}};
\end{tikzpicture}
}}.
\end{equation}
One last operation that we depict graphically is the antisymmetrization of, say, one spin-1/2 state, belonging to a blob $A$, with another spin-1/2 state, belonging to a different blob $B$ (antisymmetrizing two states in the same blob gives zero, as states within a blob are symmetrized). The antisymmetrization in question has to respect the symmetrization within blobs, so it actually involves a sum of antisymmetrizations over all pairs of spin-1/2 states, where the first element of the pair belongs to $A$, and the second to $B$ --- the resulting state is denoted by a directed segment connecting the blobs, so that, \eg, (in simplified notation)
\begin{eqnarray}
\text{
\raisebox{-1.8ex}{
\begin{tikzpicture}
\coordinate (x) at (0,.03);
\coordinate (y) at (1,.03);
\draw[black] (x) circle (9pt);
\filldraw[black] (-.1,-.1) circle (1pt) node[anchor=south]{\small{1}};
\filldraw[black] (.1,-.1) circle (1pt) node[anchor=south]{\small{2}};
\draw[black, thick, directed] (.3,.03) -- (.7,.03);
\draw[black] (y) circle (9pt);
\filldraw[black] (.9,-.1) circle (1pt) node[anchor=south]{\small{3}};
\filldraw[black] (1.1,-.1) circle (1pt) node[anchor=south]{\small{4}};
\end{tikzpicture}
}}
& \sim &
1234-3214+1234-4231+1234-1324+1234-1432
\nonumber
\\
 & = &
 (1 \wedge 3) \, 24+(1\wedge 4) \, 23 + (2 \wedge 3) \, 14 + (2 \wedge 4) \,13
 \, ,
 \label{twoblobsanti}
\end{eqnarray}
where, in the second line, it is understood that the state $\ket{\hat{\mathbf{n}}_i}$ resides in the $i$-th tensor factor,
while
\begin{equation}
\label{twoblobsantisym}
\text{
\raisebox{-1.8ex}{
\begin{tikzpicture}
\coordinate (x) at (0,.03);
\coordinate (y) at (1,.03);
\draw[black] (x) circle (9pt);
\filldraw[black] (-.1,-.1) circle (1pt) node[anchor=south]{\small{1}};
\filldraw[black] (.1,-.1) circle (1pt) node[anchor=south]{\small{2}};
\draw[black, thick, directed] (.3,.03) -- (.7,.03) node[anchor=south]{$\! \! \! \! \! \! \vee$};
\draw[black] (y) circle (9pt);
\filldraw[black] (.9,-.1) circle (1pt) node[anchor=south]{\small{3}};
\filldraw[black] (1.1,-.1) circle (1pt) node[anchor=south]{\small{4}};
\end{tikzpicture}
}}
=
\text{
\raisebox{-1.8ex}{
\begin{tikzpicture}
\coordinate (x) at (0,.03);
\coordinate (y) at (1,.03);
\draw[black] (x) circle (9pt);
\filldraw[black] (-.1,-.1) circle (1pt) node[anchor=south]{\small{1}};
\filldraw[black] (.1,-.1) circle (1pt) node[anchor=south]{\small{2}};
\draw[black, thick, directed] (.3,.03) -- (.7,.03);
\draw[black] (y) circle (9pt);
\filldraw[black] (.9,-.1) circle (1pt) node[anchor=south]{\small{3}};
\filldraw[black] (1.1,-.1) circle (1pt) node[anchor=south]{\small{4}};
\end{tikzpicture}
}}
+
\text{
\raisebox{-1.8ex}{
\begin{tikzpicture}
\coordinate (x) at (0,.03);
\coordinate (y) at (1,.03);
\draw[black] (x) circle (9pt);
\filldraw[black] (-.1,-.1) circle (1pt) node[anchor=south]{\small{3}};
\filldraw[black] (.1,-.1) circle (1pt) node[anchor=south]{\small{4}};
\draw[black, thick, directed] (.3,.03) -- (.7,.03);
\draw[black] (y) circle (9pt);
\filldraw[black] (.9,-.1) circle (1pt) node[anchor=south]{\small{1}};
\filldraw[black] (1.1,-.1) circle (1pt) node[anchor=south]{\small{2}};
\end{tikzpicture}
}}
\sim
(1\wedge 3)  (2\wedge 4) +(1 \wedge 4)(2 \wedge 3)
\, .
\end{equation}
\begin{myexample}{Irreducible components of $\mathbf{4}^{\vee 3}$ in the constituent spin-1/2 picture}{Icincsp}
We show here how the above graphical representation of symmetry operations captures in a very compact way rather complicated operations. We focus on the decomposition in~(\ref{s32ss2}), $\mathbf{4}^{\vee 3}=
\mathbf{10} \oplus \mathbf{6} \oplus \mathbf{4}$, and seek to express a particular state in the \rhs{} of that expression, say, the spin-3/2 state with $S_z=3/2$, in terms of constituent spin-1/2 states. We work out the relation between the induced and the BD bases, and find, in standard notation,
\begin{align}
\label{s3232ib}
\ket{\boldsymbol{\Psi}}=\left( (0,0,0,0,0,0,0,0,0,0),(0,0,0,0,0,0),(1,0,0,0) \right)^T_{\text{BD}}
&=
3\sqrt{\frac{2}{35}} \hat{e}_{114}
+2 \sqrt{\frac{2}{35}} \hat{e}_{222}-\frac{3}{\sqrt{35}} \hat{e}_{123}
\, ,
\end{align}
with, \eg,
\begin{equation}
\label{e114exp}
\hat{e}_{114} \sim \ket{+ + + + + + - - \, \, - } +\ket{+ + + - - - + + \, \, + }+\ket{- - - + + + + + \, \, + }
\, ,
\end{equation}
and similar expressions for the embedding of $\hat{e}_{222}$, $\hat{e}_{123}$ in $\mathcal{H}_{1/2}^{\otimes 9}$ --- when written out explicitly, (\ref{s3232ib}) involves 84 terms. Now, $\ket{\boldsymbol{\Psi}}$ is a nine spin-1/2 state, that transforms like $\ket{\frac{3}{2},\frac{3}{2}}$, so six of the nine spins must form three spin-0 antisymmetric pairs while the remaining three spins, properly symmetrized, give rise to the spin-3/2 --- the question is how to factorize the above 84 terms in $\ket{\boldsymbol{\Psi}}$ so as to make this explicit. The answer may be obtained starting with the state $(+_1 \wedge -_4)(+_5 \wedge -_2) (+_6 \wedge -_7)+_3 +_8 +_9$ and then symmetrizing $(123)$ among themselves, similarly $(456)$ and $(789)$, and, finally, adding all (six) permutations that exchange the symmetrized blobs, \eg, $(123) \leftrightarrow (456)$ or $(123) \rightarrow (456) \rightarrow (789) \rightarrow (123)$, \etc. In our graphical language, the above process corresponds to
\begin{equation}
\label{s32graph2}
\ket{\boldsymbol{\Psi}} \sim
\text{
\raisebox{-2.5ex}{
\begin{tikzpicture}
[inner sep=0.0mm,minimum size=8mm,node distance=3mm]
			\node[circle,draw]  (blob1)									{\small{$+ \!   - \! +$}};
			\node[circle,draw]  (blob2)	[right=of blob1] 	{\small{$+ \!   - \! +$}};
			\node[circle,draw]  (blob3)  [right=of blob2]		{\small{$- \!   + \! +$}};
			\draw [double, double distance=.6mm, line width=.4mm, -]
				(blob1.east) -- node[anchor=south]{$\vee$} (blob2.west);
			\draw [line width=.4mm, -]
				(blob2.east) -- node[anchor=south]{$\vee$} (blob3.west);
\end{tikzpicture}
}},
\end{equation}
where, consistent with the use of $\sim$, we omit arrows the reversal of which results in an overall sign flip. 
Other decorations of the blobs with $\pm$ give the rest of the spin-3/2 states, so the blob pattern
\raisebox{-.3ex}{%
\begin{tikzpicture}
[inner sep=0.0mm,minimum size=3mm,node distance=2.2mm]
			\node[circle,draw]  (blob1)									{\small{}};
			\node[circle,draw]  (blob2)	[right=of blob1] 	{\small{}};
			\node[circle,draw]  (blob3)  [right=of blob2]		{\small{}};
			\draw [double, double distance=.3mm, line width=.2mm, -]
				(blob1.east) -- node[anchor=south]{\tiny{$\vee$}} (blob2.west);
			\draw [line width=.2mm, -]
				(blob2.east) -- node[anchor=south]{\tiny{$\vee$}} (blob3.west);
\end{tikzpicture}
}
characterizes the spin-3/2 multiplet of $\mathbf{4}^{\vee 3}$.
\end{myexample}

The above considerations generalize to $\mathcal{H}_{s}^{\vee k}$: every irreducible spin multiplet is characterized by a graph, consisting of $k$ blobs, each containing $2s$ vertices, with various line segments connecting the blobs. In particular, the $s_{\text{max}}$ multiplet is characterized by $k$ blobs without any connections --- thus, all $2ks$ constituent spins 1/2  are available to be symmetrized, and the corresponding principal constellation is the union of the Majorana constellations of the factor states. This same set of stars may be divided into $k$ subsets of $2s$ each in  $\frac{(2sk)!}{(2s)!^k (k!)}$ ways --- this then is the number of $\vee$-factorizable states that share the same principal constellation. As a result, given the principal constellation of a state $\ket{\boldsymbol{\Psi}}$, one may determine all $\vee$-factorizable states with that same principal constellation, and then check whether $\ket{\boldsymbol{\Psi}}$ is one of them --- if it is not, then $\ket{\boldsymbol{\Psi}}$ is not $\vee$-factorizable.
Our treatment above of $\mathbf{4}^{\vee 3}$ may be  further formalized using standard techniques in classical invariant theory, in particular, Gordan's method for computing covariants in terms of \emph{digraphs} (short for \emph{directed graphs}), the \emph{$\Omega$-process}, and \emph{transvectants} --- see, \eg,~\cite{Olv:99}. 

As a final demonstration of the usefulness of the constituent spin-1/2 picture, consider now the antisymmetrization of $k$ blobs of $2s+k-1$ vertices each, with a single edge connecting every pair of blobs, \eg, for $k=4$,
\begin{equation}
\label{MurnaghanGraph}
\text{
\raisebox{-2.5ex}{
\begin{tikzpicture}[auto]
[inner sep=0.0mm,minimum size=3mm]
			\node[circle,draw]  (blob1)									{};
			\node[circle,draw]  (blob2)	[right=of blob1] 	{};
			\node[circle,draw]  (blob3)  [below=of blob2]	{};
			\node[circle,draw]  (blob4)  [left=of blob3]		{};
			\draw [line width=.4mm,  -]
				(blob1.east) -- node[anchor=south]{\small{$\wedge$}} (blob2.west);
			\draw [line width=.4mm,  -]
				(blob2.south) -- node[anchor=west]{\small{$\wedge$}} (blob3.north);
			\draw [line width=.4mm,  -]
				(blob3.west) -- node[anchor=north]{\small{$\wedge$}} (blob4.east);
			\draw [line width=.4mm,  -]
				(blob4.north) -- node[anchor=east]{\small{$\wedge$}} (blob1.south);
			\draw [line width=.4mm,  -]
				(blob1.south east) -- node[anchor=south, pos=.3] {\small{$\wedge$}} (blob3.north west);
			\draw [line width=.4mm,  -]
				(blob4.north east) -- node[anchor=north, pos=.82] {\small{$\wedge$}} (blob2.south west);
\end{tikzpicture}
}}.
\end{equation}
Note that the wedges indicate antisymmetrization of the blobs (\ie, adding all their permutations with an appropriate sign) while the edges stand, as usual,  for antisymmetrization of pairs of vertices. Because of the two antisymmetrizations, the state is symmetric \wrt{} exchange of blobs, while, within each blob, $k-1$ vertices participate in spin-0 pairs, leaving $2s$ vertices free. Thus, an antisymmetric state of $k$ spins $(s+(k-1)/2)$ transforms identically to a symmetric state of $k$ spins $s$ --- this is Murnaghan's isomorphism, from an ``atomic'' point of view.
\section{Geometric measure of entanglement for $\vee$-factorizable states}
\label{Gmoefvfs}
Consider the $k$-partite $\vee$-factorizable state $\ket{\boldsymbol{\Psi}}=\ket{\psi_0} \vee \ket{\psi_1} \vee \ldots \vee\ket{\psi_{k-1}}$ and define its geometric measure of entanglement $E(\ket{\boldsymbol{\Psi}})$ as its Fubini-Study distance from the set of factorizable states, \ie, states of the form $\ket{\boldsymbol{\Phi}}=\ket{\phi_0} \otimes \ket{\phi_1} \otimes \ldots \otimes \ket{\phi_{k-1}}$. As shown in~\cite{Hub.Kle.Wei.Gon.Guh:09}, the state $\ket{\boldsymbol{\Phi}}$ that minimizes the distance from $\ket{\boldsymbol{\Psi}}$ lies in the diagonal embedding of $\mathcal{H}$ in $\mathcal{H}^{\otimes k}$, \ie, it is of the form $\ket{\boldsymbol{\Phi}}=\ket{\phi} \otimes  \ldots \otimes \ket{\phi}$ ($k$ factors), with $\ket{\phi} \in \mathcal{H}$, so that
\begin{equation}
\label{EGdef}
E(\ket{\boldsymbol{\Psi}})=\min_{\ket{\phi} \in \mathcal{H}} \arccos \left| {}^{\otimes k} \! \braket{\phi}{\boldsymbol{\Psi}} \right|
\, .
\end{equation}
A related quantity is $\tilde{E}$, defined as
\begin{equation}
\label{EGtdef}
\tilde{E}(\ket{\boldsymbol{\Psi}})
=\
\min_{\ket{\phi} \in \mathcal{H}} \left( 1- \left| {}^{\otimes k} \! \braket{\phi}{\boldsymbol{\Psi}} \right|^2 \right)
\, .
\end{equation}
It is also clear that the minimizing $\ket{\phi}$, for both $E$, $\tilde{E}$, must lie in the $r$-plane $\Pi$ spanned by the $\ket{\psi_I}$. We call $r$ the \emph{rank} of $\ket{\boldsymbol{\Psi}}$ and deduce the extremum conditions for $\ket{\phi}$, in the full-rank case $r=k$, in section~\ref{Tcolifs}. We also point out that when $k<2s+1$ the problem may be mapped to one of spin $(k-1)/2$,  which can lead to significant simplification --- we treat this case in section~\ref{Tcoldfs}. 
\subsection{The case of linearly independent factor states}
\label{Tcolifs}
We assume here that the factor states $\ket{\psi_I}$ are linearly independent, which is the generic case, when $k\leq 2s+1$.  Note that in this section capital indices range over the set $\{0,1,\ldots,k-1\}$, while small case latin indices range over $\{1,\ldots,k-1\}$. Define the matrix $G$ of inner products among the $\ket{\psi_I}$, and its inverse, with elements given by
\begin{equation}
\label{GIJdef}
G_{IJ} \equiv \braket{\psi_I}{\psi_J}
\, ,
\qquad
G^{IJ} \equiv \braket{\psi^I}{\psi^J}
\, ,
\end{equation}
where $\{\ket{\psi^I}\}$ is the dual basis in $\Pi$, with $\braket{\psi^I}{\psi_J}=\delta^I_{\phantom{I}J}$ and $\delta^I_{\phantom{I}J}=\delta^{\phantom{J}I}_J$ is the Kronecker delta. Note that, as usual,
\begin{equation}
\label{GGIdef}
\ket{\psi^I}=\ket{\psi_J}G^{JI}
\, ,
\qquad
\ket{\psi_I}=\ket{\psi^J}G_{JI}
\, ,
\qquad
G_{IJ}G^{JK}=\delta_{I}^{\phantom{I}K}
\, ,
\qquad
\ket{\psi_I}\bra{\psi^I}=\mathbb{1}
\, ,
\end{equation}
where repeated capital indices are summed over their range, $\delta_{I}^{\phantom{I}K}$ denotes the Kronecker delta, and the last equation is valid since we restrict our attention to $\Pi$ . Expanding $\ket{\phi}$ in the basis of the $\ket{\psi_I}$ we get
\begin{equation}
\label{expphipsi}
\ket{\phi}=\phi^I \ket{\psi_I}
\, ,
\quad \text{with} \quad
\phi^I\equiv \braket{\psi^I}{\phi}
\, ,
\quad
\text{and}
\quad
\braket{\phi}{\phi}=\bar{\phi^I} G_{IJ} \phi^J=\bar{\phi}_IG^{IJ} \phi_J
\, .
\end{equation}
Up to $\ket{\phi}$-independent normalization factors, the function we wish to maximize is
\begin{align}
f(z,\bar{z})
&=
\frac{|\phi_0\phi_1\ldots \phi_{k-1}|^2}{\braket{\phi}{\phi}^{k}}
\\
&=
\frac{|z_1 z_2\ldots z_{k-1}|^2}{(\bar{z}_I G^{IJ}z_J)^{k}}
\, ,
\label{fdefdist}
\end{align}
where
\begin{equation}
\label{vardefs1}
\phi_I\equiv \braket{\psi_I}{\phi}=G_{IJ}\phi^J
\, ,
\quad
z_i \equiv \frac{\phi_i}{\phi_0}
\, ,
\quad
z_0\equiv 1
\, ,
\end{equation}
and the power in the denominator of $f$ guarantees that $f$ is invariant under the rescaling $\phi_I \rightarrow \lambda \phi_I$, so that $f$ descends to $\mathbb{P}$. The $z_i$, $i=1,\ldots,k-1$, are the standard coordinates on $\mathbb{P}$, in the chart $U_0$  where $\phi_0 \neq 0$. Setting $\partial f/\partial z_i=0=\partial f/\partial \bar{z}_i$ gives the system of quadratic equations
\begin{align}
\label{quadsystemeqs}
\bar{z}_I G^{IJ} z_J-k\bar{z}_M G^{Mi} z_i
&=
0
\, ,
\\
\bar{z}_I G^{IJ} z_J-k\bar{z}_i G^{iM} z_M
&=
0
\, ,
\end{align}
where $\braket{\phi}{\phi} \neq 0$ was used, $z_i \neq 0$ was assumed (since for $z_i=0$, $f$ attains its minimum), and, here and below in this section, there is no summation over $i$ --- one gets one pair of complex conjugate equations for each $i=1,\ldots,k-1$. Defining $z^I\equiv G^{IM}z_M$ (note the order of the indices in $G$), the above can be written as
\begin{align}
\label{quadsystemeqs2}
\bar{z}_I z^I-k \bar{z^i} z_i
&=
0
\, ,
\\
\bar{z}_I z^I-k \bar{z}_i z^i
&=
0
\, ,
\end{align}
implying that, for all $i=1,\ldots,k-1$,
\begin{align}
 \bar{z^i} z_i
 &=
 \bar{z}_M G^{Mi} z_i
 \nonumber
 \\
 &=
 \left( G^{0i}+\bar{z}_1 G^{1i}+\ldots +\bar{z}_{k-1} G^{k-1,i} \right) z_i
 \nonumber
 \\
 &=
 G^{00}+\bar{z}_1 G^{10}+\ldots +\bar{z}_{k-1} G^{k-1,0}
 \nonumber
 \\
 &\equiv
 \bar{z^0}
 \, .
 \label{zizieq}
 \end{align}
 To get an idea of the geometrical content of~(\ref{zizieq}), denote by $\ket{\phi_I}$ the (vector) component of $\ket{\phi}$ along $\ket{\psi_I}$,
 \begin{equation}
 \label{phiI}
\ket{\phi_I}=\ket{\psi_I}\braket{\psi^I}{\phi}
\, ,
\quad
\text{(no summation over $I$)}
\, ,
\end{equation}
 and compute its projection back onto $\ket{\phi}$,
 \begin{align}
 \braket{\phi_I}{\phi}
 &=
 \braket{\phi}{\psi^I}\braket{\psi_I}{\phi}
 \nonumber
 \\
 &=
 \bar{\phi}^I \phi_I
 \quad
\text{(no summation over $I$)}
\, .
 \end{align}
 Thus, a necessary condition for maximizing $f(z,\bar{z})$ is that all vector components $\ket{\phi_I}$ contribute equally to $\ket{\phi}$ --- this analytical requirement seems to quantify the intuitive expectation that the $\ket{\phi}$ that maximizes $f$ should be some sort of ``barycenter'' of the $\ket{\psi_I}$. However appealing this interpretation might be, it should be kept in mind that the above condition also characterizes other critical points of $f$, not just the maximum. The solutions of~(\ref{zizieq}) only determine well-behaved critical points of $f$ --- note that  the minimum of $f$ is degenerate, as $f$ becomes zero for every $\ket{\phi}$ in the union of the orthogonal complements of each $\ket{\psi_I}$, the image of the latter in the corresponding projective space being known as the \emph{antipodal submanifold of} $\pket{\Psi}$~\cite{Bes:78}.
\begin{myexample}{Closest diagonal state for a tripartite symmetric state, with linearly independent spin-1 factors}{Gmetsfs}
Consider the state $\ket{\Psi} \sim \ket{\psi_0}\vee \ket{\psi_1}\vee \ket{\psi_2}$, with
\begin{equation}
\label{Psi123ex}
\ket{\psi_0}=\frac{1}{\sqrt{2}}(1,0,1)^T
\, ,
\qquad
\ket{\psi_1}=\frac{1}{\sqrt{2}}(1,0,i)^T
\, ,
\qquad
\ket{\psi_2}=\frac{1}{\sqrt{3}}(1,1,1)^T
\, .
\end{equation}
The factor states $\ket{\psi_I}$ are linearly independent, so $k=r=3$ in this example.
Working in the chart $U_0$, the extremum conditions (\ref{zizieq}) give the system of quadratic relations ($z_j\equiv x_j+i y_j$)
\begin{align*}
2 x_1^2+\sqrt{6}x_2+2y_1^2
&=
4
\, ,
\\
3x_2^2+x_1+3y_2^2-y_1
&=
4
\, ,
\\
2x_1+2y_1+\sqrt{6} y_2
&=
0
\, ,
\\
x_1+y_1+2\sqrt{6} y_2
&=
0
\, ,
\end{align*}
which admits four real solutions. Computing the eigenvalues of the Hessian of $f(z,\bar{z})$ at each solution, one identifies that its maximum is obtained for (normalizing the result)
\begin{equation}
\label{fmaxex}
\ket{\phi}=(0.604,0.391-i \, 0.391,0.574)^T
\, ,
\end{equation}
so that $\ket{\Phi}=\ket{\phi} \otimes \ket{\phi} \otimes \ket{\phi}$ is the diagonal product state closest to $\ket{\Psi}$.
\end{myexample}
\subsection{Mapping to a lower spin problem when the factor states are linearly dependent}
\label{Tcoldfs}
We deal now with the case $r<k$, and assume, without loss of generality, that the first $r$ factor states, $\ket{\psi_0}, \ldots ,\ket{\psi_{r-1}}$, are linearly independent. Define an orthonormal basis $\{ \ket{e_0},\ldots,\ket{e_{r-1}} \}$ in the $r$-plane $\Pi$ they span as follows: put $\ket{e_0}=\ket{\psi_0}$, then take $\ket{e_1}$ in the space spanned by $\ket{\psi_0}$, $\ket{\psi_1}$, $\ket{e_2}$ in the space spanned by the first three $\ket{\psi_i}$'s, and continue like this, until completing the basis in $\Pi$.  By construction, the $\ket{\psi_i}$, $i=0, 1, \ldots, k-1$, have the following form in this basis,
\begin{align}
\ket{\psi_0}
&=
(1,0,0,\ldots)^T
\nonumber
\\
\ket{\psi_1}
&=
(a,b,0,\ldots,0)^T
\label{psicompse}
\\
\vdots \quad & \qquad \quad \vdots
\qquad
\nonumber
\\
\ket{\psi_{r-1}}
&=
(\psi^{(r-1)}_0,\psi^{(r-1)}_1,\ldots,\psi^{(r-1)}_{r-1})^T
\nonumber
\\
\vdots \quad & \qquad \quad \vdots
\nonumber
\\
\ket{\psi_{k-1}}
&=
(\psi^{(k-1)}_0,\psi^{(k-1)}_1,\ldots,\psi^{(k-1)}_{r-1})^T
\nonumber
\end{align}
\ie, $\ket{\psi_i}$, with $0 \leq i \leq r-1$, has its last $r-i-1$ components equal to zero, while, for $r \leq i \leq k-1$, the components can have any value. The basis is now extended (still orthonormal) to the whole of $\mathcal{H}$, which results in $2s+1-r$ trailing zeros being added to the above components. The basis $\ket{e_I}$ in $\mathcal{H}$ thus constructed, may be mapped to the standard one of $S_z$ eigenvectors by a unitary transformation $U$, so that $U\ket{e_0}=\ket{s,s}$, $U\ket{e_1}=\ket{s,s-1}$, \etc, so that the images $\ket{\psi'_I}=U\ket{\psi_I}$ have the same components, in the standard $S_z$-eigenbasis, as those that appear in the \rhs{} of~(\ref{psicompse}). The state $\ket{\phi'}$ that maximizes $f$ in~(\ref{fdefdist}) for the $\ket{\psi'_I}$ will clearly also have $2s+1-r$ trailing zeros, since all $\ket{\psi'_I}$ do. Then the trailing zeros may be ignored, and the problem is reduced to one of spin $j=(r-1)/2$. Given a solution for $\ket{\phi'}$, the trailing zeros are added back, and the resulting ket is mapped by $U^{-1}$ to the solution of the original problem for the $\ket{\psi_I}$'s. Note that the complexity of the problem is determined primarily by the rank $r$ and the number of factor states $k$, not by their spin $s$.
\begin{myexample}{Geometric measure of entanglement for a bipartite, rank $r=2$, spin-$1$ state}{Gmebr2s123}
Consider the bipartite spin-1, $r=2$,  symmetric state 
$\ket{\boldsymbol{\Psi}}=\ket{\psi_1} \vee \ket{\psi_2}/\sqrt{2}$, with
\begin{equation}
\label{s1r2ex}
\ket{\psi_1}=(1,0,0)^T
\, ,
\qquad
\ket{\psi_2}=(0,1,1)^T/\sqrt{2}
\, .
\end{equation}
The factor states provide, by themselves, an orthonormal basis in the plane they define, which may be extended to  the whole of $\mathcal{H}_1$ by the inclusion of $\ket{\psi_3}=(0,-1,1)^T/\sqrt{2}$. 
The unitary transformation 
\begin{equation}
U= \left( 
\begin{array}{c c c}
1 & 0 & 0 \\
0 & \frac{1}{\sqrt{2}} & \frac{1}{\sqrt{2}}
\\
0 & -\frac{1}{\sqrt{2}} & \frac{1}{\sqrt{2}}
\end{array} 
\right) 
\, ,
\end{equation}
maps $\ket{\psi_1}$ to itself (\ie, $\ket{\psi'_1}=\ket{\psi_1}$), and $\ket{\psi_2}$ to $\ket{\psi'_2}=(0,1,0)^T$. Ignoring the trailing zero in $\ket{\psi'_{1,2}}$, the problem becomes one of spin-1/2 factor states, with the two primed states
$\ket{\psi'_{1,2}}$ being equal to $\ket{\hat{\mathbf{z}}\pm}$, respectively, and $\ket{\boldsymbol{\Psi}'}$ being the spin-1 $S_z=0$  state. The diagonal states closest to $\ket{\boldsymbol{\Psi}'}$ are of the form 
$\ket{\boldsymbol{\Phi}'}=\ket{\phi'} \otimes \ket{\phi'}$, with $\ket{\phi'}=(1,e^{i \alpha})^T/\sqrt{2}$, $0 \leq \alpha <2\pi$. Putting the trailing zero back to $\ket{\phi'}$ and applying $U^{-1}$ one gets 
\begin{equation}
\label{phipPhi}
\ket{\phi}=U^{-1}\ket{\phi'}=
\left( \frac{1}{\sqrt{2}} , \frac{e^{i \alpha}}{2} , \frac{e^{i \alpha}}{2} \right)^T
\, ,
\end{equation}
and the diagonal states closest to $\ket{\boldsymbol{\Psi}}$ are of the form $\ket{\boldsymbol{\Phi}}=\ket{\phi} \otimes \ket{\phi}$, their distance to $\ket{\boldsymbol{\Psi}}$ being, for all values of the parameter $\alpha$,  $E(\ket{\boldsymbol{\Psi}})=\arccos|\braket{\boldsymbol{\Phi}}{\boldsymbol{\Psi}}|=\pi/4$.
\end{myexample}
\subsection{Gramian of a $\vee$-factorizable state and geometric measure of entanglement}
As argued in section~\ref{Gmoefvfs}, the geometric entanglement measures $E(\ket{\boldsymbol{\Psi}})$, 
$\tilde{E}(\ket{\boldsymbol{\Psi}})$, of a spin-$s$ $\vee$-factorizable state
$\ket{\boldsymbol{\Psi}}=\ket{\psi_0} \vee \ket{\psi_1} \vee \ldots \vee\ket{\psi_{k-1}}$ are completely determined by the
Gramian matrix $G$, defined in Eq.~(\ref{GIJdef}). We focus here on $\tilde{E}(\ket{\boldsymbol{\Psi}})$, results analogous to those reported later in this section hold true for $E(\ket{\boldsymbol{\Psi}})$. In the extreme diagonal case, where all factor states of $\ket{\boldsymbol{\Psi}}$ are equal, $\ket{\psi_0} = \ldots =\ket{\psi_{k-1}}=\ket{\phi}$, $\tilde{E}(\ket{\boldsymbol{\Psi}})$ and $\det G$ are both zero. It is clear though that $\tilde{E}(\ket{\boldsymbol{\Psi}})$  is sensitive to finer details of the set of factor states $\{\ket{\psi_i}\}$, for example, when only two of the latter are equal, $\det G$ is still zero, while $\tilde{E}(\ket{\boldsymbol{\Psi}})$ is not --- it seems therefore reasonable to look for a relationship between $\tilde{E}(\ket{\boldsymbol{\Psi}})$ and the various minors of $G$. Taking into account that such functional dependence ought to be symmetric in the factor states, leads us to the main focus of this section:  to study the relationship between $\tilde{E}(\ket{\boldsymbol{\Psi}})$ and the eigenvalues $\lambda_i$, $i=0,1,\ldots,k-1$ 
of $G$, the latter being symmetric functions of the minors of $G$. 

We start by pointing out some elementary facts about the spectrum of the Gram matrix $G$:
\begin{enumerate}
\item
$G$ is a hermitean, positive semidefinite matrix, so $\lambda_i \geq 0$.
\item
$G$ has $k-r$ eigenvalues equal to zero, where $r \leq 2s+1$ is the rank of $\ket{\boldsymbol{\Psi}}$.
\item
The diagonal elements of $G$ are equal to 1, so that the trace of $G$, and, hence, the sum of its eigenvalues, is equal to the number of factor states, $\sum_{i=0}^{k-1} \lambda_i=k$, implying that the number of independent eigenvalues is $r-1$.  
\end{enumerate}

We begin our study with the simplest case, $s=1/2$, for which $\ket{\boldsymbol{\Psi}}$ is a symmetric state of $k \geq 2$ qubits, \ie, a spin-$k/2$ state, uniquely determined by its  principal (Majorana) constellation. For  $\ket{\boldsymbol{\Psi}}$ generic, there are $2$ eigenvalues of $G$ that are different
from zero, and they satisfy the relation $\lambda_1+ \lambda_2=k$, so only one of them is independent ---  we assume that $\lambda \equiv \lambda_1\leq \lambda_2$, so that $0 \leq \lambda \leq k/2$, and study the relation between $\lambda$ and $\tilde{E}$. The real dimension of the projective space $\mathbb{P}$  is $2k$, while the orbit $\mathcal{O}_{\pket{\psi}}$ of a generic state $\pket{\psi}$ under rigid rotations (\ie, under the diagonal action of $SU(2)$) is 3-dimensional, so that the quotient \emph{shape} space $\mathcal{S} =\mathbb{P}/\mathcal{O}$ has dimension $2k-3$ --- since both $\tilde{E}$ and $\lambda_i$ are constant on $\mathcal{O}$,  they descend to $\mathcal{S}$.

\begin{figure}[h]
\includegraphics[width=.35\linewidth]{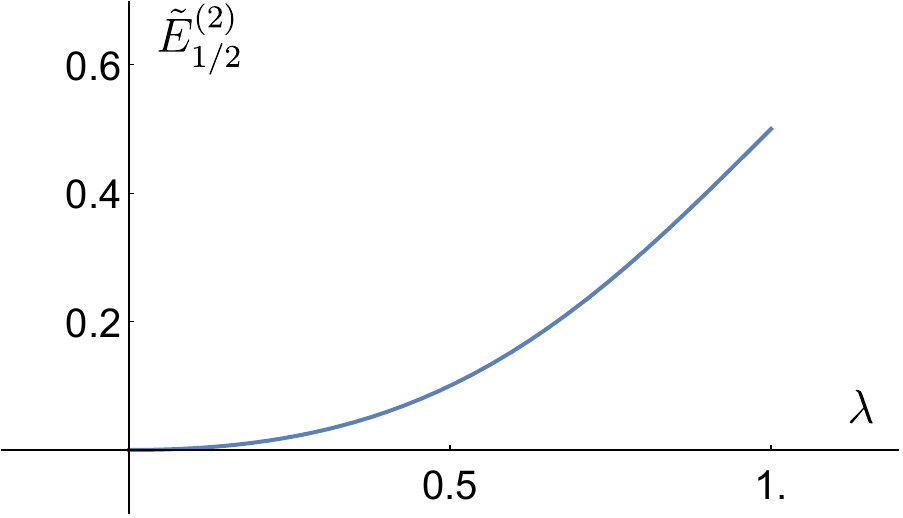}
\caption{%
Parametric plot $(\lambda(\alpha),\tilde{E}_{1/2}^{(2)}(\alpha))$, where $\alpha$  is the half-angle between the Majorana stars of a spin-1 state, $0 \leq \alpha \leq \pi/2$ (see Eq.~(\ref{Elambdak2}). The value $\alpha=0$ corresponds to a coherent state, with both $\lambda$ and $\tilde{E}_{1/2}^{(2)}$ equal to zero. For $\alpha=\pi/2$ one gets (appropriately orienting) the $S_z=0$ state, with $\lambda=1$ and $\tilde{E}_{1/2}^{(2)}$ reaching its maximum value $1/2$.
}
\label{fig:lambdavstEs12}
\end{figure}

For $k=2$ (\ie, spin-1 states), $\mathcal{S}$ is 1-dimensional, and may be parametrized by the angle $2\alpha$ between the two Majorana stars. We find 
\begin{equation}
\label{Elambdak2}
\lambda=1-\cos \alpha
\, ,
\qquad
\tilde{E}_{1/2}^{(2)}=  1-\left( 1+\tan^4 \frac{\alpha}{2} \right)^{-1} 
\, ,
\end{equation}
where $\tilde{E}_s^{(k)}$ refers to the case of $k$ factor states of spin $s$ each --- for the eigenvalues of $G$ we use here and in what follows a simplified notation, denoting them by $\lambda$, $\lambda'$, \etc. 
A plot of $\tilde{E}_{1/2}^{(2)}(\alpha)$ \emph{vs} $\lambda(\alpha)$ appears in Fig.~\ref{fig:lambdavstEs12}.
In this case, there is a 1-to-1 relation between the two quantities. It should be kept in mind that to each point in the curve shown in Fig.~\ref{fig:lambdavstEs12} there corresponds a three-dimensional subset of the projective space, consisting of the orbit, under rotations, of a particular constellation shape, parametrized by $\alpha$. 

\begin{figure}[h!]
\begin{tabular}{c c c c c c c c c}
 A & B & C & C$^{\prime}$ & D & E & F & G & $\text{A}'$
 \\
\scalebox{0.2}{ \includegraphics{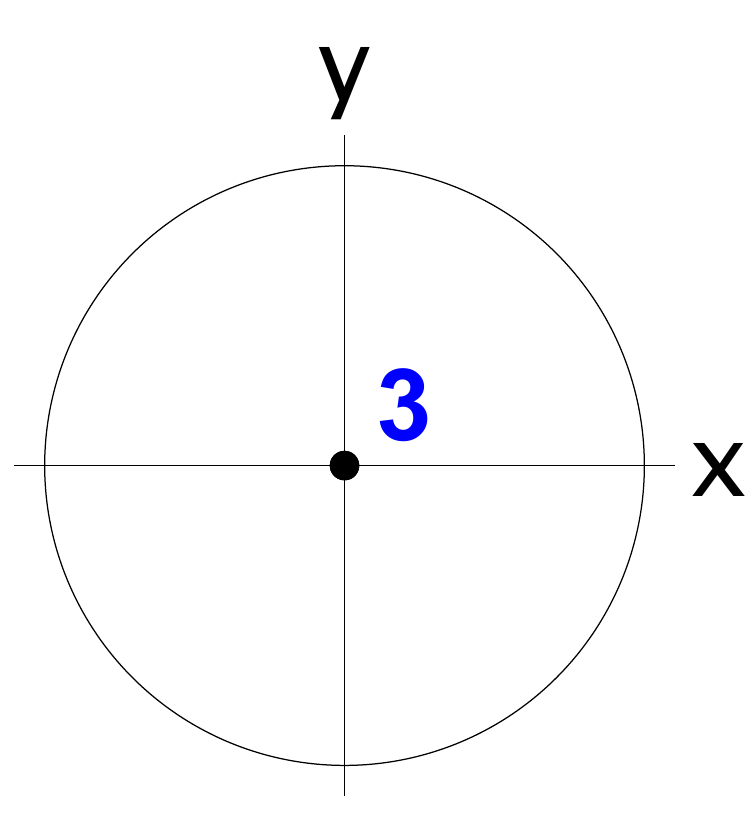}}
&
\scalebox{0.2}{ \includegraphics{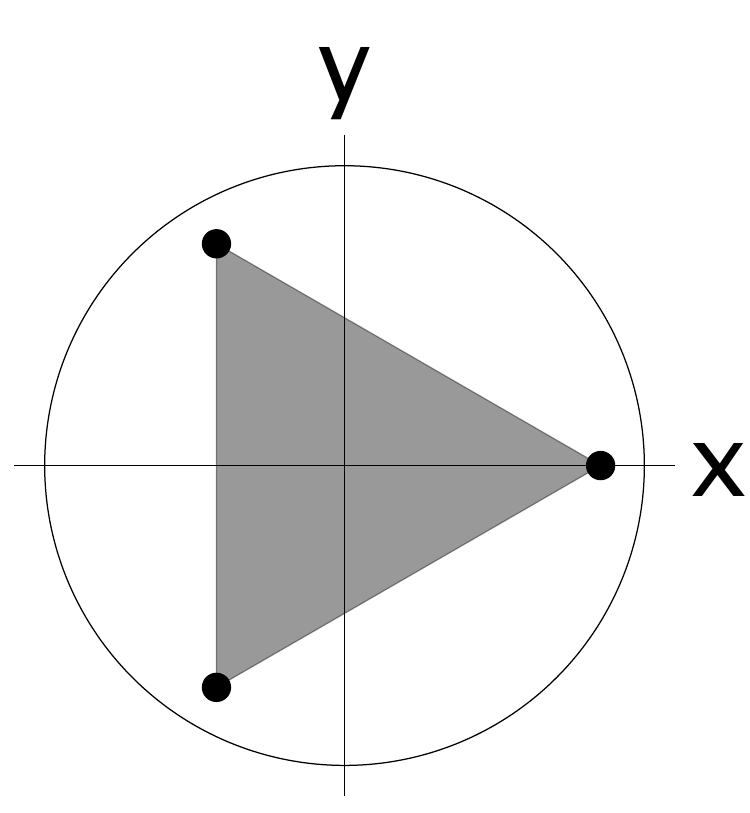}}
&
\scalebox{0.2}{ \includegraphics{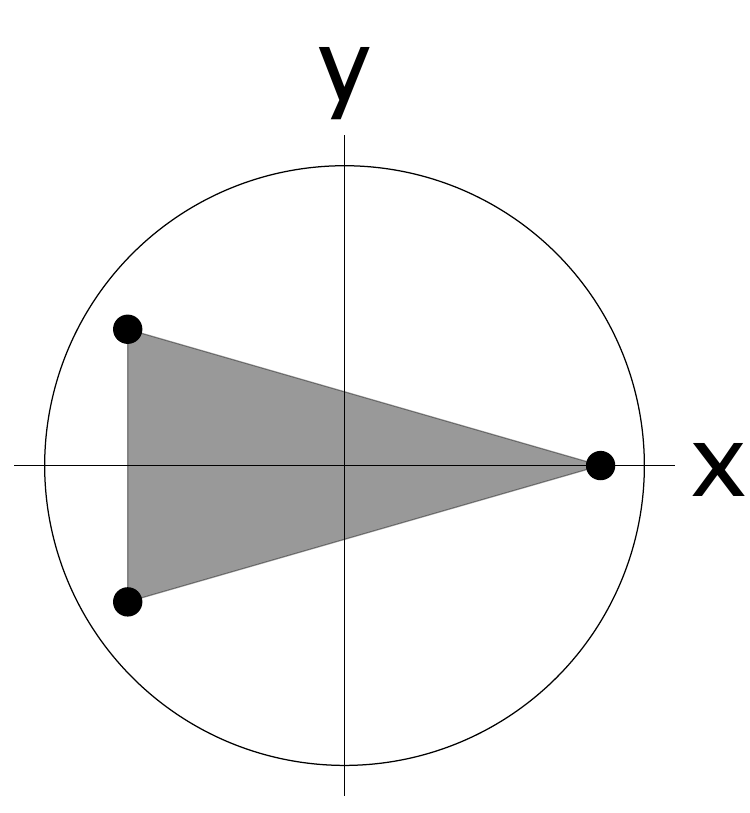}}
&
\scalebox{0.2}{ \includegraphics{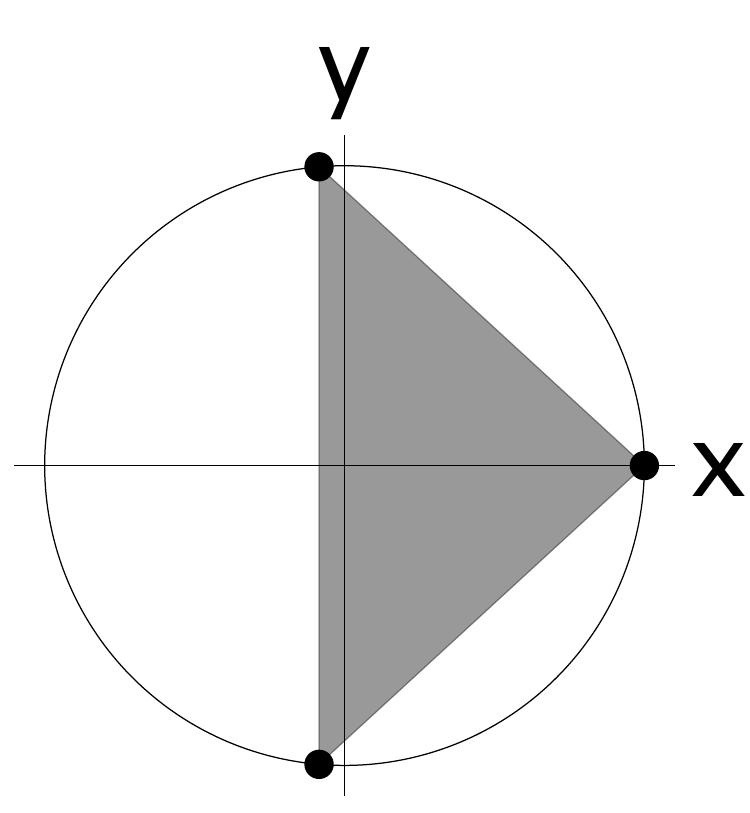}}
&
\scalebox{0.2}{ \includegraphics{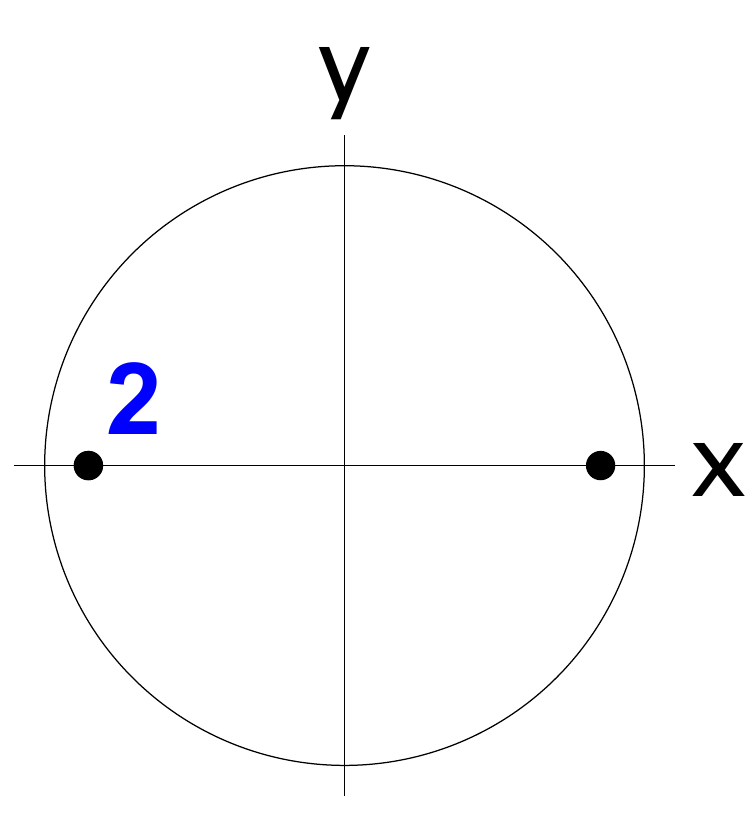}}
&
\scalebox{0.2}{ \includegraphics{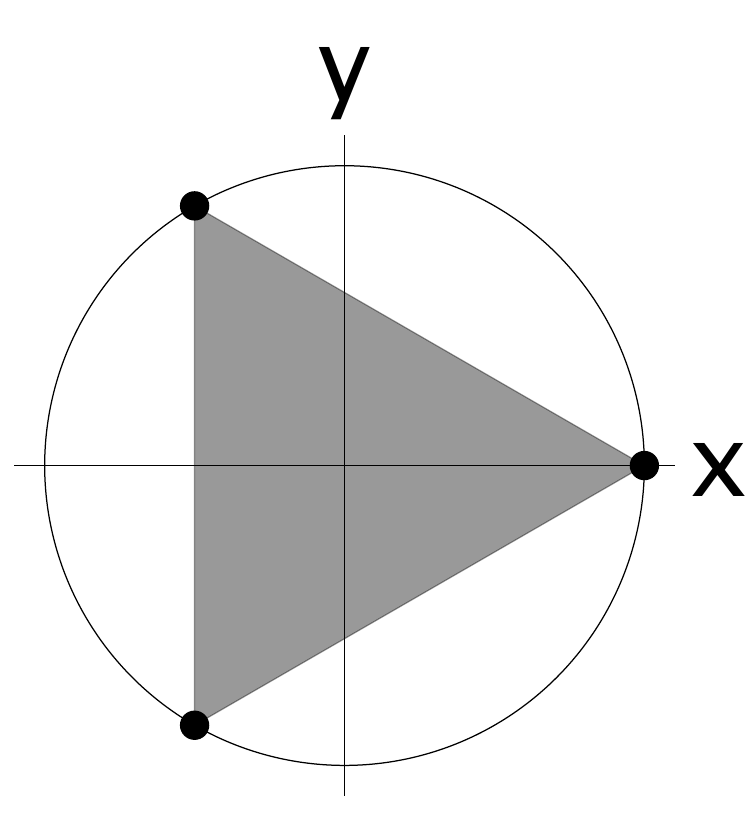}}
&
\scalebox{0.2}{ \includegraphics{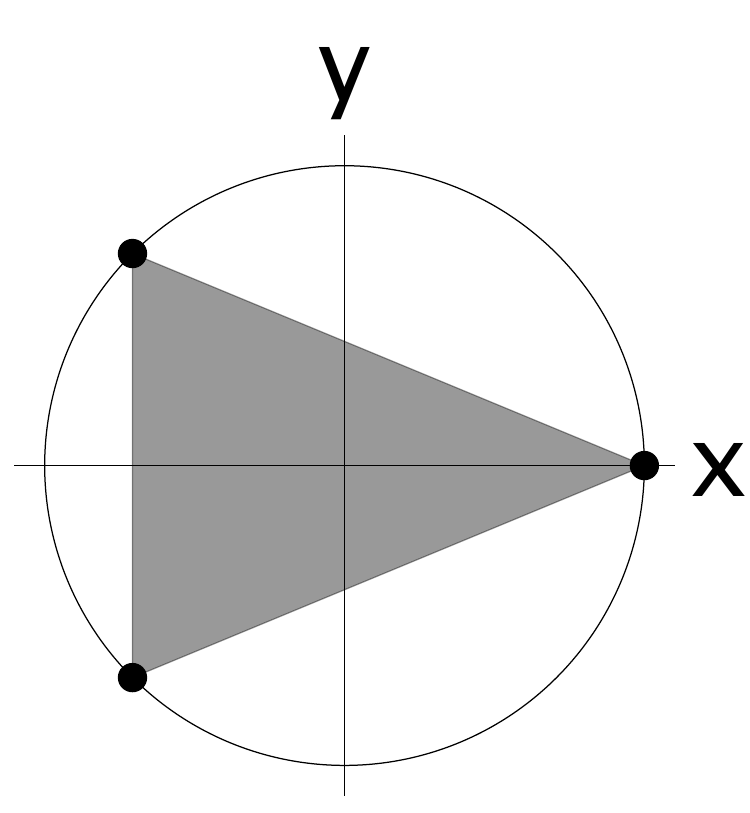}}
&
\scalebox{0.2}{ \includegraphics{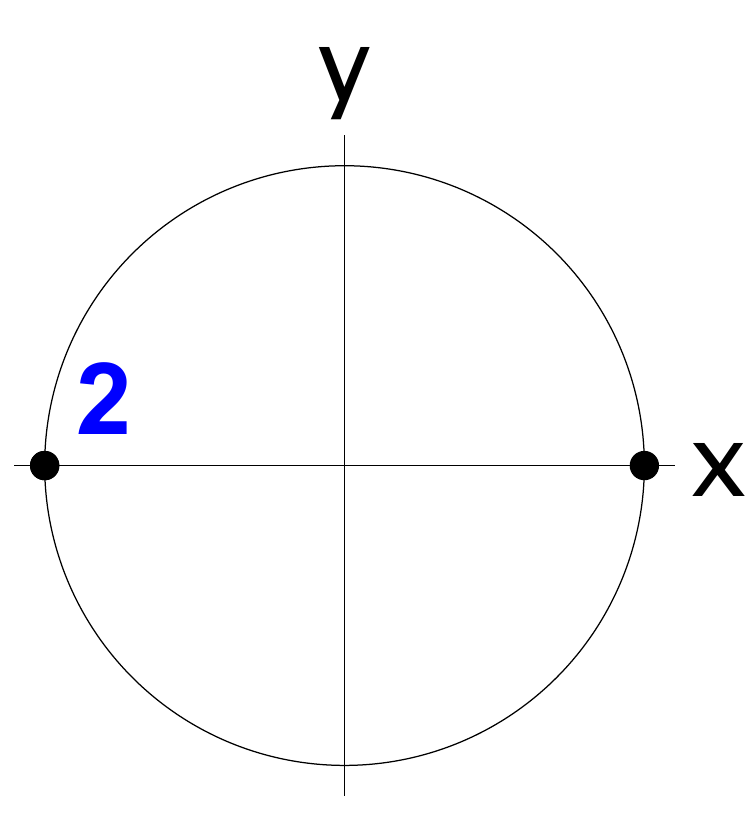}}
&
\scalebox{0.2}{ \includegraphics{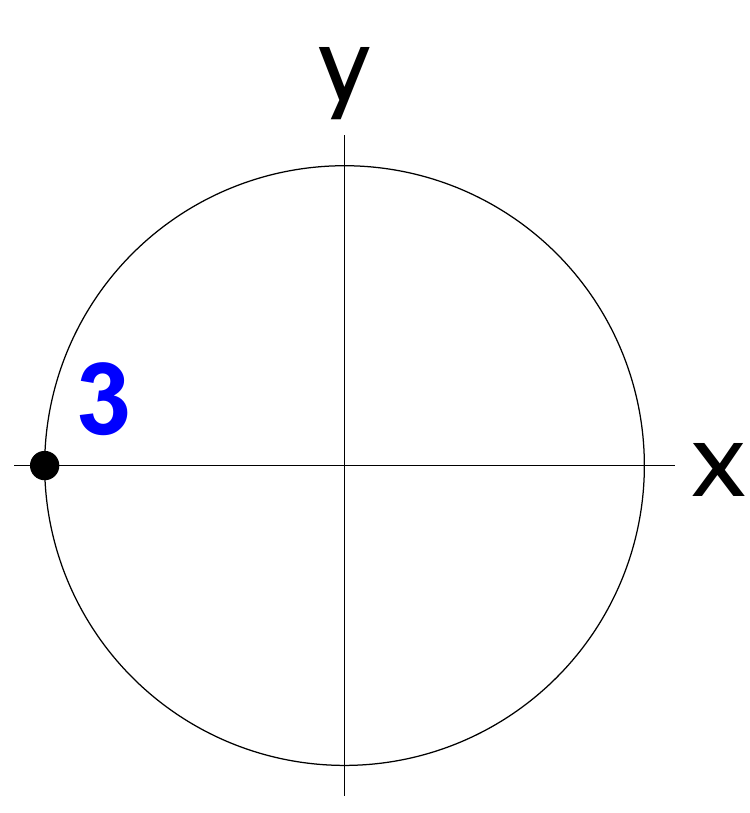}}
\\
\multicolumn{4}{l}{\scalebox{0.7}{\includegraphics{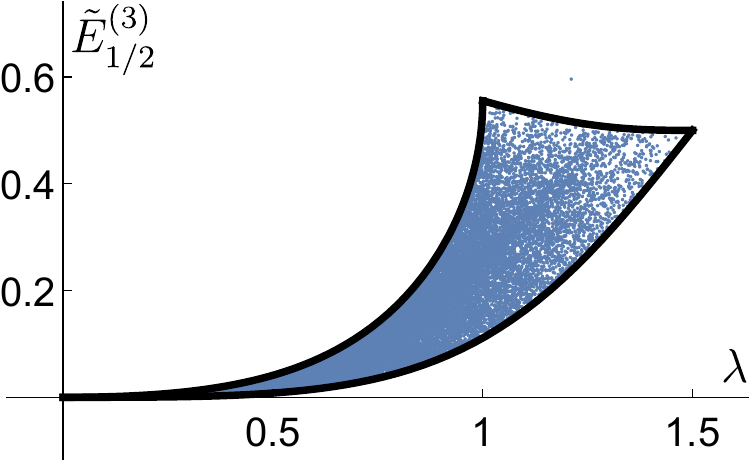}}}
& & 
\multicolumn{4}{r}{\scalebox{0.5}{\includegraphics{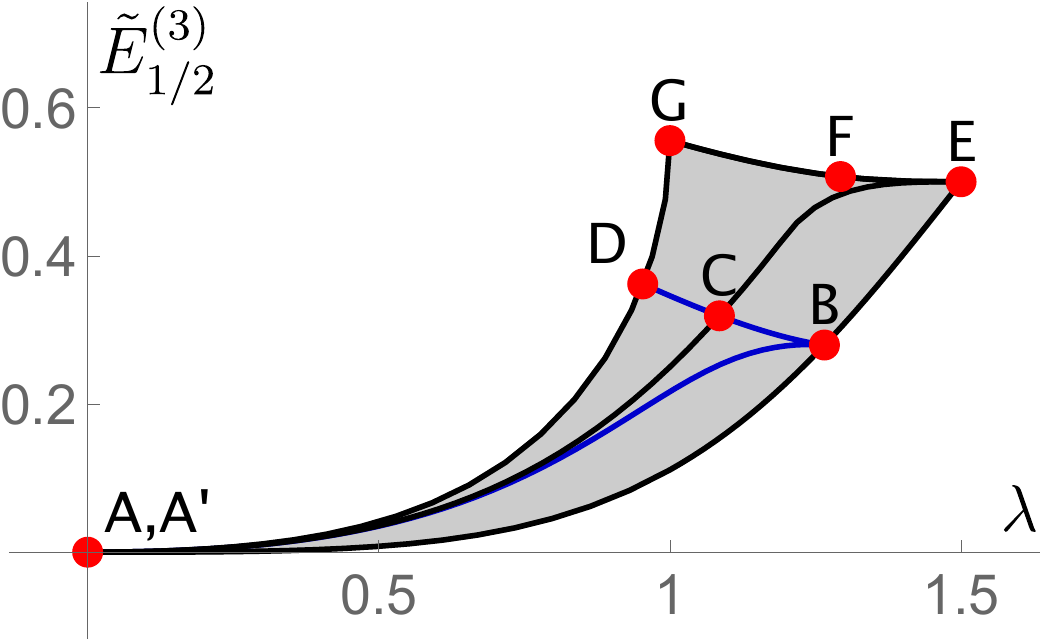}}}
\end{tabular}
\caption{\label{s32.gram}
\emph{Top:} Stereographic projection to the complex plane of the Majorana constellation of  a number of symmetric 3-qubit states (equivalently, spin-3/2 states), belonging to a family of horizontal isosceles triangles, with the three stars at  $(\theta,\phi)$, $(\theta,0)$, $(\theta,-\phi)$ in standard spherical coordinates (numbers next to dots denote degeneracy).  A, $\text{A}'$:  coherent states along $\hat{z}$, $-\hat{x}$, respectively; G: a W state along $\hat{x}$; E:  a maximal equilateral triangle (a 1-anticoherent state~\cite{Zim:06}).
\emph{Bottom left:} Scatter plot of the geometric measure of entanglement $\tilde{E}_{1/2}^{(3)}$ \emph{vs} the lowest nonzero eigenvalue $\lambda$ of the Gram matrix $G$,   for 30,000 randomly generated states. 
\emph{Bottom right:} Same plot as on the left. The capital letters on the plot correspond to the states shown at the top.  The black curves AG, GE, EA, correspond to geodesics in the projective space, connecting the vertices A, G, E. The curve ABCD shown corresponds to a family of isosceles triangles with fixed $\theta= 9\pi/20$ and  angle $\phi$ varying in the interval $[0,\pi]$. The curve starts, for  $\phi=0$, at A (coherent state) regardless of $\theta$, reaches the right border of the shaded region (curve AE) for $\phi=2\pi/3$ (equilateral triangle) and bounces back, reaching the left border (curve AG) for $\phi=\pi$.}
\end{figure}

Fig.~\ref{s32.gram} refers to  3-qubit states.
%
In the scatter plot on the bottom left the $(\lambda,\tilde{E}_{1/2}^{(3)})$-data of 30,000 randomly chosen states in the projective space are shown. Superimposed are the images of geodesics in the projective space, in the Fubini-Study metric, connecting the vertices --- we conjecture that they form the boundary of the plot.
The highlighted points A, B, $\ldots$, G, in the plot at the bottom right, correspond to the states shown in the top row (more precisely, what is shown there is the stereographic projection, on the complex plane,  of their Majorana constellation). Note that the vertices of the curvilinear triangular figure  correspond to either the separable (coherent) states A, $\text{A}'$, or the highly entangled E (a GHZ state) and  G (a W state)~\cite{Enr.Win.Zyc:16}.  Just like in the previous case, of two qubits, many different states in the projective space get mapped to the same point in the plot. In the case at hand, of three qubits, simple dimension-counting shows that the preimage of each point in the plot is a 4-dimensional subset of $\mathbb{P}$. Thus, apart from the expected 3-dimensional rotation orbit, there is a 1-dimensional continuum of different shapes that have the same $(\lambda,E_{1/2}^{(3)})$ --- a particular example is the states denoted by C, $\text{C}'$ in the top row, both of which project to C in the right bottom. An interesting question that arises is whether one can find a simple, 2-parameter family of states that covers the plot. In this regard, we find that
 the shaded region in the bottom right of Fig.~\ref{s32.gram} can be fully explored by the set of states with constellation given by an isosceles triangle, $C_3(\theta,\phi)=\{ (\theta,\phi) , (\theta,-\phi) ,(\theta,0) \}$, where
$(\theta, \phi)\in[0, \pi/2] \times [0,\pi]$ are the spherical angles of the stars. The left boundary
ADG of Fig.~\ref{s32.gram} corresponds to $\phi=\pi$ (two stars in coincidence), while the right boundary ABE has $\phi=2\pi/3$ (equilateral triangles).  The black curve ACEFG is given by
states with constellations of the type $C_3(\pi/2,\phi$) (isosceles triangles in the equatorial plane), with $\phi \in [0,2\pi/3]$ mapping to ACE and $\phi \in [2\pi/3, \pi]$ to  EFG. The above family of triangles covers the entire plot, singly within ACEFGDA, and doubly within ABEFCA --- the restricted family $(\theta,\phi) \in [0,\pi/2] \times [2\pi/3,\pi]$ covers the plot bijectively, except at A.

A final point worth mentioning is that some of the geodesics in $\mathbb{P}$ mentioned above, the images of which seem to bound the plot, are horizontal, in the following sense: at each point, their tangent vector is orthogonal to all tangent vectors corresponding to rigid rotations --- the latter span the vertical subspace of the tangent space of $\mathbb{P}$. Starting from the coherent state along $\hat{z}$ at A, the horizontal geodesic that projects to the curve ABE ends on the GHZ state, oriented as shown in E on the top row. From there, the horizontal geodesic that projects to the curve EFG, ends on the W state, oriented as shown in G on the top row. However, starting from this orientation of the W state, the horizontal geodesic that ends on a coherent state, oriented along $\hat{x}$,  does not seem to bound the scatter plot. Rather, a geodesic that does seem to do the job, ends on the coherent state $\text{A}'$, oriented along $-\hat{x}$, and projects to the curve GDA shown in Fig.~\ref{s32.gram} --- this latter geodesic is not horizontal. 

We next take up the four-qubit case. 
\begin{figure}[h!]
\begin{tabular}{c c c c c c}
  {A} & {B} & {C} & {D} & {E} & {A${}'$}
 \\
\scalebox{0.25}{ \includegraphics{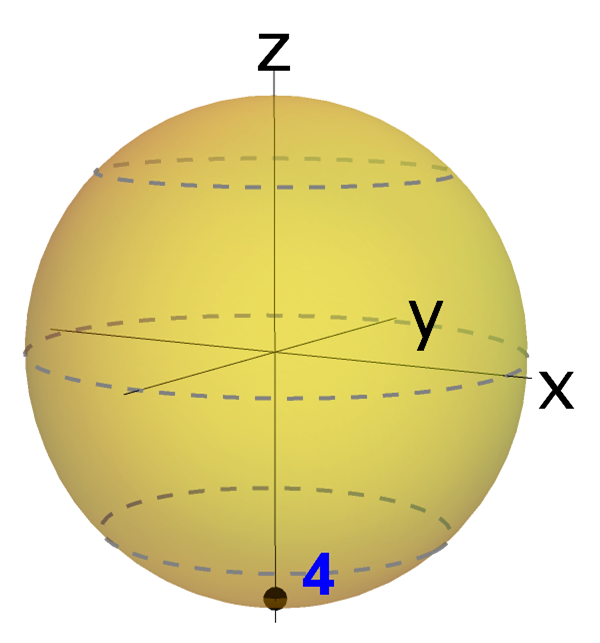}}
&
\scalebox{0.25}{ \includegraphics{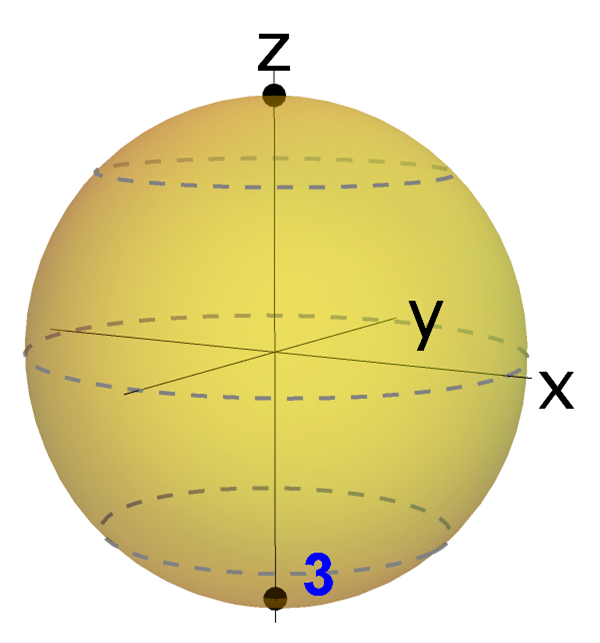}}
&
\scalebox{0.25}{ \includegraphics{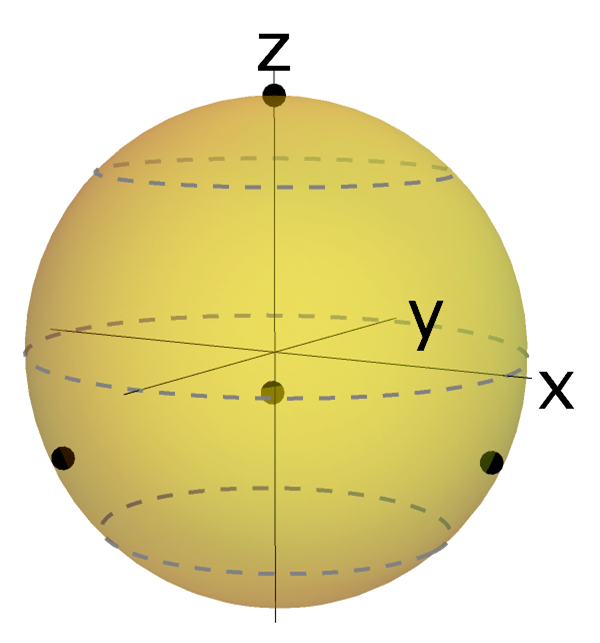}}
&
\scalebox{0.25}{ \includegraphics{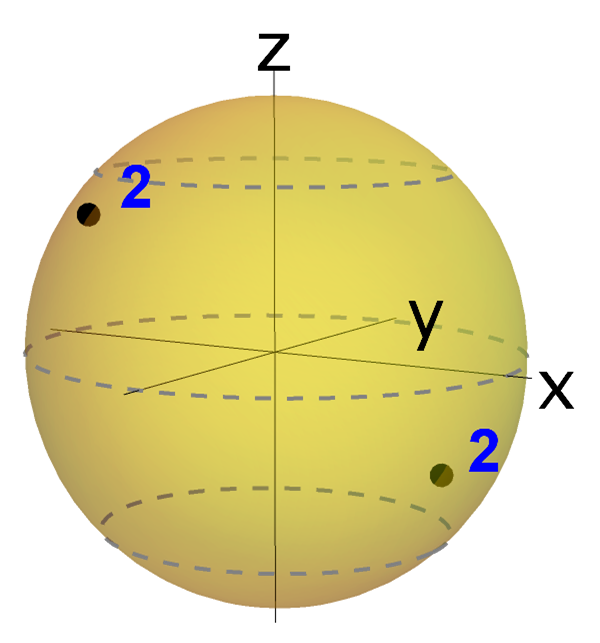}}
&
\scalebox{0.25}{ \includegraphics{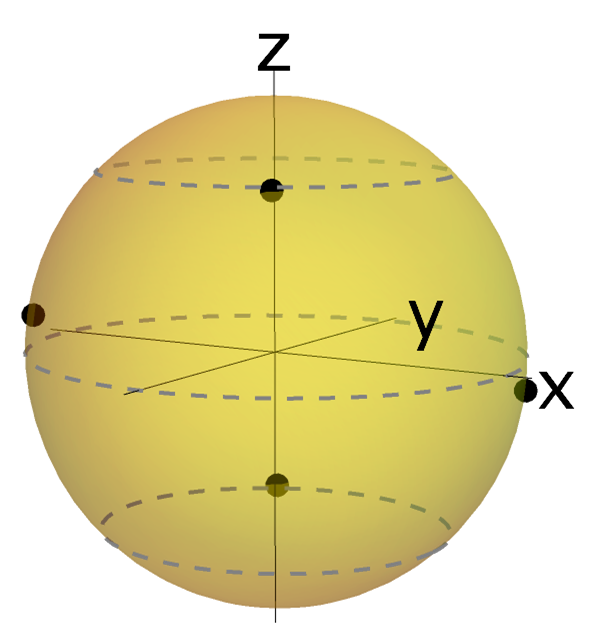}}
&
\scalebox{0.25}{ \includegraphics{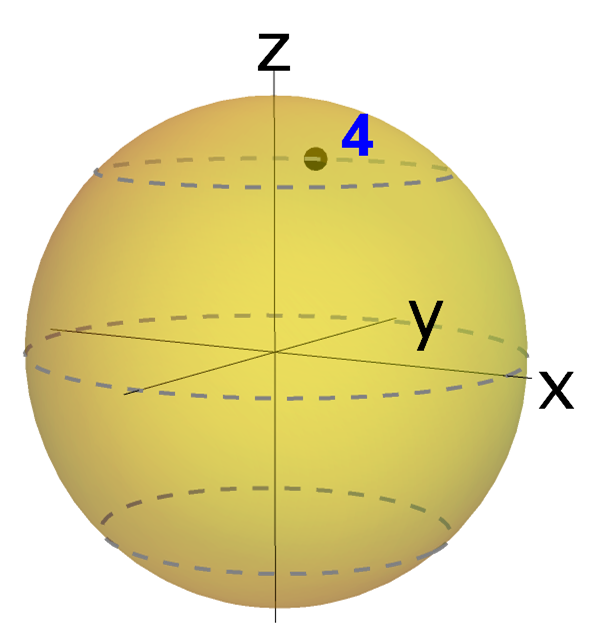}}
\\
\multicolumn{6}{c}{\scalebox{0.65}{\includegraphics{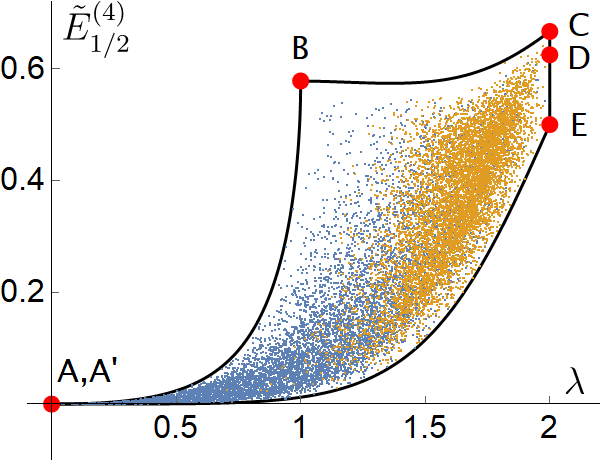}}}
\end{tabular}
\caption{
\label{4s.gram}
\emph{Top}: Constellations of various symmetric 4-qubit states (equivalently, spin-2 states), which are projected on the plot at the bottom:
 A: a coherent state along $-\hat z$,
B: W state with three stars in the south pole and one in the north pole, C: a regular tetrahedral
state with one star at the north pole, D:  two antipodal pair of stars (a rotated $S_z=0$ state), E: a GHZ state,
$\text{A}'$: another coherent state.
\emph{Bottom}: 
Scatter plot of the geometric measure of entanglement $\tilde{E}_{1/2}^{(4)}$ \emph{vs} the lowest nonzero eigenvalue $\lambda$ of the Gram matrix $G$, for 15,000 randomly generated symmetric 4-qubit states. 
The two different colors correspond to different measures in the projective space: yellow for the $SU(4)$-invariant measure (7,500 points), blue for uniform measure on the sphere for each of the stars (7,500 points). The heavy points in red are the images of the  states at the top. The curves in black correspond to geodesics in the projective space, connecting particular representatives of the vertices --- we conjecture them to be the boundary of the plot. 
}
\end{figure}
The curvilinear triangle of Fig.~\ref{s32.gram} is now replaced by the quadrangle shown at the bottom in Fig.~\ref{4s.gram}. We make a number of observations:
\begin{itemize}
\item
The scatter plot at the bottom seems to be bounded by the images of geodesics in the projective space, all of which are horizontal, except for AB. 
\item
The states shown in the top row are mapped to the border of the plot --- they include  coherent states (A, $\text{A}'$), a W state (B), a tetrahedral state (C), an $S_z=0$ state, rotated to some obscure direction (D), and a GHZ state (E) --- it is as though any spin-2 state that is ``somebody'', ends up in the border of the plot. 
\item
The segment CDE is vertical.
\item
The preimage, in $\mathbb{P}$, of a generic point in the plot is 5-dimensional (hence 2-dimensional in shape space $\mathcal{S}$). 
\item
In order to adequately define the borders, we were led to use two different measures in $\mathbb{P}$ --- the corresponding points show up in different colors in the plot.  
\item
We have not been able to find a single, simple family of states that covers the plot bijectively --- it is easy though to cover it, with some overlap, using two distinct families, \eg,  $C_4(\theta,\theta_1)=\{ (\theta,0) , (\theta,2\pi/3) ,(\theta, 4\pi/3), (\theta_1, 0) \}$, with $\theta \in [0,\pi/2]$, $\theta_1 \in [0,\pi]$, and $C'_4(\theta, \theta_{1})=\{ (\theta,0) , (\theta,\pi) ,(\theta_{1}, \pi/2),(\theta_{1}, 3\pi/2) \}$, with $\theta$, $\theta_1 \in [0,\pi]$.
\end{itemize}

Finally,  we treat the case of three qutrits. There are now three eigenvalues of the Gramian, summing up to 3. We denote by $\lambda$, $\lambda'$ the two smallest ones, and scatter plot them \emph{vs} $\tilde{E}_{1}^{(3)}$ in figure~\ref{k3s1.rep}. Some remarks, pertaining to that figure, follow:
\begin{figure}[h!]
\begin{tabular}{c c c c}
 A & B & C & D
 \\
\scalebox{0.25}{ \includegraphics{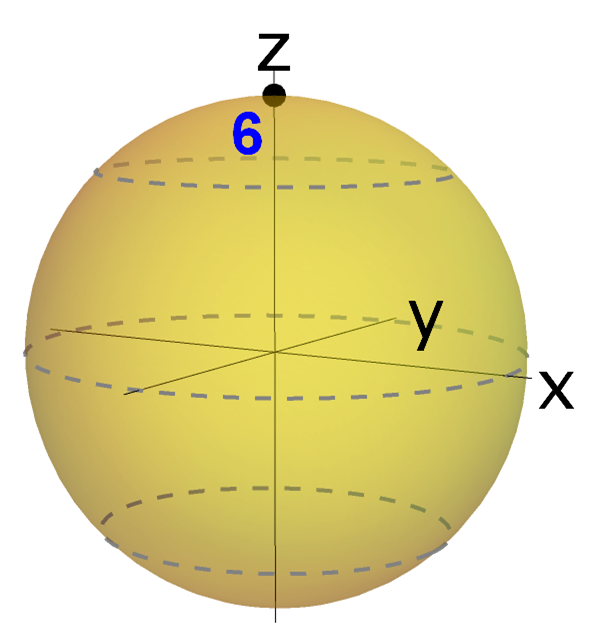}}
&
\scalebox{0.25}{ \includegraphics{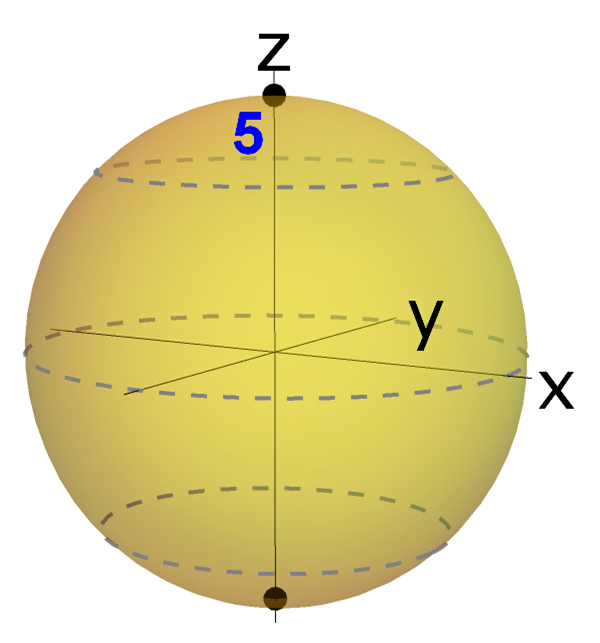}}
&
\scalebox{0.25}{ \includegraphics{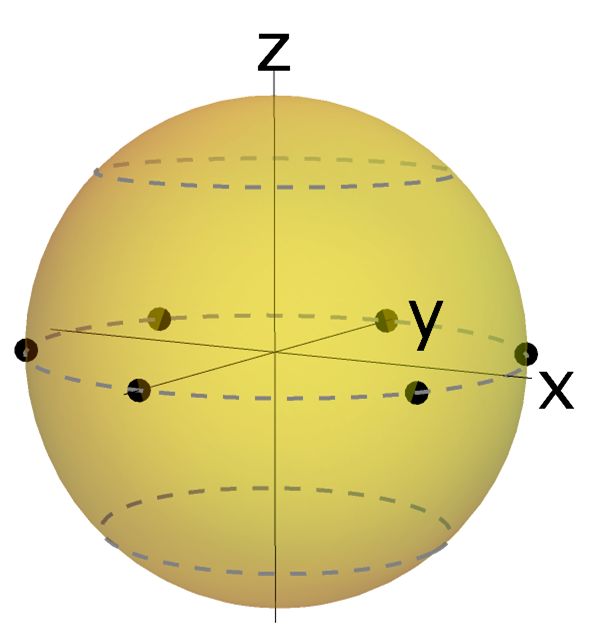}}
&
\scalebox{0.25}{ \includegraphics{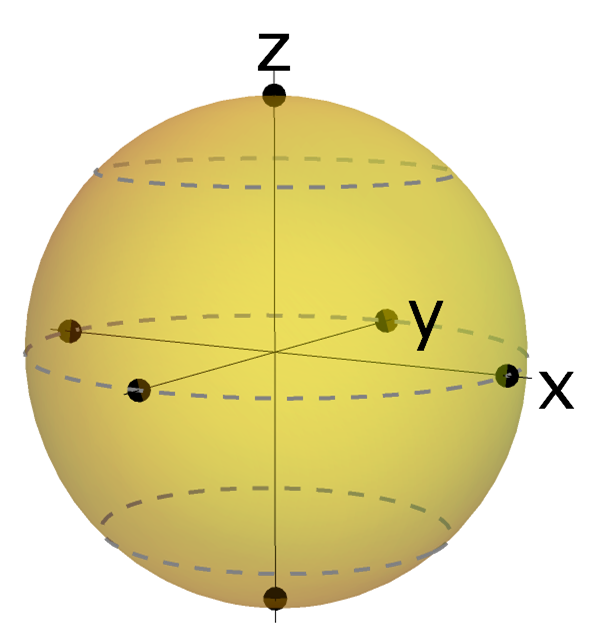}}
\\
\multicolumn{4}{c}{\scalebox{0.5}{\includegraphics{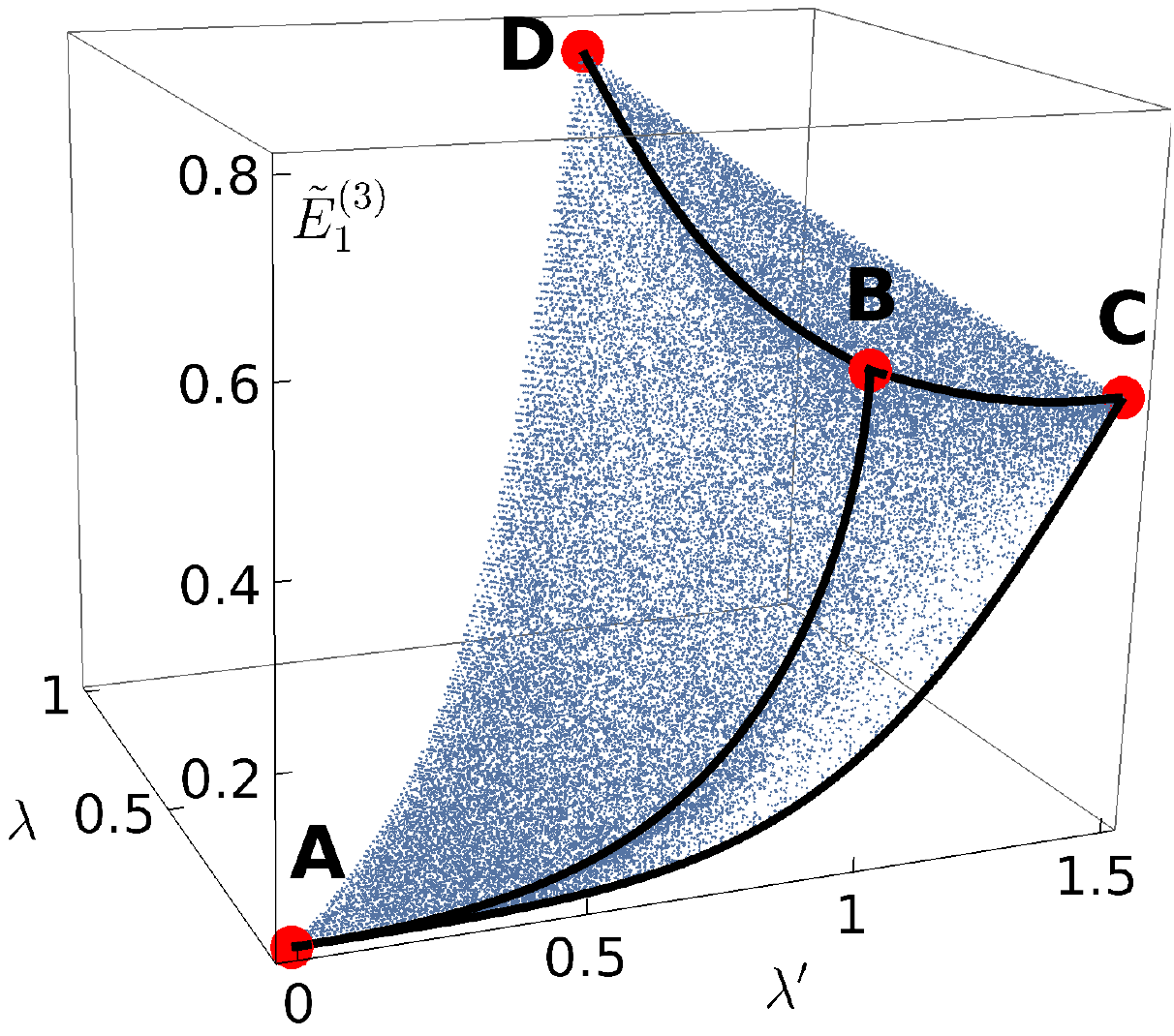}}}
\end{tabular}
\caption{\label{k3s1.rep}
  \emph{Top}:
Principal (spin-3) constellations of various symmetric 3-qutrit states, that map to the heavy red points in the plot at the bottom --- they are all characterized by the fact that their secondary (spin-1) component vanishes.
\emph{Bottom}: 3D scatter plot of the geometric measure of entanglement $\tilde{E}_{1}^{(3)}$ \emph{vs} the lowest two nonzero eigenvalues $\lambda$, $\lambda'$,  of the Gram matrix $G$, for 57,500 randomly generated symmetric 3-qutrit states. 
The heavy points in red are the images of the  states at the top. The curves in black, connecting red points,  correspond to geodesics in the projective space, connecting particular representatives of the vertices --- we conjecture them to form the boundary of the faces of the plot.}
\end{figure}
\begin{itemize}
\item
The plot resembles a curvilinear 3-simplex, the vertices of which are labelled by A, B, C, D. 
\item
The $SU(2)$-irrep decomposition in this case is $\mathcal{H}_1^{\vee 3}=\mathbf{7} \oplus \mathbf{3}$ (\ie, spin-3 and spin-1). 
States that differ by the diagonal action of $SU(3)$ project to the same point in the plot --- this replaces the freedom to rigidly rotate  the Majorana constellation in the $s=1/2$ case. The diagonal action of $SU(3)$ changes, in general, the $SU(2)$-irrep decomposition, and, hence, the multiconstellation of the state. We use this to select a representative state, in the preimage of each point in the plot, with vanishing spin-1 component --- the state then is completely characterized by a spin-3 constellation (the spectator constellation consists of a single star in the north pole). The spin-3 constellations of such representative states, that project to the vertices of the plot, are shown in the top row of the figure. Explicitly, they are (ignoring normalization)
\begin{alignat*}{3}
A 
& 
\colon 
(1,0,0)^T \vee (1,0,0)^T \vee (1,0,0)^T
& 
\text{coherent state along } \hat{z}
\\
B 
&
\colon 
(1,0,0)^T \vee (1,0,0)^T \vee (0,1,0)^T
& 
\text{W state along } \hat{z}
\\
C 
& 
\colon 
(1,0,1)^T \vee (e^{-i \pi/3},0,e^{i \pi/3})^T \vee (e^{-i2 \pi/3},0,e^{i2 \pi/3})^T
& 
\text{GHZ state in the equatorial plane}
\\
D 
& 
\colon 
(1,0,1)^T \vee (1,0,-1)^T \vee (0,1,0)^T
& 
\text{octahedral state (maximally entangled)}
\end{alignat*}
\item
By virtue of the results in Sect.~\ref{Tcoldfs}, the curvilinear triangle ABC, corresponding to $\lambda=0$, is identical to the one in Fig.~\ref{s32.gram}.
\item
We have been able to identify the borders of the figure shown in black as the images of horizontal geodesics in projective space. It would be interesting to relate these states, in particular the ones that project to the curve BC, to the 1-parameter family of maximally entangled states considered in~\cite{Tam.Wei.Par:09}.
\end{itemize}
 \section{Symmetric quantum rotosensors}
\label{Sqr}
Our multiconstellation representation of symmetric $k$-partite spin-$s$ states is based on the $SU(2)$ action on the underlying Hilbert space $\mathcal{H}_s^{\vee k}$. The latter is stratified in invariant subspaces, and the projections of a multipartite state $\ket{\boldsymbol{\Psi}}$  on  these subspaces determine the various spin components of the state in the BD basis. This is not the only way to slice the total Hilbert space in subspaces, though. Apart from the $SU(2)$ action mentioned above, the full $SU(2s+1)$ group also acts on $\mathcal{H}_s^k$, and a treatment parallel to ours in this work could be based on this latter action. Our motivation for focusing on $SU(2)$ stems from the central importance, both theoretical and experimental,  of the behavior of a quantum state under rotations, and a natural arena for our methods is the plethora of problems in quantum metrology, related to rotation detection by quantum means. We refer to quantum states suitable for rotation detection generically as \emph{quantum rotosensors}~\cite{Chr.Her:17,Gol.Jam:18,Mar.Wei.Gir:19}, distinguishing some as optimal, according to a variety of criteria. Our focus here will be on optimal detection in three particular cases: infinitesimal rotations around a given axis and uniformly averaged over all axes, and finite rotations by an angle $\eta$, uniformly averaged over all axes. The averaging version might sound a bit artificial, but is actually quite relevant in experiments,  \eg, when a sample of atoms starts out in a common quantum spin state, and, passing through a sufficiently varying magnetic field, ends up with uniform orientation (see, \eg, the discussion in~\cite{Chr.Her:17}).
\subsection{Optimal rotosensors for small rotations about a given axis}
\label{Orsrga}
\begin{figure*}[h!]
\begin{tabular}{c@{\hskip 1cm} c}
 \scalebox{0.4}{\includegraphics{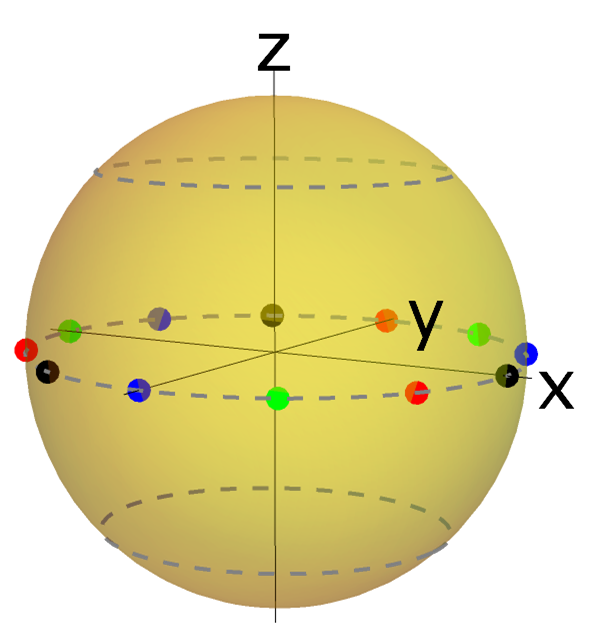}}
& \scalebox{0.55}{\includegraphics{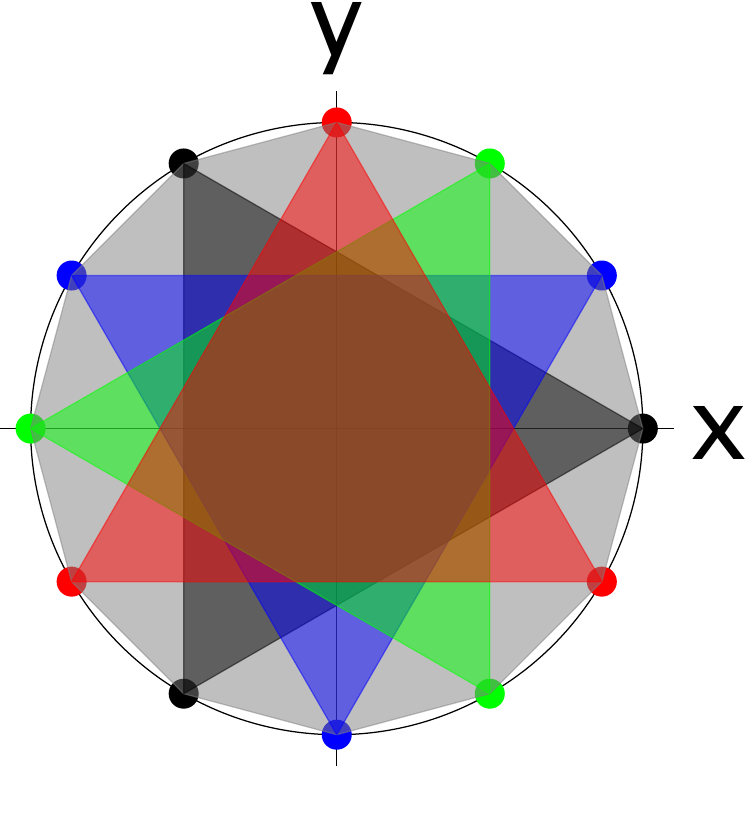}}  
\end{tabular}
\caption{\label{polygon.trian}
Plots of the principal constellation of  $\ket{\bm{\Psi}_{\text{GHZ}}}$ for $s=3/2$, $k=4$ on the sphere (left) and, viewed from above,  on the equator (right). $\ket{\bm{\Psi}_{\text{GHZ}}}$ is  the $\vee$-product of 4 spin-3/2 states $\ket{\phi_{\alpha}}$ with constellations given by equilateral triangles on the equator denoted by different colors. The principal constellation of $\ket{\bm{\Psi}_{\text{GHZ}}}$ is a dodecagon formed by the union of all the stars of the $\ket{\phi_{\alpha}}$, drawn in gray in the figure. $\ket{\bm{\Psi}_{\text{GHZ}}}$ has no other constellations since all but the top spinor vanish in its BD decomposition.}
\end{figure*}

As shown in~\cite{Bra.Cav:94} the optimal rotosensor for small rotations about $\hat{n}$  is the  state $\ket{\bm{\Psi}}$ that maximizes the Quantum Fisher Information (QFI) of the generator $\hat{n} \cdot \mathbf{S}$. Geometrically, the action of $SU(2)$ on $\mathcal{H}_s^k$ induces one on the corresponding projective space, $ \mathbb{P}(\mathcal{H}_s^k) \equiv \mathbb{P}_s^k$, where states are represented by density matrices,
\begin{equation}
\label{SU2actps}
 g  \triangleright \rho_{\ket{\boldsymbol{\Psi}_0}}=\mathfrak{D}^{(s,k)}_\text{BD}(g) \rho_{\ket{\boldsymbol{\Psi}_0}} \mathfrak{D}^{(s,k)}_\text{BD}(g)^{-1}
\, ,
\end{equation}
with $g \in SU(2)$, $\rho_{\ket{\boldsymbol{\Psi}_0}} \in \mathbb{P}_s^k$ and $\ket{\boldsymbol{\Psi}_0}$ is expanded in the BD basis (we drop, for the rest of this section, the index BD --- it is understood that all vectors and matrices are referred to this basis).  We may, without loss of generality, take $\hat{n}=\hat{z}$, and consider the curve $g_t=e^{-i t S_z}$ in $SU(2)$, which gives rise to the curve 
\begin{equation}
\label{ciPSz}
\rho_{\ket{\boldsymbol{\Psi}_t}}
=
g_t \triangleright \rho_{\ket{\boldsymbol{\Psi}_0}}
=
 e^{-i t S_z^{(s,k)}} \rho_{\ket{\boldsymbol{\Psi}_0}} e^{i t S_z^{(s,k)}}
\end{equation}
in $\mathbb{P}_s^k$, with tangent vector, at $t=0$,
\begin{equation}
\label{tvcgt}
V_z \equiv \frac{\partial}{\partial t} \left. \rho_{\ket{\boldsymbol{\Psi}_t}} \right|_{t=0}=i [S^{(s,k)}_z,\rho_{\ket{\boldsymbol{\Psi}_0}}]
\, .
\end{equation}
The Fubini-Study metric $F$ in $\mathbb{P}_s^k$, evaluated on two tangent vectors $A$, $B$, at $\rho_{\ket{\boldsymbol{\Psi}}}$, gives
\begin{equation}
\label{FSPks}
F_{\ket{\boldsymbol{\Psi}}}(A,B)=\frac{1}{2} \Tr (AB)
\, ,
\end{equation}
so that the modulus squared of $V_z$, evaluated at $\rho_{\ket{\boldsymbol{\Psi}}} \equiv \pket{\boldsymbol{\Psi}}$, turns out to be
\begin{equation}
\label{msvz}
|V_z|^2
=
F_{\pket{\boldsymbol{\Psi}}}(V_z,V_z)
=
\bra{\boldsymbol{\Psi}}{S_z^{(s,k)}}^2 \ket{\boldsymbol{\Psi}}
-
\bra{\boldsymbol{\Psi}}{S_z^{(s,k)}} \ket{\boldsymbol{\Psi}}^2
=
\left( \Delta S^{(s,k)}_z \right)^2_{\pket{\boldsymbol{\Psi}}}
\, .
\end{equation}
Substituting the BD components of $\ket{\boldsymbol{\Psi}}$ from (\ref{PsiDdef}), we find
\begin{equation}
\label{QFI.Pure}
|V_z|^2
 =
\sum_{j} |z_{j}|^2 \bra{\psi^{(j)}} (S_z^{(j)})^2 \ket{\psi^{(j)}} 
-
\left( \sum_{j} |z_{j}|^2 \bra{\psi^{(j)} } S_z^{(j)} \ket{\psi^{(j)}} \right)^2 
\, ,
\end{equation}
where the index $j$ runs over the irreducible spin-$j$ blocks of $\ket{\bm{\Psi}}$. Note that $|V_z|$ is independent of the phases of the $z_j$.
It is easily seen that the state that maximizes \eqref{QFI.Pure} is the GHZ state \cite{Gre.Hor.Zei:89}
\begin{equation}
\label{GHZ.symk}
\ket{\bm{\Psi}_{\text{GHZ}}} 
=
\frac{1}{\sqrt{2}} 
\left( 
\ket{s,s}^{\otimes k} + e^{i \theta} \ket{s,-s}^{\otimes k}
\right)
 \, ,
\end{equation} 
for which all $z_j$ are zero, except for $z_{s_{\text{max}}}$, which is 1, where $s_{\text{max}}=ks$. This is so because $|V_z|^2$ is the difference of two non-negative terms, and the GHZ clearly maximizes the first, and minimizes the second. The top spinor $\ket{\psi^{(s_{\text{max}})}}$ in $\ket{\boldsymbol{\Psi}}$ has a regular $2ks$-agon at the equator as Majorana constellation, and changing the phase $\theta$ in (\ref{GHZ.symk}) results in a rotation of that figure around the $z$-axis. Interestingly, $\ket{\bm{\Psi}_{\text{GHZ}}}$ is $\vee$-factorizable, $\ket{\bm{\Psi}_{\text{GHZ}}}= \bigvee_{\alpha} \ket{\phi_{\alpha}}$, where each spin-$s$ constituent $\ket{\phi_{\alpha}}$ has as Majorana constellation a regular $(2s)$-agon on the equator, each rotated by an angle  $\pi/(sk)$ about the $\hat{z}$-axis \wrt{} the previous one, with $\alpha=0,1,\dots,k-1$. To see this, note that 
\begin{equation}
\label{phialphadef}
\ket{\phi_{\alpha}} \propto \ket{s,s} + e^{2\pi i \alpha /k}\ket{s,-s} 
\, ,
\end{equation}
and, therefore,  $ \bigvee_{\alpha=0}^{k-1} \ket{\phi_{\alpha}} $ is a linear combination of $\vee-$products of $r$  $\ket{s,s}$ factors and $(k-r)$ $\ket{s,-s}$ ones,
\begin{equation}
\bigvee_{\alpha=0}^{k-1} \ket{\phi_{\alpha}} 
\propto 
\sum_{r=0}^k A_r 
\underbrace{\ket{s,s}\vee \dots \vee \ket{s,s}}_{r} 
\vee
\underbrace{\ket{s,-s}\vee \dots \vee \ket{s,-s}}_{k-r} 
 \, .
\end{equation}
The coefficients $A_r$ are obtained (up to a global factor) by the following expansion
\begin{equation}
\prod_{\al=0}^{k-1} \left( x+ e^{2\pi i \al /k} \right) = \sum_{k=0}^{k} A_r x^r 
\, , 
\end{equation}
where, conveniently, the \lhs{}  is equal to $x^k + (-1)^{k-1}$. Hence, $A_0=(-1)^{k-1}$, $A_k=1$ and $A_r=0$ for $0<r<k$, leading to
\begin{equation}
\bigvee_{\alpha=0}^{k-1} \ket{\phi_{\alpha}} \propto 
 \ket{s,s}^{\otimes k} + (-1)^{k-1} \ket{s,-s}^{\otimes k}
=\ket{ks,ks} + (-1)^{k-1} \ket{ks,-ks}
 \, ,
\end{equation}
which is the GHZ state \eqref{GHZ.symk}, in a particular orientation. As an example, we consider the GHZ state (\ref{GHZ.symk}) for $s=3/2$, $k=4$, and draw its principal constellation, as well as the Majorana constellations of its factor states $\ket{\phi_\alpha}$, $\alpha=0,1,2,3$, in Fig. \ref{polygon.trian}, where each color represents the stars of a particular $\ket{\phi_{\alpha}}$.
Since both the Hermite and Murnaghan isomorphisms $h$, $m$, are implemented by the identity matrix in the BD basis, both $h(\ket{\boldsymbol{\Psi_{\text{GHZ}}}})$ and $m(\ket{\boldsymbol{\Psi}_{\text{GHZ}}})$ are also GHZ states. For the former we find
\begin{equation}
\label{hGHZ}
h(\ket{\bm{\Psi}_{\text{GHZ}}})
\propto 
\ket{k/2,k/2}^{\otimes 2s} + e^{i \theta} \ket{k/2,-k/2}^{\otimes 2s}
\, ,
\end{equation}
with its $\vee$-factorization involving $k$-gons on the equator as factor states, while for the latter, it is easily shown that
\begin{equation}
m\left( \ket{\bm{\Psi}} \right) =
\frac{1}{\sqrt{2}} \big( 
\left| \tilde{s},\tilde{s} \right\rangle
\wedge 
\left| \tilde{s},\tilde{s}-1  \right\rangle
\wedge \dots
\wedge
\left| \tilde{s},\tilde{s} -(k-1) \right\rangle
+
\left| \tilde{s},-\tilde{s} \right\rangle
\wedge 
\left| \tilde{s},-\tilde{s}+1  \right\rangle
\wedge \dots
\wedge
\left| \tilde{s},-\tilde{s} +(k-1) \right\rangle
 \big)
 \, ,
 \label{GHZ.wedge}
\end{equation}
where $\tilde{s}=s+\frac{k-1}{2}$, which is not necessarily $\wedge-$factorizable (see Example \ref{H1V2Mur} below).
\begin{myexample}{Uniqueness of the GHZ state as the optimal rotosensor}{H1V2infk}
We show here that $\ket{\boldsymbol{\Psi}}_{\text{GHZ}}$, for $s=1$, $k=2$, is the only state that maximizes $\Delta S^{(1,2)}_z$.
We have $\Hs^{\vee 2}_1 \sim \mathbf{3} \oplus \mathbf{1}$, so that, for any state $\ket{\bm{\Psi}} \in \mathcal{H}_1^{\vee 2}$ we get the decomposition $\ket{\bm{\Psi}} = z_2 \ket{\psi^{(2)}} \oplus z_0 \ket{\psi^{(0)}}$. The GHZ state, on the other hand, also factorizes as $\ket{\boldsymbol{\Psi}}_{\text{GHZ}}=\ket{\phi_1} \vee \ket{\phi_2}$ (see (\ref{phialphadef})).

The spin-0 basis state is
\begin{align}
\label{psizbs}
\ket{\psi_0} 
&
=  
\left( -\sqrt{2} \hat{e}_{13} + \hat{e}_{22} \right)/\sqrt{3} 
\\
&
= \left(\ket{0,0} -\ket{1,-1} - \ket{-1,1} \right)/\sqrt{3}
\, ,
\end{align}
(the second line in the $(m_1,m_2)$ notation) and 
$z_0 = \braket{\psi_0}{ \bm{\Psi} } \propto \braket{\phi^A_2}{ \phi_1}=0$,
where $\ket{\phi^A_2}$ is the state antipodal to $\ket{\phi_2}$.

 It is clear that a state  $\ket{\bm{\Psi}}^T =(\ket{\psi^{(2)}}^T,\ket{\psi^{(0)}}^T)$ that satisfies 
\begin{equation}
\bra{\psi^{(2)}}S_z^2\ket{\psi^{(2)}} = 4 
\, ,
\quad \quad
\bra{\psi^{(2)}}S_z\ket{\psi^{(2)}} = 0 
\, ,
\quad \quad
z_0  = 0 
\, ,
\label{1eq.k2s1}
\end{equation}
maximizes $|V_z|$ in~(\ref{QFI.Pure}). The first of these equations implies that only $S_z=2$ components can be present in the spin-2 component, the second says that they should have equal moduli, so that $\ket{\psi^{(2)}} \propto (\ket{2,2} + e^{-i\al} \ket{2,-2})$, while the third implies that the $j=2$ spinor is the only one present --- this determines the state to be the GHZ one, with the phase $\alpha$ effecting a rotation around  the $z$-axis. The generalization of this result to any $(s,k)$ is straightforward. 
\end{myexample}

\begin{myexample}{Optimal rotosensors, Murnaghan isomorphism, and $\wedge$-factorizability}{H1V2Mur}
We consider in some detail the Murnaghan isomorphism $m \colon \mathcal{H}_1^{\vee 2} \rightarrow \mathcal{H}_{3/2}^{\wedge 2}$. Since the state $m(\ket{\bm{\Psi}_{\text{GHZ}}}) \equiv \ket{\tilde{\bm{\Psi}}_{\text{GHZ}}} \in \Hs_{3/2}^{\wedge 2}$ has identical BD structure as  $\ket{\bm{\Psi}_{\text{GHZ}}} \in \mathcal{H}_1^{\vee 2} $, it also  maximizes $|V_z|$ in \eqref{QFI.Pure},
\begin{equation}
\ket{\tilde{\bm{\Psi}}_{\text{GHZ}}} 
=
\left(  
\tilde{z}_2 \ket{\tilde{\psi}^{(2)}} , \tilde{z}_0 \ket{\tilde{\psi}^{(0)}} \right)^T
=
\left( 
\big( c_{(2,2)}, c_{(2,1)} , c_{(2,0)} , c_{(2,-1)} , c_{(2,-2)} \big) , \big( c_{(0,0)} \big) 
\right)^{T} 
=
\left( 
\big( \frac{1}{\sqrt{2}}, 0,0,0, \frac{1}{\sqrt{2}} \big) , \big( 0 \big) \right)^{T} 
\, .
\label{GHZ.mur}
\end{equation}
It is easily seen that $\ket{\tilde{\bm{\Psi}}_{\text{GHZ}}}$ is not $\wedge-$factorizable: the condition for this is the Pl\"ucker relation on the components of $\ket{\tilde{\bm{\Psi}}_{\text{GHZ}}}$  \cite{Sha.Rem:13,Chr.Guz.Han.Ser:21}, 
\begin{equation}
2c_{(2,2)} c_{(2,-2)}-2c_{(2,1)}c_{(2,-1)}+c_{(2,0)}^2 - c_{(0,0)}^2 = 0 
\, ,
\label{fact.wedge}
\end{equation}
which is clearly not satisfied by the $c_{(s,m)}$'s of~\eqref{GHZ.mur}.

The Pl\"ucker relation can be expressed as 
\begin{equation}
\left| 
\tilde{z}_2 \braket{ ({\tilde{\psi}}^{(2)} )^A }{ \tilde{\psi}^{(2)} } 
\right|^2 
=
|\tilde{z}_0|^2 
 \, ,
\end{equation}  
where $\ket{\tilde{\psi}^A}$ denotes the state antipodal to $\ket{\tilde{\psi}}$. Substituting this into~\eqref{QFI.Pure} yields a new equation, involving only $\ket{\tilde{\psi}^{(2)}}$, to which we have found numerically two different shape solutions, both of which reach the maximum value $|V_z|^2=5/2$,
\begin{align}
\ket{\tilde{\Psi}} 
&
=
\left( 
\frac{1}{2}
\big( -1,1,0,-1,1 \big) , 
\big(0 \big) 
\right)^T 
=
\ket{\chi_1} \wedge \ket{\chi_2}
\, ,
\nn
\\
\ket{\tilde{\Psi}'} 
&
=
\left( \frac{1}{2}\big( 1,1,0,1,1 \big) , \big(0\big) \right)^T
=
\ket{\chi_1} \wedge \ket{\chi_3}
\, ,
\label{Twost.32}
\end{align}
where
\begin{equation}
\label{chi123}
\ket{\chi_1}=\frac{1}{\sqrt{2}} 
\left( 
\ket{\frac{3}{2},\frac{3}{2}} - \ket{\frac{3}{2},-\frac{3}{2}} 
\right)
\, ,
\qquad
\ket{\chi_2}=\frac{1}{\sqrt{2}} 
\left( 
\ket{\frac{3}{2},\frac{1}{2}} + \ket{\frac{3}{2},-\frac{1}{2}} 
\right)
\, ,
\qquad
\ket{\chi_3}=\frac{1}{\sqrt{2}} 
\left( 
\ket{\frac{3}{2},\frac{1}{2}} - \ket{\frac{3}{2},-\frac{1}{2}} 
\right)
\, .
\end{equation}
 Both states have only principal constellation and their spin-2 components both satisfy 
\begin{equation}
\braket{(\psi^{(2)})^A }{ \psi^{(2)}}  
=
0 
 \, , 
\quad
\bra{\psi^{(2)}} S_z \ket{\psi^{(2)}} = 0 
\, , 
\quad 
\bra{\psi^{(2)}} S_z^2 \ket{\psi^{(2)}} = \frac{5}{2} 
\, .
\end{equation}
Their principal constellations 
and those of $\ket{\chi_i}$  are plotted in Fig. \ref{Wedge.H32.2}.
\begin{figure*}[t]
 \scalebox{0.3}{\includegraphics{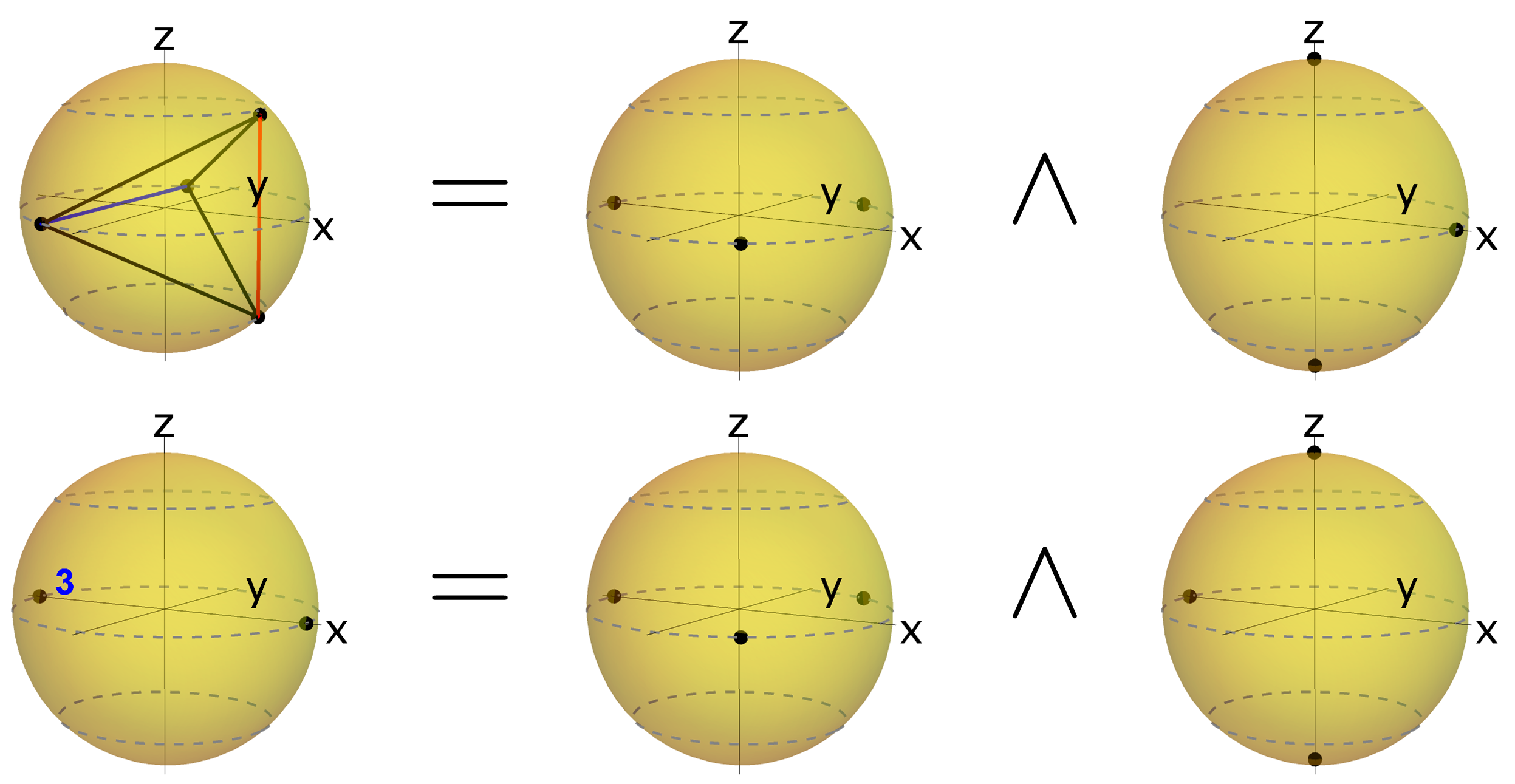}}
\caption{
\label{Wedge.H32.2}
The principal constellations of the  $\wedge$-factorizable optimal rotosensors $\ket{\tilde{\Psi}}$, $\ket{\tilde{\Psi}'} $ of Eq.~\eqref{Twost.32} (\lhs{}, top and bottom),
and the constellations of their factor spin-$3/2$ states $\ket{\chi_i}$ of Eq.~\eqref{chi123} (\rhs, top and bottom). The principal constellation of $\ket{\tilde{\Psi}}$ (top left) is not a regular tetrahedron, only the edges shown in black in the figure  have equal lengths. 
}
\end{figure*}
\end{myexample}
%
%
%
%
\subsection{Optimal rotosensor for an infinitesimal rotation around an averaged axis}
We generalize the spin-$s$ rotosensor notion treated in~\cite{Gol.Jam:18} to $\mathcal{H}_s^{\vee k}$. The quantity to be maximized is $|V_{\hat{n}}|^2$, with $V_{\hat{n}}=i[\hat{n} \cdot \mathbf{S}, \rho_{\ket{\boldsymbol{\Psi}}}]$, uniformly averaged over the rotation axis $\hat{n}$, \ie, we seek the state $\ket{\boldsymbol{\Psi}}$ maximizing
\begin{equation}
I(\ket{\bm{\Psi}})
=
\int_{S^2} \td \hat{n} |V_{\hat{n}}|^2
\propto
\frac{1}{3}
\left(
\left(\Delta S_x  \right)^2 + \left( \Delta S_y \right)^2 + \left( \Delta S_z \right)^2 
\right)_{\ket{\boldsymbol{\Psi}}}
\, ,
\label{QFI.av}
\end{equation}
Working in the BD basis we find
\begin{equation}
\label{QFIinf.av}
I(\ket{\Psi}) 
= 
\frac{1}{3}  
\left( 
\sum_{r=1}^t 
|z_r|^2 j_r (j_r+1) 
- \sum_{i=x,y,z} 
\left( 
\sum_{r=1}^t
|z_r|^2 \bra{\psi^{(j_r)}} S_i^{(j_r)} \ket{\psi^{(j_r)}} \right)^2
\right)  
\, ,
\end{equation}
where $r$ enumerates the irreducible blocks of $\ket{\boldsymbol{\Psi}}$ and $t$ is their total number (we omit, for notational simplicity, a further index $\alpha$ that distinguishes among same-spin components). It is easily seen that one may independently maximize the first term and minimize the second, using  any state such that all $z_r$ are zero except for $z_{s_{\text{max}}}=z_{ks}$, which should be of modulus 1, and with principal spinor such that $\bra{\psi^{(s_{\text{max}})}} \hat{n} \cdot \bm{S} \ket{\psi^{(s_{\text{max}})}} = \bm{0}$, the latter condition termed \emph{1-anticoherence} by Zimba~\cite{Zim:06}.

\begin{myexample}{2-symmetric qutrit states that maximize $I(\ket{\bm{\Psi}})$}{H1V2infa}
A general state $\ket{\bm{\Psi}} \in \Hs_1^{\vee 2}$, in the BD basis, is given by  $\ket{\bm{\Psi}}= \left( z_2 \ket{\psi^{(2)}} , z_0 \ket{\psi^{(0)}} \right)^T$. Every state $\ket{\bm{\Psi}}$ such that $z_0=0$, $|z_2|=1$ and $\ket{\psi^{(2)}}$ is  1-anticoherent maximizes \eqref{QFIinf.av}. The set of all 1-anticoherent spin-2 states, up to rigid rotations,  are of the form~\cite{Bag.Bas.Mar:14}
\begin{equation}
\label{2spin.1ant}
\ket{\psi_2(\nu)} = \sqrt{\frac{3}{6+2|\nu|^2}} \big( 1 , 0 , \sqrt{\frac{2}{3}} \nu , 0, 1  \big) \, ,
\end{equation} 
where the complex variable $\nu$ has domain 
$D= \left\{ 
\nu \in \mathbb{C} | \, \text{Re}(\nu) \geq 0 , \text{Im} (\nu) \geq 0 \text{ and } | \nu - 1 | \leq 2 
\right\} $. If we, additionally,  insist on $\ket{\boldsymbol{\Psi}}$ be $\vee$-factorizable, only the value $\nu=0$ gives a solution, corresponding to a GHZ state.
\end{myexample}

\subsection{Optimal rotosensor for a finite rotation around an averaged axis}
Finally, we take up the notion of optimal rotosensors put forth in~\cite{Chr.Her:17}, and extend the results found there and in~\cite{Mar.Wei.Gir:19} to $\mathcal{H}_s^{\vee k}$. The relevant quantity here is the transition probability between the state $\ket{\bm{\Psi}}$ and $\Rr (\eta , \bm{n})\ket{\bm{\Psi}}$, also called the fidelity  \cite{Mar.Wei.Gir:19}
\begin{equation}
\label{fid.axes}
F_{\ket{\bm{\Psi}}} (\eta , \hat{n}) 
\equiv 
| \ev{\bm{\Psi} |\Rr(\eta , \hat{n})| \bm{\Psi} }|^2 
= 
\left(
\sum_{r=1} |z_{r}|^2 \ev{\psi^{(j_r)} |\Rr^{(j_r)}(\eta, \hat{n}) | \psi^{(j_r)}} 
\right)^2 
\, ,
\end{equation}
with the slightly simplified notation $\Rr^{(j_r)}(\eta, \hat{n}) \equiv D^{(j_r)}(\Rr(\eta, \hat{n}))$. 
As in the previous two cases, the complex phases of the  $z_{j}$ are not relevant. The optimal rotosensor for a rotation by an angle $\eta$ around an averaged axis is the state that minimizes the average of the fidelity over the rotation axis
\begin{equation}
\label{fid.ave}
\Fc_{\ket{\bm{\Psi}}}(\eta) 
\equiv 
\frac{1}{4 \pi} \int_{S^2} F_{\ket{\bm{\Psi}}} (\eta , \hat{n}) \diff \hat{n} 
\, .
\end{equation}
For the irrep $\Rr^{(j)}$ we use the formula  (\cite{Var.Mos.Khe:88}, equation (18), p.46)
\begin{equation}
\Rr^{(j)}
= 2\sqrt{\pi} \sum_{L=0}^{2j} 
\frac{(-i)^L}{\sqrt{2j+1}} \chi_L^{(j)}(\eta) 
\sum_{M=-L}^L 
Y_{LM}^* (\theta, \phi) T^{(j)}_{LM} 
\, ,
\end{equation}
where $\chi_L^{(j)}(\eta)$ are the generalized characters of the irrep $\Rr^{(j)}$ \cite{Var.Mos.Khe:88} and the spherical harmonics are normalized according to
\begin{equation}
\int_{S^2} Y_{LM}(\theta,\phi) Y^*_{L' M'}(\theta,\phi) d\bm{n} 
= 
\delta_{L L'} \delta_{M M'} 
\, ,
\end{equation}
and find, integrating over the sphere,
\begin{equation}
\Fc_{\ket{\bm{\Psi}}}(\eta) 
= 
\sum_{q,r} 
\frac{|z_{q} z_{r}|^2}{\sqrt{(2q+1)(2r+1)}} 
\left( 
\sum_{L=0}^{\min(2q,2r)} 
\sum_{M=-L}^L 
\chi_L^{(q)}(\eta) \chi_L^{(r)} (\eta)
 \bra{\psi^{(q)}} 
 T_{LM}^{(q)} 
 \ket{\psi^{(q)}} 
 \bra{\psi^{(r)}} 
 T_{LM}^{(r) \da} 
 \ket{\psi^{(r)}}
\right)\, ,
\end{equation}
where $q$ and $r$ range over the BD blocks of $\ket{\bm{\Psi}}$.
\begin{myexample}{Average fidelity for 2-symmetric qutrit states}{H1V3infa}
In this case we find
\begin{align}
\Fc_{\ket{\bm{\Psi}}}(\eta) 
=
|z_0|^4
+
\frac{2|z_0 z_2|^2 \chi_0^{(2)}(\eta)}{5} 
+ 
\frac{|z_2|^4}{5} 
\left(
\frac{\left( \chi_0^{(2)}(\eta) \right)^2}{5}
+ 
\sum_{L=1}^4
 \left( \chi_L^{(2)}(\eta) \right)^2 w_L^2
\right) 
\, ,
\label{QFIav.k2s1}
\end{align}
where 
\begin{equation}
w_L^2 
\equiv 
\sum_{M=-L}^L
\bra{\psi^{(2)}} T_{LM}^{(2)} \ket{\psi^{(2)}} \bra{\psi^{(2)}} T_{LM}^{(2) \da} \ket{\psi^{(2)}} 
\, .
\end{equation}
It is also illustrative to compute $\Fc_{\ket{\bm{\Psi}}}(\eta)$ by expanding \eqref{fid.axes} in the BD basis and using $\Rr^{(0)}=I$,
\begin{equation}
\Fc_{\ket{\bm{\Psi}}}(\eta) 
= 
|z_0|^4
+
\frac{2}{5}|z_0 z_2|^2\chi_0^{(2)}(\eta)
+ |z_2|^4 \Fc_{\ket{\psi^{(2)}}}(\eta) 
\, ,
\end{equation}
where $\Fc_{\ket{\psi^{(2)}}}(\eta)$ is the last term in \eqref{QFIav.k2s1}, and has also been calculated in \cite{Mar.Wei.Gir:19} using anticoherent measures and studied numerically in \cite{Chr.Her:17}. 
\end{myexample}
%
\section{Conclusions}
\label{Conc}
In this work we have introduced the Majorana representation for pure
symmetric states of $k$ spins-$s$ (or pseudo-spins-$s$, \ie, quantum systems with
finite, $2s+1$-dimensional Hilbert space --- the $SU(2)$ action can still be relevant in this case, see, \eg,~\cite{Bon.Sch.Haa:71}). This is achieved by
decomposing the states into irreps of $SU(2)$ and then using the standard
Majorana representation for the pure state of a single spin $j$
\cite{Maj:32}.  Each irrep component gives rise to its own Majorana
constellation that transforms rigidly under rotation, but is agnostic
about its relative phase and amplitude.  The relative amplitudes, with
respect to the other $SU(2)$ components, are fixed by an
additional ``spectator'' constellation, defined through transforming
the respective constellations to a canonical orientation, similar to
what was developed for mixed states of multiqubit systems
\cite{Ser.Bra:20} and for antisymmetric states
\cite{Chr.Guz.Han.Ser:21}.  The general  procedure was illustrated with a number of
examples, in particular the case of symmetric states of up to four
qutrits.  Our results not only enabled us to visualize complex states
of these quantum systems, but also to investigate some of their
properties.

The decomposition into $SU(2)$ irreps allowed us to gain a
better understanding of the Hermite and Murnaghan isomorphisms. By
considering spins-$s$ as symmetric states of $2s$ spins-1/2, it became evident 
that the symmetrized states of $k$ spins-$s$ are equivalent
to those of $2s$ spins-$k/2$, in the sense of their decomposition into $SU(2)$
irreps. By combining this isomorphism with Murnaghan's 
isomorphism, that relates symmetric states of $k$ spins-$s$ to totally
anti-symmetric states of $k$ spins-$s+(k-1)/2$, and a map from the
$k$-planes that represent totally anti-symmetric $k$-partite
factorizable spin-$s$ states to the $2s+1-k$ planes that are their
orthogonal complements, we obtained a unified view of the Hermite and
Murnaghan isomorphisms. 

We calculated the Quantum Fisher Information
relevant for measuring rotations, for spins that
are physical spins or angular momentum, rather than ``pseudo-angular
momenta'' \ie, abstract $n$-level systems. The state that maximizes
the QFI for rotation about a given axis is a generalized GHZ state,
namely a superposition of all $k$ spins-$s$ pointing up and down. This
state has a multiconstellation with only the principal
component (max value of total spin) non-zero and represented by a
regular $2ks$-agon in the $xy$-plane.  Via the Murnaghan isomorphism
there is a totally antisymmetric state of $k$
spins-$s+(k-1)/2$ that contains exactly the same $SU(2)$ components and
hence leads to the same QFI.  Reasoning in terms of the Majorana
constellations is very helpful here.

We also investigated the geometrical entanglement of multiqudit
states and showed in particular that the GHZ-state of $k$
spins $s$ is $\vee$-factorizable for any $s>1/2$. More generally, we looked into the
relation between the geometric entanglement $E$ and the
eigenvalues of the Gramian matrix constructed from the single-qudit
states through which the state in question is obtained via
symmetrization.  While there is no one-to-one relation between
geometrical entanglement and these eigenvalues, as explicitly shown
\eg, for three qubits, where only one independent eigenvalue
$\lambda$ exists, the Majorana constellations help to identify
vertices of the plot of $E$ versus $\lambda$.  The projections, to the $(\lambda,E)$-plane, 
of geodesics connecting these vertices in projective space,  delimit the range of possible pairs
$(\lambda,E)$.

Having a visualization of complex quantum states with the nice
properties of the Majorana representation helps to identify potentially
interesting states, to reduce the number of parameters in optimization
problems by profiting from the rigid rotation of the constellations
under $SU(2)$ transformations, and more generally to develop
an intuitive understanding of  geometric  insights for quantum
states. We hope that our generalization of the Majorana representation
to symmetric states of qudits will contribute to the further exploration of
these objects.  
\section*{Acknowledgements}
CC, LH, and ESE would like to acknowledge partial financial support from the DGAPA PAPIIT project IN111920 of UNAM. ESE would also like to acknowledge support from a postdoctoral fellowship of CONACyT.
\clearpage

\begin{thebibliography}{10}

\bibitem{Maj:32}
E.~Majorana, ``Atomi orientati in campo magnetico variabile,'' {\em Nuovo
  Cimento}, vol.~9, pp.~43--50, 1932.

\bibitem{Pen.Rin:90}
R.~Penrose and W.~Rindler, {\em Spinors and {S}pace-time, Vol.{} 1}.
\newblock Cambridge University Press, 1990.

\bibitem{Pen:90}
R.~Penrose, {\em The Emperor's New Mind}.
\newblock Oxford University Press, 1990.

\bibitem{Pen:94}
R.~Penrose, {\em Shadows of the Mind}.
\newblock Oxford University Press, 1994.

\bibitem{Pen:07}
R.~Penrose, {\em The {R}oad {T}o {R}eality}.
\newblock Vintage Press, 2007.

\bibitem{Pen:up}
R.~Penrose, ``Orthogonality of general spin states.'' Unpublished notes.

\bibitem{Per:86}
A.~Perelomov, {\em Generalized coherent states and their applications}.
\newblock Springer, 1986.

\bibitem{Rad:71}
J.~M. Radcliffe, ``Some properties of coherent spin states,'' {\em J. Phys. A:
  Gen. Phys.}, vol.~4, pp.~313--324, 1971.

\bibitem{Are.Cou.Gil.Tho:72}
F.~T. Arecchi, E.~Courtens, R.~Gilmore, and H.~Thomas, ``Atomic coherent states
  in quantum optics,'' {\em Phys. Rev.}, vol.~A 6, p.~2211, 1972.

\bibitem{Gil:08}
R.~Gilmore, {\em Lie {G}roups, {P}hysics, and {G}eometry}.
\newblock Cambridge University Press, 2008.

\bibitem{Chr.Guz.Ser:18}
C.~Chryssomalakos, E.~Guzm\'an-Gonz\'alez, and E.~Serrano-Ens\'astiga,
  ``Geometry of spin coherent states,'' {\em J.{} Phys.{} A: Math.{} Theor.},
  vol.~51, no.~16, p.~165202, 2018.
\newblock \texttt{arXiv:1710.11326}.

\bibitem{Leb:91}
P.~Leb{\oe}uf, ``Phase space approach to quantum dynamics,'' {\em J. Phys.},
  vol.~A~24, p.~4575, 1991.

\bibitem{Bog.Boh.Leb:92}
E.~Bogomolny, O.~Bohigas, and P.~Leb{\oe}uf, ``Distribution of roots of random
  polynomials,'' {\em Phys. Rev. Lett.}, vol.~68, p.~2726, 1992.

\bibitem{Han:96}
J.~H. Hannay, ``Chaotic analytic zero points: exact statistics for those of a
  random spin state,'' {\em J.{} Phys.{} A: Math.{} Gen.}, vol.~29, no.~5,
  pp.~L101--L105, 1996.

\bibitem{Zim:06}
J.~Zimba, ``Anticoherent spin states via the {M}ajorana representation,'' {\em
  Electronic Journal of Theoretical Physics}, vol.~3, no.~10, pp.~143--156,
  2006.

\bibitem{Bag.Dam.Gir.Mar:15}
D.~Baguette, F.~Damanet, O.~Giraud, and J.~Martin, ``{Anticoherence of spin
  states with point-group symmetries},'' {\em Phys.{} Rev.{} A}, vol.~92,
  p.~052333, Nov. 2015.

\bibitem{Bag.Mar:17}
D.~Baguette and J.~Martin, ``Anticoherence measures for pure spin states,''
  {\em Phys.{} Rev.{} A}, vol.~96, p.~032304, 2017.

\bibitem{Gir.Bra.Bra:08}
O.~Giraud, P.~Braun, and D.~Braun, ``Classicality of spin states,'' {\em
  Phys.{} Rev.{} A}, vol.~78, p.~042112, 2008.

\bibitem{Boh.Bra.Gir:16}
F.~{Bohnet-Waldraff}, D.~Braun, and O.~Giraud, ``Quantumness of spin-1
  states,'' {\em Physical Review A}, vol.~93, no.~1, 2016.

\bibitem{Gir.Bra.Bra:10}
O.~Giraud, P.~Braun, and D.~Braun, ``Quantifying quantumness and the quest for
  queens of quantum,'' {\em New J.{} Phys.}, vol.~12, p.~063005, 2010.

\bibitem{Bjo.Kli.Hoz.Gra.Leu.San:15}
G.~Bj{\"o}rk, A.~B. Klimov, P.~de~la Hoz, M.~Grassl, G.~Leuchs, and L.~L.
  S{\'a}nchez-Soto, ``Extremal quantum states and their {M}ajorana
  constellations,'' {\em Phys.{} Rev.{} A}, vol.~92, p.~031801, 2015.

\bibitem{Gol.Kli.Gra.Leu.San:20}
A.~Z. Goldberg, A.~B. Klimov, M.~Grass, G.~Leuchs, and L.~L. S{\'a}nchez-Soto,
  ``Extremal quantum states,'' {\em AVS Quantum Sci.}, vol.~2, p.~044701, 2020.

\bibitem{Mar:11}
D.~J.~H. Markham, ``Entanglement and symmetry in permutation symmetric
  states,'' {\em Phys. Rev.}, vol.~A 83, p.~042332, 2011.

\bibitem{Bjo.Gra.Hoz.Leu.San:15}
G.~Bj{\"o}rk, M.~Grassl, P.~de~la Hoz, G.~Leuchs, and L.~L. S{\'a}nchez-Soto,
  ``Stars of the quantum universe: extremal constellations on the {P}oincar\'e
  sphere,'' {\em Phys.{} Scr.}, vol.~90, p.~108008, 2015.

\bibitem{Mat.Kri.God.Lam.Sol.Bas:10}
P.~Mathonet, S.~Krins, M.~Godefroid, L.~Lamata, E.~Solano, and T.~Bastin,
  ``Entanglement equivalence of $n$-qubit symmetric states,'' {\em Phys. Rev.},
  vol.~A 81, p.~052315, 2010.

\bibitem{Ben.Zyc:17}
I.~Bengtsson and K.~\.{Z}yczkowski, {\em Geometry of Quantum States (2nd
  {E}d.)}.
\newblock Cambridge University Press, 2017.

\bibitem{Mar.Gir.Bra.Bra.Bas:10}
J.~Martin, O.~Giraud, P.~A. Braun, D.~Braun, and T.~Bastin, ``Multiqubit
  symmetric states with high geometric entanglement,'' {\em Phys.{} Rev.{} A},
  vol.~81, p.~062347, 2010.

\bibitem{Aul.Mar.Mur:10}
M.~Aulbach, D.~Markham, and M.~Murao, ``The maximally entangled symmetric state
  in terms of the geometric measure,'' {\em New J. Phys.}, vol.~12, p.~073025,
  2010.

\bibitem{Aul:12}
M.~Aulbach, ``Classification of entanglement in symmetric states,'' {\em Int.
  J. of Quantum Inf.}, vol.~10, no.~7, p.~1230004, 2012.

\bibitem{Zyc.Slo:01}
K.~\.Zyczkowski and W.~S{\l}omczy{\'n}ski, ``The {M}onge metric on the sphere
  and geometry of quantum states,'' {\em J. Phys. A: Math. Gen}, vol.~34,
  pp.~6689--6722, 2001.

\bibitem{Gan.Kus.Zyc:12}
W.~Ganczarek, M.~Ku{\'s}, and K.~\.{Z}yczkowski, ``Barycentric measure of
  quantum entanglement,'' {\em Phys.{} Rev.{} A}, vol.~85, no.~3, p.~032314,
  2012.

\bibitem{Ser.Bra:20}
E.~Serrano-Ens\'astiga and D.~Braun, ``Majorana representation for mixed
  states,'' {\em Physical Review A}, vol.~101, p.~022332, Feb. 2020.

\bibitem{Chr.Guz.Han.Ser:21}
C.~Chryssomalakos, E.~Guzm\'an-Gonz\'alez, L.~Hanotel, and
  E.~Serrano-Ens\'astiga, ``Stellar representation of multipartite
  antisymmetric states,'' {\em Commun.{} Math.{} Phys.{}}, 2021.
\newblock \texttt{arXiv:1909.02592}.

\bibitem{Mur:38}
F.~D. Murnaghan, {\em The theory of group representations}.
\newblock Johns Hopkins Press, Baltimore, 1962.

\bibitem{Mur:51}
F.~D. Murnaghan, ``A generalization of {H}ermite's law of reciprocity,'' {\em
  Proc.{} Natl.{} Acad.{} Sci.}, vol.~37, pp.~439--441, 1951.

\bibitem{Wei.Gol:03}
T.~Wei and P.~M. Goldbart, ``Geometric measure of entanglement and applications
  to bipartite and multipartite quantum states,'' {\em Phys. Rev. A}, vol.~68,
  p.~042307, 2003.

\bibitem{Hub.Kle.Wei.Gon.Guh:09}
R.~Hubener, M.~Kleinmann, T.-C. Wei, C.~Gonz\'alez-Guill\'en, and O.~G{\"u}hne,
  ``Geometric measure of entanglement for symmetric states,'' {\em Phys. Rev.
  A}, vol.~80, p.~032324, 2009.

\bibitem{Enr.Win.Zyc:16}
M.~Enr{\'{\i}}quez, I.~Wintrowicz, and K.~{\.{Z}}yczkowski, ``Maximally
  entangled multipartite states: A brief survey,'' {\em Journal of Physics:
  Conference Series}, vol.~698, p.~012003, Mar. 2016.

\bibitem{Chr.Her:17}
C.~Chryssomalakos and H.~Hern\'andez-Coronado, ``Optimal quantum rotosensors,''
  {\em Phys.{} Rev.{} A}, vol.~95, 2017.
\newblock article No.: 052125.

\bibitem{Mar.Wei.Gir:19}
J.~Martin, S.~Weigert, and O.~Giraud, ``Optimal {D}etection of {R}otations
  about {U}nknown {A}xes by {C}oherent and {A}nticoherent {S}tates,'' {\em
  {Quantum}}, vol.~4, p.~285, June 2020.

\bibitem{Ful.Har:04}
W.~Fulton and J.~Harris, {\em Representation {T}heory: {A} {F}irst {C}ourse}.
\newblock Springer, 2004.

\bibitem{Mol:97}
T.~Molien, ``Invarianten der linearen {S}ubstitutionsgruppen,'' {\em
  Sitzungsber.{} K{\"o}nigl.{} Preuss.{} Akad.{} Wiss.}, vol.~FdM28/115,
  pp.~1152--1156, 1897.

\bibitem{Wey:68}
H.~Weyl, ``Zur {D}arstellungstheorie und {I}nvariantenabz{\"a}hlung der
  projektiven, der komplex- und der {D}rehungsgruppe,'' {\em Ges.{} Abh.},
  vol.~Bd.{} III, pp.~1--25, 1968.

\bibitem{Pop.Vin:94}
V.~L. Popov and E.~B. Vinberg, ``Invariant theory,'' in {\em Algebraic Geometry
  IV} (A.~N. Parshin and I.~R. Shafarevich, eds.), Springer, 1994.

\bibitem{Spr:80}
T.~A. Springer, ``On the invariant theory of {$SU_2$},'' {\em Indagaciones
  Mathematicae (Proceedings)}, vol.~83, no.~3, pp.~339--345, 1980.

\bibitem{Bro.Coh:79}
A.~E. Brouwer and A.~M. Cohen, ``The {P}oincar{\'e} series of the polynomials
  invariant under ${SU}_2$ in its irreducible representation of degree $\leq$
  17,'' Tech. Rep. ZW 134/79, 1979.
\newblock url: \texttt{ir.cwi.nl/pub/6809}.

\bibitem{Syl:73}
J.~J. Sylvester, {\em Collected Math.{} Papers}.
\newblock Chelsea, 1973.

\bibitem{Pol.Sfe:16}
A.~P. Polychronakos and K.~Sfetsos, ``Composition of many spins, random walks
  and statistics,'' {\em Nucl.{} Phys.{} B}, vol.~913, pp.~664--693, 2016.

\bibitem{Her:54}
M.~Hermite, ``Th\'eorie de fonctions homog\`enes \`a deux ind\'etermin\'ees,''
  {\em The Cambridge and Dublin Mathematical Journal}, vol.~IX, pp.~172--217,
  1854.

\bibitem{Zac:92}
C.~K. Zachos, ``Altering the symmetry of wave functions in quantum algebras and
  supersymmetry,'' {\em Mod.{} Phys.{} Lett.{} A}, vol.~07, no.~18,
  pp.~1595--1600, 1992.

\bibitem{Cur.Kor.Zac:90}
T.~L. Curtright, T.~S. {Van Kortryk}, and C.~K. Zachos, ``Spin
  multiplicities,'' {\em Phys.{} Lett.{} A}, vol.~381, no.~5, pp.~422--427,
  1990.

\bibitem{Gya.Bar:18}
J.~A. Gyamfi and V.~Barone, ``On the composition of an arbitrary collection of
  {SU}(2) spins: an enumerative combinatoric approach,'' {\em Journal of
  Physics A: Mathematical and Theoretical}, vol.~51, p.~105202, Feb 2018.

\bibitem{Lit:50}
D.~Littlewood, {\em The {T}heory of {G}roup {C}haracters and {M}atrix
  {R}epresentations of {G}roups}.
\newblock Oxford University Press, 1950.

\bibitem{Wyb:70}
B.~G. Wybourne, {\em Symmetry principles and atomic spectroscopy}.
\newblock Wiley-Interscience, 1970.

\bibitem{Var.Mos.Khe:88}
D.~Varshalovich, A.~Moskalev, and V.~Khersonskii, {\em Quantum {T}heory of
  {A}ngular {M}omentum}.
\newblock World Scientific, 1988.

\bibitem{Gel.Kap.Zel:94}
I.~M. Gelfand, M.~M. Kapranov, and A.~V. Zelevinsky, {\em Discriminants,
  {R}esultants and {M}ultidimensional {D}eterminants}.
\newblock Birkh\"auser, 1994.

\bibitem{Wyb:69}
B.~G. Wybourne, ``Hermite's reciprocity law and the angular-momentum states of
  equivalent particle configurations,'' {\em J.{} Math.{} Phys.}, vol.~10,
  pp.~467--471, 1969.

\bibitem{Mur:62}
F.~D. Murnaghan, {\em The unitary and rotation groups}.
\newblock Spartan Books, Washington D.{} C.{}, 1962.

\bibitem{Olv:99}
P.~J. Olver, {\em Classical Invariant Theory}.
\newblock London Mathematical Society Student Texts 44, Cambridge University
  Press, 1999.

\bibitem{Bes:78}
A.~L. Besse, {\em Manifolds all of whose geodesics are closed}.
\newblock Springer-Verlag, 1978.

\bibitem{Tam.Wei.Par:09}
S.~Tamaryan, T.-C. Wei, and D.~Park, ``Maximally entangled three-qubit states
  via geometric measure of entanglement,'' {\em Phys.{} Rev.{} A}, vol.~80,
  p.~052315, 2009.

\bibitem{Gol.Jam:18}
A.~Z. Goldberg and F.~V. James, ``Quantum-limited euler angle measurements
  using anticoherent states,'' {\em Phys. Rev. A}, vol.~98, p.~032113, 2018.

\bibitem{Bra.Cav:94}
S.~L. Braunstein and C.~M. Caves, ``Statistical distance and the geometry of
  quantum states,'' {\em Phys. Rev. Lett.}, vol.~72, p.~3439, May 1994.

\bibitem{Gre.Hor.Zei:89}
D.~M. Greenberger, M.~A. Horne, and A.~Zeilinger, ``Going beyond {B}ell's
  theorem,'' in {\em {B}ell's theorem, quantum theory and conceptions of the
  universe}, pp.~69--72, Springer, 1989.

\bibitem{Sha.Rem:13}
I.~R. Shafarevich and A.~O. Remizov, {\em Linear {A}lgebra and {G}eometry}.
\newblock Springer, 2013.

\bibitem{Bag.Bas.Mar:14}
D.~Baguette, T.~Bastin, and J.~Martin, ``Multiqubit symmetric states with
  maximally mixed one-qubit reductions,'' {\em Phys.{} Rev.{} A}, vol.~90,
  p.~032314, 2014.

\bibitem{Bon.Sch.Haa:71}
R.~Bonifacio, P.~Schwendimann, and F.~Haake, ``Quantum statistical theory of
  superradiance. {I},'' {\em Phys.{} Rev.{} A}, vol.~4, p.~302, 1971.

\end{thebibliography}

\end{document}